\documentstyle[12pt,epsf]{article}
\begin{document}
\newcommand{\thickbar}{\kern1pt\hbox{\vphantom{$|$}\vrule width 1pt}\kern1pt}
\rightline{RU01-14-B}
\title{\bf\LARGE{Inverse Scattering, the Coupling Constant Spectrum,
and the Riemann Hypothesis}}
\author{N.N. Khuri\\
Department of Physics\\
The Rockefeller University, New York, New York 10021}
\date{}
\begingroup
\renewcommand{\newpage}{}
\maketitle
\endgroup
\thispagestyle{empty}
\setcounter{page}{0}
\newpage
\begin{center}
{\bf{Abstract}}
\end{center}
{\small{
It is well known that the s-wave Jost function for a 
potential, $\lambda V$, is an entire function of $\lambda$ with 
an infinite number of zeros extending to infinity.  For a 
repulsive V, and at zero energy, these zeros of the
"coupling constant", $\lambda$, will all be real and 
negative, $\lambda_n(0)<0$.  By rescaling $\lambda$, such
that $\lambda_n<-1/4$, and changing variables to s, 
with $\lambda=s(s-1)$, it follows that as a 
function of $s$ the Jost function has only zeros on the 
line $s_n=\frac{1}{2}+i\gamma_n$.  Thus finding a repulsive
V whose coupling constant spectrum coincides with the Riemann 
zeros will establish the Riemann hypothesis, but this 
will be a very difficult and unguided search.

In this paper we make a significant enlargement of 
the class of potentials needed for a 
generalization of the above idea.  We also make this new 
class amenable to construction via inverse scattering 
methods.  We show that all one needs is a one parameter 
class of potentials, $U(s;x)$, which are analytic in 
the strip, $0\leq Res\leq 1$, $Im s>T_o$, and in addition 
have an asymptotic expansion in powers of $[s(s-1)]^{-1}$, i.e.
$U(s;x)=V_o(x)+gV_1(x)+g^2V_2(x)+ ... +O(g^N)$,
with $g=[s(s-1)]^{-1}$.  The potentials $V_n(x)$ are 
real and summable.  Under suitable conditions on 
the $V_n's$ and the $O(g^N)$ term we show that the 
condition, $\int_{o}^{\infty}|f_o(x)|^2V_1(x)dx\neq 0$,
where $f_o$ is the zero energy and $g=0$ Jost function
for $U$, is sufficient to guarantee that the zeros $g_n$
are real and hence $s_n=\frac{1}{2}+i\gamma_n$, for $\gamma_n\geq T_o$.

Starting with a judiciously chosen Jost function, $M(s,k)$, 
which is constructed such that $M(s,0)$ is Riemann's $\xi(s)$ 
function, we have used inverse scattering methods to actually 
construct a $U(s;x)$ with the above properties.
By necessity we had to generalize inverse methods 
to deal with complex potentials and a non-unitary S-matrix.  This 
we have done at least for the special cases under consideration.

For our specific example, $\int_{o}^{\infty}|f_o(x)|^2V_1(x)dx=0$, 
and hence we get no restriction on $Img_n$ or $Re s_n$.  The reasons
for the vanishing of the above integral are given, and they 
give us hints on what one needs to proceed further.  The problem 
of dealing with small but non-zero energies is also discussed.}}

\newpage
\hsize=6in
\hoffset=-.5in
\baselineskip=2\baselineskip
\newcounter{chanum}
\newcounter{eqnnum1}
\renewcommand{\theequation}{\arabic{chanum}.\arabic{eqnnum1}}
\setcounter{chanum}{1}
\setcounter{eqnnum1}{1}
\noindent{\bf\Large{I. Introduction}}

Many physicists have been intrigued by the Riemann conjecture 
on the zeros of the zeta function.  The main reason for 
this is the realization that the validity of the hypothesis 
could be established if one finds a self-adjoint operator whose 
eigenvalues are the imaginary parts of the non-trivial
zeros.  The hope is that this operator could be the Hamiltonian 
for some quantum mechanical system.  Results by 
Dyson$^{1}$, and Montgomery$^{2}$ first made the situation 
more promising.  The pair distribution between neighboring zeros 
seemed to agree with that obtained for the eigenvalues of a large 
random hermitian matrix.  But later numerical work showed 
correlations between distant spacings do not agree with those of a 
random Hermitian matrix.  The search for such a Hamiltonian in 
physical problems has eluded all efforts.  Berry$^{3}$ has 
suggested the desired Hamiltonian could result from quantizing 
some chaotic system without time reversal symmetry.  This seems to 
be in better agreement with numerical work on the correlations of 
the Riemann zeros, but one is still far from even a model or 
example.  It is useful to explore new ideas.

Our choice for this paper is an idea, due to Chadan$^{4}$.  In 
this approach one tries to relate the zeros of the Riemann zeta 
function to the "coupling constant spectrum'' of the zero 
energy, $S$-wave, scattering problem for repulsive 
potentials.  We sketch this idea briefly.

The Schrodinger equation on $x\in[0,\infty)$ is
\begin{equation}
-\frac{d^2f}{dx^2}(\lambda;k;x)
+\lambda V(x)f(\lambda;k;x)=k^2f(\lambda;k;x),
\end{equation}
where $k$ is the wave number, $\lambda$ a 
parameter physicists call the coupling constant, $V(x)$ is 
a real potential satisfying an integrability condition 
as in Eq. (2.2) below, and $f$ is the Jost solution 
determined by a boundary condition at infinity, 
$(e^{-ikx}f)\rightarrow1$ as $x\rightarrow+\infty$.  
The Jost function, $M(\lambda;k)$, is defined by 
$lim_{x\rightarrow0}f(\lambda;k;x)=M(\lambda;k)$.  
It is well known that $M$ is also the Fredholm 
determinant of the Lippmann-Schwinger scattering integral 
equation for $S$-waves.  Both $f(\lambda;k;x)$  and 
$M(\lambda;k)$ are for any fixed $x\geq0$, analytic in 
the product of the half plane, $Imk>0$, and any large 
bounded region in the $\lambda$ plane.  In fact it is 
known that for any fixed $k$, $Imk\geq0$, $M(\lambda;k)$ is 
entire in $\lambda$ and of finite order.  Thus $M(\lambda;k)$ 
has an infinite number of zeros, $\lambda_n(k)$, 
with $\lambda_n(k)\rightarrow\infty$ as $n\rightarrow\infty$.

Starting with Eq. (1.1), and its complex 
conjugate with $k=i\tau$, $\tau>0$, and setting 
$\lambda=\lambda_n(i\tau)$ we obtain
\addtocounter{eqnnum1}{1}
\begin{equation}
[Im\lambda_n(i\tau)]
\int_{o}^{\infty}|f(\lambda_n(i\tau);i\tau;x)|^2V(x)dx=0.
\end{equation}
For the class of potentials we deal with 
$V=O(e^{-mx})$ as $x\rightarrow\infty$.  Thus 
we can take the limit $\tau\rightarrow 0$, and get
\addtocounter{eqnnum1}{1}
\begin{equation}
[Im\lambda_n(0)]\int_{o}^{\infty}|f(\lambda_n(0);0;x)|^2V(x)dx=0.
\end{equation}
Hence for repulsive potentials, $V(x)\geq0$, all the 
zeros $\lambda_n(0)$ are real.  For any $\tau,\tau>0$, the same 
is true for all $\lambda_n(i\tau)$.  But $\lambda_n(i\tau)$ must 
be negative since the potential $[\lambda_n(i\tau)V]$ will have 
a bound state at $E=-\tau^2$, and that could not happen 
if $V\geq0$ and $\lambda_n(i\tau)>0$.  Hence by 
continuity, $\lambda_n(0)$, for all $n$, is real and 
negative.$^5$  The zero energy coupling constant 
spectrum, $\lambda_n(0)$, lies on the negative real 
line for $V\geq0$.  Chadan's idea is very simple.  He 
introduces a new variable,$s$, and defines
\addtocounter{eqnnum1}{1}
\begin{equation}
\lambda\equiv s(s-1).
\end{equation}
Thus one can write
\addtocounter{eqnnum1}{1}
\begin{equation}
M(\lambda,0)=M(s(s-1);0)\equiv\chi(s).
\end{equation}
It is easy to see now that for $|Ims|>1$, the 
zeroes,$s_n$, of $\chi(s)$ are all such that
\addtocounter{eqnnum1}{1}
\begin{equation}
s_n=\frac{1}{2}+i\gamma_n;\;\;\;\lambda_n(0)\equiv s_n(s_n-1).
\end{equation}

The problem is actually somewhat simplified by 
noting that first we do not need the condition $\lambda_n<0$ as 
long as we restrict ourselves to the 
strip $0\leq Res\leq 1$, and $Ims>1$.  Second, it 
is sufficient to prove that the integral in Eq. (1.3) 
does not vanish.  Thus, one does not need a fully 
repulsive potential for the Riemann problem.

One might comment that it is very difficult to 
find a potential with $\lambda_n(0)=s_n(s_n-1)$ 
and $s_n=\frac{1}{2}\pm i\gamma_n$, $s_n$ being 
the Riemann zeros.  But it is probably as difficult 
as finding a hermitian operator whose eigenvalues 
are $\gamma_n$.  Indeed the latter may be impossible 
without introducing chaotic systems.

The results mentioned above also apply when 
$V=V_o+\lambda V_1$, with only $V_1\geq0$, and $V_0,V_1$ 
both real and satisfying Eq. (2.2) and with certain 
restrictions on $V_o$.  This remark leads directly to the 
basic idea of this paper.

The objective of this paper is to show that the coupling 
constant approach can be significantly simplified and 
made amenable to inverse scattering methods.

Our first remark is that one does not need a 
potential, $V=V_o+\lambda V_1$, depending linearly on the 
coupling parameter $\lambda$.  Given a one parameter family 
of complex potentials, $U(s;x),x\in[0,\infty)$, which for 
fixed $x$ are analytic in $s$ in the strip, 
$0\leq Re s\leq 1$, $Ims>T_o>2$, we can following 
similar arguments as above, obtain, for $s=s_n$, $s_n$ being 
a zero of the zero energy Jost function,
\addtocounter{eqnnum1}{1}
\begin{equation}
\int|f(s_n;0;x)|^2[ImU(s_n;x)]dx\equiv 0,
\end{equation}
where $f$ is the zero energy Jost solution evaluated at $s=s_n$.

Next, suppose in addition to the above properties, $U$ has 
an asymptotic expansion in inverse powers of s, actually 
better, $s(s-1)$, i.e. 
\addtocounter{eqnnum1}{1}
\begin{equation}
U(s,x)=V_0(x)+gV_1(x)+g^2V_2(x)+ ... +g^NV_R^{(N)}(g;x);
\end{equation}
where
\addtocounter{eqnnum1}{1}
\begin{equation}
g\equiv\frac{1}{s(s-1)}.
\end{equation}
Under suitable conditions on the $V_n(x)$ and 
estimates of the $O(g^N)$ term, and its phase, one gets again,
\addtocounter{eqnnum1}{1}
\begin{equation}
[Img_n]\int_{o}^{\infty}|f(0;0;x)|^2V_1(x)dx=0,
\end{equation}
with $g_n=[s_n(s_n-1)]^{-1}$, the $s_n$'s are the 
zeros of $M(s,0)$ the zero energy Jost function, and $f(g;k;x)$ 
is the Jost solution with the full $U$.  The result (1.10) is 
only established for zeros with $Ims_n>T_o$, where $T_o$ is 
large enough for the $V_1$ contribution to Eq. (1.7) to 
dominate the integral in Eq. (1.7).  However, this is 
sufficient since the Riemann hypothesis has already been 
proved for zeros with $|Ims_n|<T$, where $T$ could be as large as $10^5$. 
 
Again, all we need for, $s_n=\frac{1}{2}+i\gamma_n$, is 
to have the integral in (1.10) not vanishing.  In the end 
only the properties of $V_1$ matter.  

In this paper we will use inverse scattering methods, 
albeit for complex potentials, to actually prove the 
existence of such a $U(s;x)$.  This potential, by 
construction, has the additional property that the zero 
energy Jost function is Riemann's $\xi$ function,
\addtocounter{eqnnum1}{1}
\begin{equation}
 \lim_{k\rightarrow 0}M(s;k)\equiv2\xi(s).
\end{equation}
We will also give explicit expressions for $V_o,V_1,V_2$, 
and bounds on $V_R^{(N)}$.

The difficult point turns out to be that in our specific example,
\addtocounter{eqnnum1}{1}
\begin{equation}
\int_{o}^{\infty}|f(0;0;x)|^2V_1(x)dx\equiv 0.
\end{equation}

Thus we get no information on $[Img_n]$, or 
$Re(s_n-\frac{1}{2})$.  We shall discuss what one 
needs to proceed further.  This will require working 
with small, but non-zero energy values.

We start by introducing a special class Jost 
functions, $M^\pm$, which depend on an extra 
parameter $\nu=s-\frac{1}{2}$, with the property that 
the zero energy limit, $lim_{k\rightarrow0}M^{\pm}(\nu,k)
=2\xi(\nu+\frac{1}{2})$. For fixed $\nu$, the Jost functions 
are taken to be of the
Martin$^8$ type, i.e. having cut plane analyticity 
in the momentum variable $k$.  This is the class of 
Jost functions that results when the potential is a 
superposition of Yukawa potentials.  We then use inverse 
scattering methods to prove the existence of a complex 
potential $U(\nu,\tau)$ which is determined uniquely by 
the initial $S$-matrix.  We do carry out the analysis 
for $\nu$ in the truncated critical 
strip, i.e. $-\frac{1}{2}<Re\nu<\frac{1}{2}$, 
and $Im\nu>T_o$, with $T_o>16\pi^2.$  This is of course 
the domain most relevant to the Riemann problem.  Standard 
techniques of inverse scattering are not immediately 
applicable, because $S(\nu,k)$ does not satisfy the reality 
condition, and is not unitary for complex $\nu$.  However, 
we shall see that in our specific case we can bypass these 
difficulties and carry out an inverse scattering procedure 
anyway.  We have attempted to make the paper self contained, 
and do not rely on results that need the unitarity of $S$ in the proof.

In Section II, we give a brief review of relevant scattering 
theory results intended for mathematicians not familiar 
with them.  This review also helps define our physics terminology.

Section III is devoted to the introduction of our special 
class of Jost functions, $M^{\pm}(\nu;k)$.  Following that, 
in Section IV, we briefly discuss the real $\nu$ case, which 
is a standard inverse scattering case covered by well known 
results.  This section is instructive, even though real $\nu$ 
is uninteresting for the Riemann problem.  The next step, 
section V, is to study in more detail the properties 
of $M^{\pm}$.  The main result is an asymptotic expansion 
in powers of a variable, $g\equiv(\nu^2-\frac{1}{4})^{-1}$, which gives
\addtocounter{eqnnum1}{1}
\begin{equation}
M^{(\pm)}=M_o^{(\pm)}(k)+gM_1^{(\pm)}(k)+g^2M_2^{(\pm)}+...
+g^NR_N^{(\pm)}(g;k)
\end{equation}
Here all the $M_n^{(\pm)}$ can be computed exactly 
via recursion formulae, and in addition,
they satisfy $[M_n^{(+)}(k)]^*=M_n^{(-)}(k)$, 
and $M_n^{(+)}(-k)=M_n^{(-)}(k)$.  The remainder 
functions, $R^{(\pm)}_N$, are given explicitly and 
are $O(g)$ as $g\rightarrow 0$.

In Section VI, with fixed $\nu$ in the strip, we 
determine the number and positions of zeros in the 
upper half $k$-plane.  It turns out that there is at 
most \underline{one} such zero and it lies close to the 
origin.  In fact we can give a good estimate of its position.

Section VII is devoted to the study of the 
case $|\nu|\rightarrow\infty$, i.e. $|g|\rightarrow 0$. 
Here $M^{(\pm)}(\nu,k)\rightarrow M^{(\pm)}_o(k)$ which 
is a known rational function in $k$.  This leads to an 
exactly soluble Marchenko equation and an exact result 
for the corresponding $V_o(x)$.

Section VIII is devoted to proving the existence of solutions 
of the Marchenko Equation for our specific class 
of $S$-matrices.  With the resulting Marchenko 
operator, $A(\nu;x,y)$, which is now complex, 
we proceed to define in the standard way a 
potential $U(\nu,x)$ and corresponding 
Jost solutions, $f^{(\pm)}(\nu;k,x)$, of 
the Schrodinger equation.  Finally, we check 
directly that indeed $f^{(\pm)}$ are solutions 
of the Schrodinger equation with the desired 
asymptotic properties.  The main difference from 
the standard case is that $U(\nu;x)$ is now 
complex unless $\nu$ is purely imaginary.

In Section IX we discuss the case $\nu=it$, $t$ real.  This 
is a standard inverse problem with $S(it,k)$ unitary 
for $k$ real, and the resulting $U(it;x)$ is real. 

More detailed properties of $V(\nu,x)$ are given in 
Section X.  There we give an asymptotic 
expansion, $V(\nu,x)=V_o(x)+gV_1(x)+g^2V_2(x)+...+g^NV_R^{(N)}(g,x)$ 
with all $V_n's$ real and all representable by  
superpositions of Yukawa potentials.  Also $V_n(x)$ 
is continuous and differentiable 
for $x\epsilon[0,\infty)$, and $V_n(0)$ is finite.  
For completeness we calculate $V_1(x)$ explicitly, 
and indicate how $V_n(x)$, $n>1$, can easily be 
computed.  We also give some needed properties 
of $V^{(N)}_R$ and of $[ImV^{(N)}_R]$ for small $Re\nu$.

In Section XI, we study the zeros, $\nu_n(k)$, 
of $M^-(\nu,k)$ for small fixed $k$ with $Imk\geq 0$.  
We prove that $\nu_n(0)$ are the standard Riemann zeros, 
and also that $|\nu_n(k)-\nu_n(0)|=O(k^{1/p})$ for 
small $k$.  Here $p$ is the multiplicity of the Riemann 
zero $\nu_n=\nu_n(0)$.  We also prove that any Riemann 
zero, $\nu_j$, is the limit of a zero of 
$M^{(-)}(\nu;k)$, $\nu_j(k)$, as $k\rightarrow 0$.

Finally, in Section XII we discuss the relation 
of our potential, $V(g;x)$, and its Jost solutions 
to the Riemann hypothesis.  We prove that in this 
case $\int_{o}^{\infty}|f(0;0;x)|^2V_1(x)dx=0$, and 
hence no information on the Riemann hypothesis can 
result directly from this example at zero energy.  But 
the reasons for the failure are clear, and they indicate 
the properties of a desired Jost function that will be 
sufficient to make the important step.  The fact that 
one can set $k=i\tau$, $\tau>0$ but small, and try to 
prove the hypothesis for $\nu_n(\tau)$, $\tau$ arbitrarily 
small, but $\tau\neq0$, provides a significant 
simplification of the problem.  

\newpage
\vspace{.25in}
\setcounter{chanum}{2}
\setcounter{eqnnum1}{1}
\noindent{\bf\Large{II.  A Sketch of Scattering Theory}}
\vspace{.25in}

This section is intended to facilitate the reading 
of this paper by those mathematicians (or physicists) 
who are not familiar with elementary scattering theory 
in quantum mechanics.  At the end of this section we will 
give a list of books and review papers where more 
information can be obtained.

The Schrodinger equation for s-waves is given by
\begin{equation}
-\frac{d^2f}{dx^2}+gV(x)f=k^2f, ~~k=\kappa+i\tau.
\end{equation}

Here $x\in[0,\infty),V(x)$ is real, $g$ is a parameter 
that physicists call a coupling constant.  The reason for 
introducing it will become apparent below.  One studies the 
class of real potentials, $V(x)$, which are locally 
summable functions and satisfy the condition,

\addtocounter{eqnnum1}{1}
\begin{equation}
\int_{o}^{\infty}x|V(x)|e^{\alpha x}dx=C<\infty, ~~0\leq\alpha\leq m.
\end{equation}

For scattering theory the important solutions 
of Eq. (2.1) are the so called Jost 
solutions.$^{9}$.  These are the two linearly 
independent solutions, $f^{(\pm)}(g,k,x)$ with 
boundary values at infinity given by
\addtocounter{eqnnum1}{1}
\begin{equation}
\lim_{x\rightarrow\infty}e^{\pm ikx}f^{(\pm)}=1
\end{equation}

Using the method of variation of parameters, we can 
replace Eq. (2.1) and the condition (2.3) by an integral equation:
\addtocounter{eqnnum1}{1}
\begin{equation}
f^{(\pm)}(g;k;x)=e^{\mp ikx}
+g\int_{x}^{\infty}\frac{\sin k(x'-x)}{k}V(x')f^{(\pm)}(g;k;x')dx'.
\end{equation}

Starting with the papers of Jost$^9$ and 
Levinson$^{10}$, the existence of solutions 
to Eq. (2.4) and their properties have been 
well established for $V(x)$ satisfying the condition (2.2).

The basic input needed is the upper bound on the kernel,
\addtocounter{eqnnum1}{1}
\begin{equation}
\left|\frac{\sin k(x'-x)}{k(x'-x)}\right |\leq C_1\frac{e^{|\tau||x'-x|}}
{1+|k||x'-x|}, ~~Im k\equiv \tau,
\end{equation}
where $C_1$ is $O(1)$. With this bound and the 
bound (2.2) one proves the absolute convergence of 
the iterative series of the Volterra equation (2.4) for 
any $x\geq 0$, and $k$ with $Im k> -\frac {m}{2}$ 
for $f^{(-)}$, and $Im k< \frac{m}{2}$ for $f^{(+)}$.  Also 
it is easy to prove that for any finite $g$ 
and $x\geq 0$, $f^{(+)}(g;k;x)$ is an analytic 
function of $k$ for $Im k<\frac{m}{2}$.  
Similarly, $f^{(-)}(g;k;x)$ is analytic 
in $Im k>-\frac{m}{2}$.  In addition, for $k$ in the 
analyticity domain, the power series in $g$ obtained 
by iterating Eq. (2.4) is absolutely and uniformly 
convergent for $g$ inside any finite region in 
the $g$-plane.  Thus both $f^{\pm}(g;k;x)$ are entire functions of $g$.

The scattering information is all contained in the Jost 
functions, denoted by $M^{(\pm)}$$(g;k)$ and defined by,
\addtocounter{eqnnum1}{1}
\begin{equation}
M^{(\pm)}(g;k)\equiv \lim_{x\rightarrow 0}f^{(\pm)}(g;k;x).
\end{equation}
Both limits in Eq. (2.6) exist for finite $|g|$, 
and $k$ in the respective domain of analyticity, for 
all potentials satisfying the condition (2.2).  The S-matrix is given by,
\addtocounter{eqnnum1}{1}
\begin{equation}
S(g;k)\equiv \frac{M^{(+)}(g;k)}{M^{(-)}(g;k)}.
\end{equation}
For real $g$ and $Im k>0$, $M^{(-)}(g,k)$ has no 
zeros except for at most a finite number on the 
imaginary k-axis.  These zeros, $k_n=i\tau_n$, give 
the point spectrum of the Hamiltonian of (2.1) 
with $E_n=-\tau_n^2$.  Their number cannot exceed the 
value of the integral $\int_{o}^{\infty}x|V|dx$, a result 
due to Bargmann$^{11}$.

Another important property of $M^{(-)}(g;k)$ was 
first obtained by Jost and Pais$^{12}$.  The regular 
solution of Eq. (2.1), $\phi(g;k;x)$, with $\phi(g;k;0)=0$, is
\addtocounter{eqnnum1}{1}
\begin{equation}
\phi(g;k;x)\equiv \frac{1}{2ik}[M^{(+)}(g;k)f^{(-)}
(g;k;x)-M^{(-)}(g;k)f^{(+)}(g;k;x)].
\end{equation}

The solution $\phi$ satisfies a Fredholm type integral 
equation which, for potentials satisfying (2.2), was 
studied in ref.12.  Jost and Pais demonstrate 
explicitly that $M^{(-)}(g,k)$ is identical to 
the Fredholm determinant of the scattering integral 
equation for $\phi$.  Hence for any 
fixed $k$, with $Im k>0$, the zeros of $M^{(-)}(g;k)$ 
in the $g$-plane, $g_n(k)$, give 
the ``coupling constant eigenvalues'' at which the 
homogeneous Fredholm equation has 
solutions, $\phi=g_n(k)K\phi$.  Since, $M^{(-)}(g,k)$ is 
an entire function of finite order in $g$, the 
sequence $g_1(k),g_2(k), ..., g_n(k)$ tends to 
infinity as $n\rightarrow\infty$.

For the purposes of this paper a result of Meetz$^5$ is 
instructive.  Let us
consider a potential which 
is repulsive, i.e. $V>0$ for all $x\in[0,\infty)$.  
Then for $k=i\tau, \tau>0$ the coupling constant 
spectrum, $g_n(i\tau)$, is real and negative.  This 
result is implicitly contained also in ref. 12.

In this brief review we need to make an important 
remark about complex potentials, $V\neq V^*$.  Mathematicians 
and mathematical physicists often ignore these 
potentials.  The Hamiltonian is no longer self-adjoint 
if $V\neq V^*$, with $g=1$.  But physicists, especially 
those who work on nuclear physics, do not have such a 
luxury.  There are many interesting and useful models, 
especially in nuclear physics, where $V$ is complex.  
Of course, the general and beautiful results which hold 
for real $V$ do not all apply for complex $V$.  But many 
survive, and one has just to be careful which to use and 
to establish alternative ones when needed.

There are many books that cover inverse scattering.  But for 
the purposes of this paper we recommend the book 
of Chadan and Sabatier$^{13}$, since it also 
discussed the superposition of Yukawa case and the 
Martin results.  For the standard results on inverse 
scattering the review paper by Faddeev$^{14}$ is highly recommended.
 
\newpage
\vspace{.25in}
\setcounter{chanum}{3}
\setcounter{eqnnum1}{1}
\noindent{\bf\Large{III.  A Special Class of Jost Functions}}

In this section we will combine two results whose 
progeny could not be more different to obtain a representation 
for a class of Jost functions that we shall study in detail.  The 
first is Martin's representation for the Jost functions of the 
class of potentials that can be represented as a Laplace 
transform.  The second is Riemann's formula for the 
function $\xi(s)$ defined below.

Starting fourty years ago, physicists$^{15,16}$, for 
reasons not relevant to this paper, studied the class of 
potentials that in addition to satisfying Eq. (2.2) have a 
Laplace transform representation, i.e. for all $x>0$,
\begin{equation}
V(x)=\int_{m}^{\infty}C(\alpha)e^{-\alpha x}d\alpha, ~~m>0,
\end{equation}
where $C(\alpha)$ is summable and restricted to 
satisfy $\int_{m}^{\infty}|C(\alpha)|\alpha^{-2}d\alpha<\infty$.  
This last condition guarantees that $x|V(x)|$ is integrable at $x=0$.

For these potentials Martin$^{17}$ proved that the Jost 
functions $M^{(\pm)}(k)$ have the representation
\addtocounter{eqnnum1}{1}
\begin{equation}
M^{(\pm)}(k)=1+\int_{\frac{m}{2}}^{\infty}\frac{w(\alpha)}
{\alpha\pm ik}d\alpha.
\end{equation}
Here $w$ is real and summable and is such that 
$M^{(\pm)}\rightarrow 1$ as $|k|\rightarrow\infty$.  We 
have set $g=1$ here.  Note that not any arbitrarily chosen 
summable $w(\alpha)$ is acceptable.  $M^{(-)}(k)$ must have 
no zeros for $Im k>0$ except for a finite number on the 
imaginary $k$-axis corresponding to the point spectrum.

For our purposes here we choose a specific family of 
functions $M^{(\pm)}(\nu;k)$ defined such that
\addtocounter{eqnnum1}{1}
\begin{equation}
M^{(\pm)}(\nu;0)\equiv2\xi(\nu+\frac{1}{2}),
\end{equation}
where
\addtocounter{eqnnum1}{1}
\begin{equation}
\xi(s)=\frac{1}{2}s(s-1)\pi^{-\frac{s}{2}}\Gamma(\frac{s}{2})\zeta(s)
\end{equation}
and
\addtocounter{eqnnum1}{1}
\begin{equation}
\zeta(s)=\sum_{n=1}^{\infty}n^{-s};~~Re s>1.
\end{equation}

Riemann's formula for $\xi(s)$ defines an entire function 
of order one in $s$, and is given by
\addtocounter{eqnnum1}{1}
\begin{equation}
2\xi(s)=1+s(s-1)\int_{1}^{\infty}\psi(\alpha)
[\alpha^{\frac{s}{2}-1}+\alpha^{\frac{-1}{2}-\frac{s}{2}}]d\alpha,
\end{equation}
where
\addtocounter{eqnnum1}{1}
\begin{equation}
\psi(\alpha)=\sum_{n=1}^{\infty}e^{-\pi n^2\alpha};~~\alpha\geq 1.
\end{equation}
We also have the symmetry relation $\xi(s)=\xi(1-s)$.

For convenience we define the variable, $\nu$, as
\addtocounter{eqnnum1}{1}
\begin{equation}
s\equiv \frac{1}{2}+\nu
\end{equation}

With this variable $\xi(\frac{1}{2}+\nu)$ is symmetric 
in $\nu$, and we have
\addtocounter{eqnnum1}{1}
\begin{equation}
2\xi(\frac{1}{2}+\nu)=1+(\nu^2-\frac{1}{4})
\int_{1}^{\infty}\psi(\alpha)\alpha^{-\frac{3}{4}}
[\alpha^\frac{\nu}{2}+\alpha^\frac{-\nu}{2}]d\alpha.
\end{equation}
Our starting point is to define two functions, $M^{\pm}(\nu;k)$, as

\addtocounter{eqnnum1}{1}
\begin{equation}
M^{(\pm)}(\nu;k)\equiv 1+(\nu^2-\frac{1}{4})
\int_{1}^{\infty}\frac{\psi(\alpha)\alpha^{\frac{1}{4}}
[\alpha^{\frac{\nu}{2}}+\alpha^{\frac{-\nu}{2}}]}{(\alpha\pm ik)}d\alpha
\end{equation}
This definition holds for any finite, real or 
complex, $\nu$, and for any $k$ excluding the cuts on 
the imaginary $k$-axis, $k=i\tau$, $1\leq\tau<\infty$, 
for $M^{(+)}$, and $-1\geq\tau>-\infty$, for $M^{(-)}$.

Obviously, we have
\addtocounter{eqnnum1}{1}
\begin{equation}
M^{(\pm)}(\nu;0)\equiv 2\xi(\frac{1}{2}+\nu).
\end{equation}
In addition, the fact that $\psi(\alpha)=O(e^{-\pi\alpha})$ 
as $\alpha\rightarrow+\infty$, guarantees that for any finite $|\nu|$
\addtocounter{eqnnum1}{1}
\begin{equation}
\lim_{|k|\rightarrow\infty}M^{\pm}(\nu;k)=1.
\end{equation}
This is true along any direction in the complex $k$ plane 
excluding the pure imaginary lines.  But even for 
$arg k=\pm\frac{\pi}{2}$ the limit holds using standard results.
The immediate question that faces us at this stage is: 
for which regions in the $\nu$-plane, if any, can one use 
the functions $M^{(\pm)}(\nu;k)$ defined in Eq. (3.10) as 
Jost functions and proceed to use the resulting 
S-matrix, $S(\nu;k)$, as the input in an inverse 
scattering program.

There are two issues involved.  The first, and most 
important, is to make sure that $M^{(-)}(\nu;k)$, has 
no complex zeros in $k$ for $Imk>0$, except for a finite 
number on the imaginary axis.  This is not true for 
any $\nu$.  But fortunately for the set of $\nu$'s most 
important to the Riemann hypothesis $M^{(-)}(\nu;k)$ has at 
most one zero close to the origin with $Imk>0$.  This will be 
shown in Section VI.

The second issue relates to the question of reality.  For real 
potentials $V$ and real $k$ we have the relations 
$[M^{(+)}(k)]^*=M^{(-)}(k)$, and $|S(k)|=1$.  Clearly 
for complex $\nu$ this does not hold for $M^{(\pm)}(\nu;k)$.  
However, we will prove that for those values of $\nu$ in the 
truncated critical strip, one can still carry out the inverse 
scattering program and obtain a unique and well defined 
$V(\nu;x)$ which of course now could be complex.  Since 
the old results of inverse scattering theory all use the 
fact that $|S(k)|=1$, we have to go back to square one and 
prove every step anew for the present case.  Our task is 
tremendously simplified by the fact that even 
though $S(\nu;k)$ is not unitary, we still 
have $|S(\nu;k)|=1+O(1/|\nu|^2)$, and we are 
only interested in $|\nu|>10^3$. 
\newpage
\vspace{.25in}
\setcounter{chanum}{4}
\setcounter{eqnnum1}{1}
\noindent{\bf\Large{IV. The Real $\nu$ Case}}

For potentials satisfying the 
representation (3.1), Martin$^{19}$, in addition 
to the results summarized in Eq. (3.2), developed an 
iterative scheme which enables one to reconstruct the 
measure, $C(\alpha)$ in Eq.(3.1) from the knowledge of 
the discontinuity of $S(k)$ along the branch cut on the 
imaginary axis, $k=i\tau$, $\frac{m}{2}\leq\tau<\infty$.  This 
gives an inverse scattering method that at first sight 
looks quite different from the standard 
ones of Gelfand-Levitan and Marchenko.  The relation 
between these two methods was first clarified by 
Gross and Kayser$^{20}$ and independently 
by Cornille$^{21}$.  They showed that for 
potentials of the form (3.1) the Marchenko kernel 
is a Laplace transform of the discontinuity 
of $S(k)$, and they carried out an extensive 
analysis of the relation between Martin's and 
Marchenko's method.  These results were reviewed and 
enlarged in a more recent paper by the author$^{22}$.

For $\nu$ real and $|\nu|>\frac{1}{2}$, the 
functions $M^{(\pm)}(\nu;k)$ defined in Eq. (3.10) are 
indeed bona fide Jost functions with $(M^{(+)}(\nu;k))^*
=M^{(-)}(\nu;k)$ for real k.  The positivity 
of $\psi(\alpha)$ guarantees the absence of a point 
spectrum.  The S-matrix is
\begin{equation}
S(\nu;k)\equiv\frac{M^{(+)}(\nu;k)}{M^{(-)}(\nu;k)},
\end{equation}

We define the discontinuity, $D(\nu;\tau)$ as
\addtocounter{eqnnum1}{1}
\begin{equation}
D(\nu,\tau)=\lim_{\epsilon\rightarrow 0}
[S(\nu;i\tau+\varepsilon)
-S(\nu;i\tau-\varepsilon)],~~\tau>1;
\end{equation}

From Eq. (3.10) one obtains
\addtocounter{eqnnum1}{1}
\begin{equation}
D(\nu,\tau)=\frac{\omega(\nu;\tau)}
{1+\frac{1}{\pi}\int_{1}^{\infty}
\frac{\omega(\nu,\beta)}{\beta+\tau}d\beta},
\end{equation}
with
\addtocounter{eqnnum1}{1}
\begin{equation}
\omega(\nu,\tau)=\pi(\nu^2-1/4)\psi(\tau)\tau^\frac{1}{4}
[\tau^\frac{\nu}{2}+\tau^\frac{_-\nu}{2}].
\end{equation}
For real $\nu>\frac{1}{2}$, $D(\nu,\tau)$ and 
$\omega(\nu,\tau)$ are positive for all $\tau\geq 1$.  The 
case with $\omega(\nu,\tau)\geq 0$ is the easiest to handle by 
the Martin inverse method, and it can be done explicitly.

Although having $\nu$ real and $\nu>\frac{1}{2}$ is of 
little direct interest to the Riemann problem, we give 
the results here as they might be helpful to the 
reader.  For details one should consult ref. 22.

The S-matrix, $S(\nu,k)$, uniquely determines a 
potential, $V(\nu;x)$, and its Jost solutions 
$f^{(\pm)}(\nu;k;x)$.  $V$ is given by
\addtocounter{eqnnum1}{1}
\begin{equation}
V(x)=4\sum_{n=0}^{\infty}(\frac{-1}{\pi})^{n+1}
\int_{1}^{\infty}d\alpha_o....\int_{1}^{\infty}d\alpha_n
\frac{(\prod_{j=0}^{n}D(\nu;\alpha_j)e^{-2\alpha_jx})}
{(\prod_{j=0}^{n-1}(\alpha_j+\alpha_{j+1}))}(\sum_{j=0}^{n}\alpha_j).
\end{equation}
This series for $V$ is absolutely and uniformly 
convergent for all $x\geq 0$, and $\nu>\frac{1}{2}$.  This follows 
from the positivity in (4.3),
\addtocounter{eqnnum1}{1}
\begin{equation}
\frac{1}{\pi}\int_{1}^{\infty}\frac{|D(\nu,\alpha)|}
{\alpha+\tau}d\alpha\leq \frac{\frac{1}{\pi}\int_{1}^{\infty}
\omega(\nu,\alpha)/\alpha d\alpha}{1+\frac{1}{\pi}\int_{1}^{\infty}
\frac{\omega(\nu,\beta)}{\beta}d\beta} <1.
\end{equation}

The Jost solutions, $f^{(\pm)}(\nu;k;x)$ are given by
\addtocounter{eqnnum1}{1}
\begin{equation}
f^{(\pm)}=e^{\mp ikx}+e^{\mp ikx}\sum_{n=0}^{\infty}
(\frac{-1}{\pi})^{n+1}\int_{1}^{\infty}
d\alpha_o...\int_{1}^{\infty}d\alpha_n
\frac{(\prod_{j=0}^{n}D(\nu;\alpha_j)e^{-2\alpha_jx})}
{[\prod_{j=0}^{n-1}(\alpha_j+\alpha_{j+1})][\alpha_o\pm ik]}
\end{equation}
Again this last series is absolutely and uniformly convergent 
for all $x\geq 0$, and $k$ in a compact domain inside the 
respective regions of analyticity.

One can check directly that $f^{\pm}$ given by Eq. (4.6) are 
solutions of the Schrodinger equation with $V(\nu;x)$ of 
Eq. (4.4) as potential, see ref. 22 for more details.
\newpage
\vspace{.25in}
\setcounter{chanum}{5}
\setcounter{eqnnum1}{1}
\noindent{\bf\Large{V.  Some Properties of 
$M^{(\pm)}(\nu,k)$ for $|\nu|>1$.}}

To proceed further and study $M^{(\pm)}(\nu,k)$ for 
complex $\nu$, and more specifically $\nu$ in the 
critical strip,$-\frac{1}{2}<Re\nu<\frac{1}{2};Im\nu>1$, 
the defining representation (3.10) is not fully instructive.  
This is because the behavior of $M^{\pm}$ for large $Im\nu$ 
is not adequately shown by Eq. (3.10).  Our final result in 
this section is to obtain an asymptotic expansion 
of $M^{\pm}(\nu,k)$ for fixed $k$ in inverse powers 
of $(\nu^2-\frac{1}{4})$.

We need to carry out integrations by parts on the 
integrand in Eq. (3.10) analogous to those 
performed in Titchmarsh's book$^{18}$, for Eq. (3.9). 

The following lemma will prove extremely useful.

{\underline{Lemma 5.1}}

Let $W(\alpha),\alpha\in[1,\infty)$, be a 
$C^{\infty}$ function, and $W(\alpha)=O(e^{-\alpha})$ 
as $\alpha\rightarrow\infty$, then given the integral
\begin{equation}
I(\nu)=\int_{1}^{\infty}W(\alpha)[\alpha^\frac{\nu}{2}
+\alpha^\frac{-\nu}{2}]d\alpha,
\end{equation}
one has after two integrations by parts
\addtocounter{eqnnum1}{1}
\begin{equation}
I(\nu)=\frac{1}{(\nu^2-\frac{1}{4})}\{8[W(1)
+(W^{\prime}(\alpha))_{\alpha=1}]
+\int_{1}^{\infty}W_1(\alpha)[\alpha^\frac{\nu}{2}
+\alpha^\frac{-\nu}{2}]d\alpha\},
\end{equation}
where
\addtocounter{eqnnum1}{1}
\begin{equation}
W_1(\alpha)=\frac{15}{4}W(\alpha)
+12\alpha W^{\prime}(\alpha)+4\alpha^2W''(\alpha).
\end{equation}

{\underline{Proof:}}

We rewrite Eq. (5.1) as
\addtocounter{eqnnum1}{1}
\begin{equation}
I(\nu)=\int_{1}^{\infty}d\alpha(W(\alpha)\alpha^\frac{3}{4})
[\alpha^{\frac{\nu}{2}-\frac{3}{4}}+\alpha^{\frac{-\nu}{2}-\frac{3}{4}}]
\end{equation}
Integrating by parts we get,
\addtocounter{eqnnum1}{1}
\begin{equation}
I(\nu)=\frac{2W(1)}{(\nu^2-\frac{1}{4})}
-\int_{1}^{\infty}d\alpha(\frac{d}{d\alpha}
[W(\alpha)\alpha^{\frac{3}{4}}])[{\frac{\alpha^{\frac{\nu}{2}
+\frac{1}{4}}}{\frac{\nu}{2}+\frac{1}{4}}}
-{\frac{\alpha^{\frac{-\nu}{2}
+\frac{1}{4}}}{\frac{\nu}{2}-\frac{1}{4}}}].
\end{equation}
This again can be rewritten as
\addtocounter{eqnnum1}{1}
\begin{equation}
I(\nu)=\frac{2W(1)}{(\nu^2-\frac{1}{4})}
-\int_{1}^{\infty}d\alpha\{\alpha^\frac{3}{2}(\frac{d}
{d\alpha}[W(\alpha)\alpha^{\frac{3}{4}}])\}[{\frac{\alpha^{\frac{\nu}{2}
-\frac{5}{4}}}{\frac{\nu}{2}+\frac{1}{4}}}-{\frac{\alpha^{\frac{-\nu}{2}
-\frac{5}{4}}}{\frac{\nu}{2}-\frac{1}{4}}}].
\end{equation}
Carrying out a second integration by parts we obtain,
\addtocounter{eqnnum1}{1}
\begin{eqnarray}
I(\nu)&=&\frac{1}{\nu^2-\frac{1}{4}}[2W(1)+8\{\frac{d}{d\alpha}
(W(\alpha)\alpha^{\frac{3}{4}})\}_{\alpha=1}]\\ \nonumber
&+&\frac{4}{(\nu^2-\frac{1}{4})}\int_{1}^{\infty}
d\alpha\alpha^{\frac{-1}{4}}\{\frac{d}{d\alpha}
(\alpha^{\frac{3}{2}}[\frac{d}{d\alpha}(W(\alpha)
\alpha^{\frac{3}{4}})])\}[\alpha^{\frac{\nu}{2}}
+\alpha^{\frac{-\nu}{2}}]
\end{eqnarray}
Performing the differentiations in (5.7) easily leads to Eq. (5.2).

We can apply this lemma to the integral in Eq. (3.10) 
which defines $M^{(\pm)}(\nu,k)$.  Setting
\addtocounter{eqnnum1}{1}
\begin{equation}
W^{(\pm)}(\alpha;k)\equiv\frac{\psi(\alpha)\alpha^\frac{1}
{4}}{\alpha\pm ik},
\end{equation}
and restricting $k$ to the corresponding domain of analyticity in $k$
\addtocounter{eqnnum1}{1}
\begin{equation}
P^{(+)}=\{k|Imk<1\};~~~P^{(-)}=\{k|Imk>-1\}.
\end{equation}
we get
\addtocounter{eqnnum1}{1}
\begin{equation}
M^{(\pm)}(\nu;k)=M^{(\pm)}_o(k)+\int_{1}^{\infty}d\alpha W_1^{(\pm)}
(\alpha;k)[\alpha^{\frac{\nu}{2}}+\alpha^{-\frac{\nu}{2}}],
\end{equation}
with
\addtocounter{eqnnum1}{1}
\begin{equation}
W^{(\pm)}_1=\sum_{\ell=1}^{3}\frac{\Phi^{(1)}_{\ell}(\alpha)}
{(\alpha\pm ik)^{\ell}},
\end{equation}
and
\addtocounter{eqnnum1}{1}
\begin{eqnarray}
\Phi^{(1)}_1(\alpha)&=&6\psi(\alpha)\alpha^\frac{1}{4}+14\psi^{\prime}
(\alpha)\alpha^{\frac{5}{4}}+4\psi''(\alpha)\alpha^{\frac{9}{4}},\\ \nonumber
\Phi^{(1)}_2(\alpha)&=&-14\psi(\alpha)\alpha^{\frac{5}{4}}-8\psi^{\prime}
(\alpha)\alpha^{\frac{9}{4}},\\ \nonumber
\Phi^{(1)}_3(\alpha)&=&8\psi(\alpha)\alpha^\frac{9}{4}.
\end{eqnarray}
The first term in (5.10) is independent of $\nu$, and given by
\addtocounter{eqnnum1}{1}
\begin{equation}
M^{(\pm)}_o(k)=1+\frac{a_1}{1\pm ik}+\frac{a_2}{(1\pm ik)^2},
\end{equation}
with
\addtocounter{eqnnum1}{1}
\begin{equation}
a_1=-1+8\psi(1)~~;~~a_2=-8\psi(1).
\end{equation}
In obtaining (5.14) we have used the identity
\addtocounter{eqnnum1}{1}
\begin{equation}
4\psi^{\prime}(1)+\psi(1)=-1/2.
\end{equation}
It is important to note that both $a_1$ and $a_2$ are negative and 
that $a_1+a_2=-1$.  This leads to
\addtocounter{eqnnum1}{1}
\begin{equation}
\lim_{k\rightarrow 0}M^{(\pm)}_o(k)=0.
\end{equation}
As a check on Eq.(5.10) we take the $k\rightarrow 0$ limit,
\addtocounter{eqnnum1}{1}
\begin{equation}
M^{(\pm)}(\nu;0)=\int_{1}^{\infty}d\alpha[\sum_{\ell=1}^{3}
(\Phi^{(1)}_{\ell}/\alpha^{\ell})][\alpha^\frac{\nu}{2}+\alpha^\frac{-\nu}{2}]
\end{equation}
Substituting the expressions for $\Phi_{\ell}^{(1)}$ given 
in Eq. (5.12) we get
\addtocounter{eqnnum1}{1}
\begin{equation}
M^{(\pm)}(\nu,0)=4\int_{1}^{\infty}d\alpha(\psi''(\alpha)\alpha^\frac{5}
{4}+\frac{3}{2}\psi'(\alpha)\alpha^{\frac{1}{4}})[\alpha^{\frac{\nu}{2}}
+\alpha^{\frac{-\nu}{2}}]=2\xi(\frac{1}{2}+\nu).
\end{equation}

The lemma 5.1 can be used repeatedly to give an asymptotic expansion 
of $M^{\pm}(\nu;k)$ in inverse powers of $(\nu^2-\frac{1}{4})$.  
Recursion formulae can be given to give each term from the preceding one.

Indeed given
\addtocounter{eqnnum1}{1}
\begin{equation}
I^{(\pm)}_n(\nu;k)=\frac{1}{(\nu^2-\frac{1}{4})^{n-1}}\int_{1}^{\infty}
d\alpha W^{(\pm)}_n(\alpha;k)[\alpha^{\frac{\nu}{2}}+\alpha^{\frac{-\nu}{2}}],
\end{equation}
with
\addtocounter{eqnnum1}{1}
\begin{equation}
W^{(\pm)}_n=\sum_{\ell=1}^{2n+1}\frac{\Phi^{(n)}_{\ell}(\alpha)}
{[\alpha\pm ik]^{\ell}},
\end{equation}
and $k\in P^{\pm}$, one obtains
\addtocounter{eqnnum1}{1}
\begin{equation}
I^{(\pm)}_{n+1}=\frac{1}{(\nu^2-\frac{1}{4})^n}\{8[W^{(\pm)}_n(1;k)
+(\frac{dW^{(\pm)}_n}{d\alpha})_{\alpha=1}]+\int_{1}^{\infty}d\alpha 
W^{(\pm)}_{n+1}(\alpha;k)[\alpha^{\frac{\nu}{2}}+\alpha^{\frac{-\nu}{2}}]\},
\end{equation}

and
\addtocounter{eqnnum1}{1}
\begin{equation}
W^{(\pm)}_{n+1}=\frac{15}{4}W^{(\pm)}_n+12\alpha(W^{(\pm)}_n)^{\prime}
+4\alpha^2(W^{(\pm)}_n)'';
\end{equation}
where the primes indicate differentiation with respect to $\alpha$.  
Again $W_{(n+1)}^{(\pm)}$ will be as in Eq. (5.20)
\addtocounter{eqnnum1}{1}
\begin{equation}
W_{(n+1)}^{\pm}=\sum_{\ell=1}^{2n+3}\frac{\Phi_{\ell}^{(n+1)}(\alpha)}
{[\alpha\pm ik]^{\ell}}.
\end{equation}

For each $n$ we have $1\leq\ell\leq2n+1$, and the functions 
$\Phi_{\ell}^{(n)}$ satisfy a recursion formula, which follows from (5.22).
\addtocounter{eqnnum1}{1}
\begin{eqnarray}
\Phi^{(n+1)}_{\ell}(\alpha)&=&\frac{15}{4}\Phi^{(n)}_{\ell}+12\alpha
(\Phi^{(n)}_{\ell})'-12\alpha(\ell-1)\Phi^{(n)}_{\ell-1}\\ \nonumber
&+&4\alpha^2[(\Phi^{(n)}_{\ell})''-2(\ell-1)(\Phi^{(n)}_{\ell-1})'
+(\ell-l)(\ell-2)\Phi^{(n)}_{\ell-2}]
\end{eqnarray}
All the $\Phi^{(n)}_{\ell}$ can thus be determined by iteration starting from
\addtocounter{eqnnum1}{1}
\begin{equation}
\Phi^{(0)}_1(\alpha)\equiv\psi(\alpha)\alpha^{\frac{1}{4}}.
\end{equation}

The general form of $\Phi^{(n)}_{\ell}(\alpha)$ is easily determined to be
\addtocounter{eqnnum1}{1}
\begin{equation}
\Phi^{(n)}_{\ell}(\alpha)=\sum_{j=0}^{2n+1-\ell}C^{(n)}(\ell;j)\
alpha^{\frac{1}{4}+\ell+j-1}\psi^{(j)}(\alpha).
\end{equation}
The coefficients $C^{(n)}(\ell;j)$ are real, and $C^{(0)}(1;0)=1$ 
determines all the others.  Also
\addtocounter{eqnnum1}{1}
\begin{equation}
\psi^{(j)}(\alpha)\equiv(\frac{d}{d\alpha})^j\psi(\alpha).
\end{equation}

At this point we can substitute Eq. (5.26) in (5.24) and obtain 
a recursion formula for $C^{(n)}(\ell;j)$,
\addtocounter{eqnnum1}{1}
\newpage
\begin{eqnarray}
C^{(n+1)}(\ell,j)&=& C^{(n)}(\ell;j)[\frac{15}{4}+12(\frac{1}{4}
+\ell+j-1)\\ \nonumber
&+&4(\frac{1}{4}+\ell+j-1)(\frac{1}{4}+\ell+j-2)]\\ \nonumber
&-&C^{(n)}(\ell-1,j)[12(\ell-1)+8(\ell-1)(\frac{1}{4}+\ell+j-2)]\\ \nonumber
&+&C^{(n)}(\ell ,j-1)[12+8(\frac{1}{4}+\ell+j-2)]+4C^{(n)}(\ell,j-2)\\ \nonumber
&-&8(\ell-l)C^{(n)}(\ell-1;j-1)+4(\ell-1)(\ell-2)C^{(n)}(\ell-2;j).
\end{eqnarray}
Here we have
\addtocounter{eqnnum1}{1}
\begin{equation}
1\leq\ell\leq2n+1, ~~~0\leq j\leq2n+1-\ell.
\end{equation}
For all other values of $\ell$ and $j$, $C^{(n)}(\ell,j)\equiv0$.

Starting with
\addtocounter{eqnnum1}{1}
\begin{equation}
C^{(0)}(1,0)\equiv1,
\end{equation}
we can compute all other $C^{(n)}(\ell,j)$.  For example,$C^{(1)}
(1,0)=6$, $C^{(1)}(1,1)=14$, and $C^{(1)}(1,2)=4$.  This agrees 
with the direct calculation given in Eq. (5.12).  In table I, 
we give all the coefficients $C^{(n)}(\ell,j)$ up to $n=4$.  
All the coefficients are integers.

Finally, we give the general form of the surface term in Eq. 
(5.21).  We define $M^{(\pm)}_n(k)$,
\addtocounter{eqnnum1}{1}
\begin{equation}
M^{(\pm)}_n(k)=8[W^{(\pm)}_n(1,k)+(\frac{dW^{(\pm)}_n}{d\alpha})_{\alpha=1}].
\end{equation}

From Eq. (5.20) and (5.26) we obtain after some algebra,
\addtocounter{eqnnum1}{1}
\begin{equation}
M^{(\pm)}_n(k)=\sum_{\ell=1}^{2n+2}\frac{\chi^{(n)}_{\ell}}{[1\pm ik]^{\ell}},
\end{equation}
with
\addtocounter{eqnnum1}{1}
\begin{eqnarray}
\chi^{(n)}_{\ell}&=&8\sum_{j=0}^{2n+2-\ell}C^{(n)}(\ell,j)\{(\frac{1}
{4}+\ell+j)\psi^{(j)}(1)+\psi^{(j+1)}(1)\}
\\ \nonumber
&-&8\sum_{j=0}^{2n+2-\ell}(\ell-1)C^{(n)}(\ell-1,j)\psi^{(j)}(1).
\end{eqnarray}

For the purposes of this paper it is sufficient to apply our lemma up 
to the $n=3$ level.  We introduce as a new variable, $g$,
\addtocounter{eqnnum1}{1}
\begin{equation}
g\equiv\frac{1}{(\nu^2-\frac{1}{4})}
\end{equation}

Our final result for $M^{(\pm}(\nu,k)$ with $k\epsilon P^{(\pm)}$ 
and $|\nu|>>1$,
\addtocounter{eqnnum1}{1}
\begin{equation}
M^{(\pm)}(\nu;k)=M^{(\pm)}_o(k)+gM^{(\pm)}_1(k)+g^2M^{(\pm)}_2(k)
+g^2R^{(\pm)}_2(\nu,k).
\end{equation}
Here we have
\addtocounter{eqnnum1}{1}
\begin{equation}
M^{(\pm)}_1(k)=\sum_{\ell=1}^{4}\frac{b_{\ell}}{[1\pm ik]^{\ell}},
\end{equation}
with $b_{\ell}\equiv \chi^{(1)}_{\ell}$, and
\addtocounter{eqnnum1}{1}
\begin{equation}
M^{(\pm)}_2(k)=\sum_{\ell=1}^{6}\frac{c_{\ell}}{[1\pm ik]^{\ell}},
\end{equation}
$c_{\ell}\equiv\chi^{(2)}_{\ell}$.  The remainder function $R^{(\pm)}_2$ 
is given by
\addtocounter{eqnnum1}{1}
\begin{equation}
R^{(\pm)}_2(\nu;k)=\int_{1}^{\infty}(\sum_{\ell=1}^{7}\frac{\Phi_{\ell}^
{(3)}(\alpha)}{[\alpha\pm ik]^{\ell}})[\alpha^{\frac{\nu}{2}}+\alpha^
{\frac{-\nu}{2}}]d\alpha.
\end{equation}

For $|Im\nu|>10^3$, the first two terms of Eq. (5.35) give a very good 
estimate for $M^{(\pm)}$.  We shall explore this in much more detail later. 
 One can go to higher orders in $g$, but the resulting series is 
 only asymptotic.  For our purposes here Eq. (5.35) is enough.

It is important to stress another property of $M^{(\pm)}_1$ and 
$M^{(\pm)}_2$, namely as $k\rightarrow 0$,
\addtocounter{eqnnum1}{1}
\begin{eqnarray}
M^{(\pm)}_1(0)=0, \\ \nonumber
M^{(\pm)}_2(0)=0.
\end{eqnarray}
We have already shown that $M^{(\pm)}_0(0)=0$.  To check this we give 
the explicit form of the coefficients $b_{\ell}$ in (5.36).  
Using $\chi^{(1)}_{\ell}=b_{\ell}$, Eq. (5.33), and table I, we get
\addtocounter{eqnnum1}{1}
\begin{eqnarray}
b_1&=& 60\psi(1)+300\psi'(1)+216\psi''(1)+32\psi'''(1),\\ \nonumber
b_2&=& -300\psi(1)-432\psi'(1)-96\psi''(1),\\ \nonumber
b_3&=& 432\psi(1)+192\psi'(1),\\ \nonumber
b_4&=& -192\psi(1)
\end{eqnarray}
Numerically, the $b$'s are given in table II.
\newpage
We now have
\addtocounter{eqnnum1}{1}
\begin{eqnarray}
M_1^{(\pm)}(0)&=&\sum_{\ell=1}^{4}b_{\ell}\\ \nonumber &=&32[\psi'''(1)
+\frac{15}{4}\psi''(1)+\frac{15}{8}\psi'(1)].
\end{eqnarray}
But
\addtocounter{eqnnum1}{1}
\begin{equation}
\psi'''(1)+\frac{15}{4}\psi''(1)+\frac{15}{8}\psi'(1)=0
\end{equation}
This identity follows from the relation$^{18}$
\addtocounter{eqnnum1}{1}
\begin{equation}
\sqrt{\alpha}(2\psi(\alpha)+1)=2\psi(1/\alpha)+1.
\end{equation}
Differentiating (5.40) once and setting $\alpha=1$ immediately gives Eq. 
(5.15).  Differentiating three times leads to Eq. (5.42).

Indeed there is an infinite sequence of identities like Eq.(5.42), 
always starting with $\psi^{2n+1}(1)$, odd derivatives, which result 
from differentiating Eq. (5.40) $(2n+1)$ times.  Thus, again
\addtocounter{eqnnum1}{1}
\begin{equation}
M^{(\pm)}_2(0)=\sum_{\ell=1}^{6}c_{\ell}=0,
\end{equation}
depends on the next identity:
\addtocounter{eqnnum1}{1}
\begin{equation}
\psi^{(5)}(1)+\frac{45}{4}\psi^{(4)}(1)+\frac{235}{4}\psi^{(3)}(1)
+\frac{975}{8}\psi^{(2)}(1)+\frac{1635}{32}\psi^{(1)}(1)=0.
\end{equation}
Of course only the first two coefficients in (5.45) are unique, 
since we can always add a multiple of the $l.h.s.$ of (5.42) to (5.45).

The vanishing of $M^{\pm}_j(0)$, $j=0,1,2$, is indeed necessary since
\addtocounter{eqnnum1}{1}
\begin{equation}
g^2R^{(\pm)}_2(\nu;0)=g^2\int_{1}^{\infty}d\alpha(\sum_{\ell=1}^{7}\frac
{\Phi_{\ell}^{(3)}(\alpha)}{\alpha^{\ell}})[\alpha^{\frac{\nu}{2}}
+\alpha^{\frac{-\nu}{2}}]=2\xi(\frac{1}{2}+\nu).
\end{equation}
which is the result of carrying out four more differentiations by 
parts on the formula for $\xi(\frac{1}{2}+\nu)$ given on page 254 
of reference 18.

In table II, we give the numerical values of $\chi_\ell^{(n)}$, 
for $n=1,2,$ and $3$, and $b_{\ell}\equiv\chi^{(1)}_{\ell}$ while 
$c_{\ell}\equiv\chi^{(2)}_{\ell}$.

For the convenience we summarize the results of this section:
\addtocounter{eqnnum1}{1}
\begin{equation}
M^{(\pm)}(\nu,k)=\sum_{n=0}^{N}g^nM_n^{(\pm)}(k)+g^NR_N^{(\pm)}(\nu,k);
\end{equation}
where
\addtocounter{eqnnum1}{1}
\begin{equation}
g=(\nu^2-1/4)^{-1},
\end{equation}

\addtocounter{eqnnum1}{1}
\begin{equation}
M^{(\pm)}_n(k)=\sum_{\ell=1}^{2n+2}\frac{\chi_{\ell}^{(n)}}{[1\pm ik]^{\ell}},
\end{equation}
and $\chi_{\ell}^{(n)}$ are real numbers given in Eq. (5.33).  
In addition we have
\addtocounter{eqnnum1}{1}
\begin{equation}
\sum_{\ell=1}^{2n+2}\chi_{\ell}^{(n)}\equiv 0,
\end{equation}
which guarantees that $M_n^{(\pm)}(k)\rightarrow 0$ as $k\rightarrow 0$. 
 For real $k$, we have $[M_n^{(+)}(k)]^*=M^{(-)}_n(k)$.

Finally the remainder term $R^{(\pm)}_N$ is given explicitly by,
\addtocounter{eqnnum1}{1}
\begin{equation}
R^{(\pm)}_N(\nu,k)=\int_{1}^{\infty}d\alpha(\sum_{\ell=1}^{2N+3}\frac
{\Phi^{(N+1)}_{\ell}(\alpha)}{(\alpha\pm ik)^{\ell})}[\alpha^{\nu/2}
+\alpha^{-\nu/2}],
\end{equation}
with
\addtocounter{eqnnum1}{1}
\begin{equation}
\Phi^{(n)}_{\ell}(\alpha)=\sum_{j=0}^{2n+1-\ell}C^{(n)}(\ell;j)\alpha^
{1/4+\ell+j-1}\psi^{(j)}(\alpha),
\end{equation}

The $C^{(n)}(\ell,j)$ are integers determined by a recursion formula 
given in Eq. (5.28), with $C^{(0)}(1,0)\equiv1$, and $\psi^{(j)}
(\alpha)$ are the $j$th derivatives of $\psi(\alpha)$, Eq. (5.27).

For $k=0$, we have
\addtocounter{eqnnum1}{1}
\begin{equation}
M^{(\pm)}(\nu,0)=2\xi(\nu+1/2).
\end{equation}
From Eqs. (5.47) and (5.51) we then have for any integer $n\geq 0$,
\addtocounter{eqnnum1}{1}
\begin{equation}
2\xi(\nu+\frac{1}{2})=g^n\int_{1}^{\infty}d\alpha(\sum_{\ell=1}^{2n+3}
\frac{\Phi^{(n+1)}_{\ell}(\alpha)}{(\alpha^{\ell})}[\alpha^{\nu/2}
+\alpha^{-\nu/2}].
\end{equation}
For $n=0$, this formula is given in ref.(18) on page 225.  
The results for larger $n$ can be obtained by successive 
integrations by parts.

\newpage
\vspace{.25in}
\setcounter{chanum}{6}
\setcounter{eqnnum1}{1}

\noindent{\bf\Large{VI.  The Zeroes of $M^{(-)}(\nu;k)$ for $Imk>0$, 
and fixed $\nu$}}

To study the Riemann hypothesis we need only to focus on the 
truncated critical strip, ${\cal{S}}(T_o)$,
\begin{equation}
{\cal{S}}(T_o)=\{\nu|Im\nu>T_o,~~-\frac{1}{2}<Re\nu<\frac{1}{2}\}
\end{equation}
Since the Riemann Hypothesis has already been rigorously established 
up to $Im\nu=O(10^6)$, we can simplify the calculations of this paper 
tremendously by taking $T_o$ to be large.  Initially, we take $T_o\cong 10^3$.

The following lemma will be quite useful.

{\underline{Lemma 6.1}}

For any $\nu\in{\cal{S}}(T_o)$, and $k$ such that $Imk>-\frac{1}{4}$, we have
\addtocounter{eqnnum1}{1}
\begin{equation}
|M^{(-)}(\nu,k)-M^{(-)}_o(k)|\leq\frac{C_2}{T^2_o},
\end{equation}
and
\addtocounter{eqnnum1}{1}
\begin{equation}
C_2\leq10^3.
\end{equation}

{\underline{Proof:}}

Taking the expansion of $M^{(-)}(\nu;k)$ in powers of $g=(\nu^2-
{\frac{1}{4}})^{-1}$ to first order we have
\addtocounter{eqnnum1}{1}
\begin{equation}
M^{(-)}(\nu,k)-M^{(-)}_o(k)=gM^{(-)}_1(k)+gR^{(-)}_1(\nu,k),
\end{equation}
where $M^{(-)}_1{(k)}$ is given by Eq. (5.36), and
\addtocounter{eqnnum1}{1}
\begin{equation}
R^{(-)}_1(\nu;k)=\int_{1}^{\infty}d\alpha(\sum_{\ell=1}^{5}\frac
{\Phi^{(2)}_{\ell}(\alpha)}{[\alpha-ik]^{\ell}})[\alpha^{\frac{\nu}
{2}}+\alpha^{\frac{-\nu}{2}}].
\end{equation}
Here $\Phi^{(2)}_{\ell}$ is given by Eq. (5.26) and table I.

First, we have for $Im k>-\frac{1}{4}$,
\addtocounter{eqnnum1}{1}
\begin{equation}
|M^{(-)}_1(k)| \leq\sum_{\ell=1}^{4}\frac{|b_{\ell}|}{|1-ik|^{\ell}}
\leq[\sum_{\ell=1}^{4}|b_{\ell}|.(\frac{4}{3})^{\ell}].
\end{equation}
Using table II, we get
\addtocounter{eqnnum1}{1}
\begin{equation}
|M^{(-)}_1(k)|<68, ~~~Imk>-\frac{1}{4}.
\end{equation}
The upper bound on $R^{(-)}_1$ for $Imk>-\frac{1}{4}$ is
\addtocounter{eqnnum1}{1}
\begin{equation}
|R^{(-)}_1(\nu,k)|\leq2\int_{1}^{\infty}\alpha^{\frac{1}{4}}
(\sum_{\ell=1}^{5}|\Phi^{(2)}_{\ell}(\alpha)|\alpha^{-\ell}
(\frac{4}{3})^{\ell}),
\end{equation}
where we have used $|\alpha/(\alpha-ik)|<\frac{4}{3}$ for 
$\alpha\geq 1$ and $Imk>-\frac{1}{4}$.
Using Eq. (5.26) we have 
\addtocounter{eqnnum1}{1}
\begin{equation}
|R^{(-)}_1(\nu,k)|\leq2\int_{1}^{\infty}d\alpha\{\sum_{\ell=1}^{5}
(\frac{4}{3})^{\ell}|\sum_{j=0}^{5-\ell}|C^{(2)}(\ell,j)|\alpha^{\frac{-1}
{2}+j}\psi^{(j)}{(\alpha)}|\}
\end{equation}
where we note that $C^{(n)}{(\ell,j)}=(-1)^{\ell+1}.|C^{(n)}{(\ell,j)|}$ 
as can be seen from Table I.

The series in Eq. (3.7) that defines $\psi(\alpha)$ is highly convergent 
for $\alpha\geq 1$.  Indeed the first term gives a good approximation to 
it and to its first six derivatives.  One can easily derive the bounds, 
$0\leq j\leq 7$,
\addtocounter{eqnnum1}{1}
\begin{equation}
\pi^je^{-\pi\alpha}\leq|\psi^{(j)}(\alpha)|\leq\pi^je^{-\pi\alpha}
(1+\epsilon(j)),
\end{equation}
where $\epsilon(j)$ is 
\addtocounter{eqnnum1}{1}
\begin{equation}
\epsilon(j)=e^{-3\pi}[1+(2)^{2j}],
\end{equation}

For $j\leq4$, $\epsilon(j)<0.021$.  Thus it is sufficient for 
the purposes of this estimate to use $\psi^{(j)}
(\alpha)\cong(-1)^j\pi^je^{-\pi\alpha}$.  Substituting this 
in (6.9) and carrying out the $\alpha$ integration we get
\addtocounter{eqnnum1}{1}
\begin{equation}
|R^{(-)}_1(\nu,k)|\leq\frac{2(1.1)}{\sqrt{\pi}}\sum_{\ell=1}^{5}
(\frac{4}{3})^{\ell}|\sum_{j=0}^{5-\ell}|C^{(2)}
(\ell,j)|(-1)^j\Gamma(j+\frac{1}{2};\pi)|\equiv C'_2,
\end{equation}
where $\Gamma(j,\beta)$ is the incomplete gamma function.  
From Table I, it is now easy to check our bound of $C_2\equiv
 C^{\prime}_2+68<200$.  This completes the proof of lemma 6.1.

It should be apparent to the reader that one could use more 
refined methods to obtain a much better bound on $R^{(-)}_1$. 
 We do not do this at this stage.  Our most important task is 
 to study the Riemann conjecture for $Im\nu>T_o$ with $T_o$ 
 taken below the maximum for which the hypothesis has been 
 rigorously established.  In a future paper, we will try to 
 find the lowest value of $T_o$ for which our method works.

The function $M_o^{(-)}(k)$ given by Eq. (5.13) is a rational function of $k$
\addtocounter{eqnnum1}{1}
\begin{eqnarray}
M^{(-)}_o(k)&=&1+\frac{a_1}{1-ik}+\frac{(-1-a_1)}{(1-ik)^2},\\ \nonumber
&=&\frac{-k[k+i(2+a_1)]}{(1-ik)^2}.
\end{eqnarray}
Here
\addtocounter{eqnnum1}{1}
\begin{equation}
a_1=-1+8\psi(1)=-0.6543
\end{equation}
Obviously, $M^{(-)}_o$ has two zeros, $k_1=0$, and 
$k_2=-i(2+a_1)=-i(1+8\psi(1))$.  Thus $Im k_2<-1$.  
Hence, $M^{(-)}_o(k)$ has only one zero in the half-plane, $Im k\geq 0$.

Focusing on the domain $Im k>-\frac{1}{4}$, and $|k|>\frac{1}{4}$ we get, 
with $k=\kappa+i\tau$,
\addtocounter{eqnnum1}{1}
\begin{equation}
|M_o^{(-)}(k)|=\frac{\sqrt{\kappa^2+\tau^2}\sqrt{\kappa^2+(\tau+\eta)^2}}
{\kappa^2+(\tau+1)^2},
\end{equation}
where $\eta=2+a_1>1$.

It is easy to find a lower bound for $|M^{(-)}_o|$ in the above domain. 
 Setting $\vec{q}=(0,-\eta)$, and $\vec{p}=(0,-1)$ we get
\addtocounter{eqnnum1}{1}
\begin{equation}
|M^{(-)}_o(k)|=\frac{|\vec{k}|.|\vec{k}-\vec{q}|}{|\vec{k}-\vec{p}
|^2}\geq\frac{|\vec{k}|^2}{|\vec{k}-\vec{p}|^2}\geq\frac{1}{25}.
\end{equation}

We can now prove the following.

{\underline{Lemma 6.2}}

For any $\nu\in{\cal{S}}(T_o)$, and $k$ such that $Im k\geq0$, 
with $|k|\geq\frac{1}{4}$, $M^{(-)}(\nu,k)$ has no zeros and has 
a lower bound
\addtocounter{eqnnum1}{1}
\begin{equation}
|M^{(-)}(\nu,k)|>(0.04-\frac{1}{T_o}).
\end{equation}
\underline{Proof:}

From lemma (6.1) we have
\addtocounter{eqnnum1}{1}
\begin{equation}
|M^{(-)}(\nu,k)|>|M^{(-)}_o(k)|-\frac{C_2}{T^2_o},
\end{equation}
with $C_2/T_o<2$.  Using (6.16) we get
\addtocounter{eqnnum1}{1}
\begin{equation}
|M^{(-)}(\nu,k)|>(0.04-\frac{1}{T_o})>\frac{1}{26}.
\end{equation}
Thus any zeros, $k_o$ of $M^{(-)}(\nu,k)$ in the upper half k-plane 
must have $|k_o|<\frac{1}{4}$.

Proceeding further we have:

{\underline{Lemma 6.3:}}

For any fixed $\nu\in{\cal{S}}(T_o)$, the maximum number of zeros 
of $M^{(-)}(\nu;k)$with $Imk\geq0$ is one.

{\underline{Proof:}}

From lemma (6.1) we get for $\nu\in{\cal{S}}(T_o)$
\addtocounter{eqnnum1}{1}
\begin{equation}
|M^{(-)}(\nu,k)-M^{(-)}_o(k)|<\frac{1}{T_o};~~|k|=\frac{1}{4},
\end{equation}
By very similar arguments we can also show that
\addtocounter{eqnnum1}{1}
\begin{equation}
|\frac{dM^{(-)}(\nu,k)}{dk}-\frac{dM^{(-)}_o}{dk}|<\frac{\lambda}
{T_o};~~|k|=\frac{1}{4},
\end{equation}
where $\lambda=O(1)$.

Let $N^{(0)}_{\frac{1}{4}}$ denote the number of zeros of $M^{(-)}_o(k)$ 
in the disc $|k|\leq\frac{1}{4}$, and $N_{\frac{1}{4}}$ be the corresponding 
number for $M^{(-)}(\nu,k)$, $\nu\epsilon{\cal{S}}(T_o)$, then
\addtocounter{eqnnum1}{1}
\begin{equation}
 N_{\frac{1}{4}}-N^{(0)}_{\frac{1}{4}}=\frac{1}{2\pi i}\oint_{C_{\frac{1}
 {4}}}dk\{\frac{[M^{(-)}(\nu,k)]'}{M^{(-)}(\nu,k)}-\frac{[M^{(-)}_o(k)]'}
 {M^-_o(k)}\},
\end{equation}
where $C_{\frac{1}{4}}$ is the circle $|k|=\frac{1}{4}$, and the prime 
indicates differentiation with respect to $k$.

Hence,
\addtocounter{eqnnum1}{1}
\begin{equation}
|N_{\frac{1}{4}}-N^{(0)}_{\frac{1}{4}}|
\leq\frac{(26)^2}{4}\{Max_{|k|=\frac{1}{4}}|M^{(-)}_o(k)[M^{(-)}(\nu,k)]'
-[M_o^{(-)}(k)]'M^{(-)}(\nu,k)|\}
\leq(\frac{169\lambda}{T_o})<1,
\end{equation}
where we have used Eqs. (6.16), (6.19), and (6.21).  Since $M^{(-)}_o(k)$ 
has only one zero in the disc so does $M^{(-)}(\nu,k)$ and lemma (6.2) 
which proves the absence of zeros with $Im k\geq 0$, and $|k|\geq\frac{1}
{4}$, completes our proof.

The next question is where is this one zero of $M^{(-)}(\nu;k)$?  
To answer this question we first order expansion for $M^{(-)}(\nu,k)$,
\addtocounter{eqnnum1}{1}
\begin{equation}
M^{(-)}(\nu,k)=M^{(-)}_o(k)+ gM_1^{(-)}(k)+g\int_{1}^
{\infty}d\alpha(\sum_{\ell=1}^{5}\frac{\Phi_{\ell}^{(2)}
(\alpha)}{(\alpha-ik)^{\ell}})[\alpha^{\frac{\nu}{2}}
+\alpha^{\frac{-\nu}{2}}]
\end{equation}
With $k=0$, we have
\addtocounter{eqnnum1}{1}
\begin{equation}
2\xi(\nu+1/2)=g\int_{1}^{\infty}d\alpha(\sum_{\ell=1}^{5}\frac{\Phi_{\ell}
^{(2)}(\alpha)}{\alpha^{\ell}}).[\alpha^{\frac{\nu}{2}}+\alpha^
{\frac{-\nu}{2}}]
\end{equation}
Subtracting these two equations we get
\addtocounter{eqnnum1}{1}
\begin{equation}
M^{(-)}(\nu,k)=2\xi(\nu+1/2)+M^{(-)}_o(k)+gM^{(-)}_1(k)
+g\int_{1}^{\infty}d\alpha(\sum_{\ell=1}^{5}\{\frac{1}{(\alpha-ik)^{\ell}}
-\frac{1}{\alpha^{\ell}}\}\Phi^{(2)}_{\ell}(\alpha)[\alpha^{\frac{\nu}{2}}
+\alpha^{\frac{-\nu}{2}}])
\end{equation}
Note that now the integral on the r.h.s. is also $O(k)$ as $k\rightarrow0$.

The $\xi$ function for large values of $|Im\nu|$ is exponentially small,
 mainly due to the $\Gamma(\frac{s}{2})$ factor in Eq. (3.4).  
 In fact given standard results on the order of $\zeta(s)$ in the 
 critical strip we have $\xi(\frac{1}{2}+\nu)=O(|\nu|^p e^{\frac{-|Im\nu|\pi}
 {4}})$ where $p=2+\delta$ and $0<\delta<1$.  Thus from Eq. (6.26)  
 and the exact expression for $M^{(-)}_o(k)$ in Eq. (6.13), we get 
 the position of the zero, $k_o$, near the origin,
\addtocounter{eqnnum1}{1}
\begin{equation}
k_o\cong\frac{-2i\xi(\frac{1}{2}+\nu)}{[(2+a_1)+O(\frac{1}{T_o^2)}]}
+O(k^2_o),
\end{equation}
where $(2+a_1)>1$.  Hence
\addtocounter{eqnnum1}{1}
\begin{equation}
Imk_o\cong\frac{-2Re\xi(\frac{1}{2}+\nu)}{(2+a_1)}.
\end{equation}
If $Re\xi(\frac{1}{2}+\nu)\geq0$, $M^{(-)}(\nu,k)$ has no zeros for $Imk>0$. 
 On the other hand if $\nu\in{\cal{S}}(T_o)$ is such that $Re\xi(\frac{1}{2}
 +\nu)<0$ there will be one zero close to the origin, but in the upper half 
 $k$ plane.

Finally, we note one important fact, namely, if $(\nu_o+1/2)$ is a zero of
 the $\xi$-function, then $M^{(-)}(\nu_o,k)$ has no zeroes for $Imk>0$, 
 and its only zero with $Imk>-\epsilon$ occurs exactly at $k=0$.

In summary $M^{(-)}(\nu;k)$ , with $\nu\epsilon\it{S}(T_o)$, has most of 
the properties of a Martin type Jost function with the exception of one, 
i.e. reality.  We list these properties:

\noindent i)  $M^{(-)}(\nu,k)$ analytic in the cut k-plane with a cut for 
$k=i\tau$, $-\infty<\tau<-1$.\\
ii) $\lim_{|k|\rightarrow\infty}M^{(-)}(\nu,k)=1$.\\
iii)$ M^{(-)}(\nu,k)$ has no zeros for $Imk>\frac{1}{4}$, 
and $\nu\varepsilon\cal{S}(T_o)$.\\
iv)  If we write
\addtocounter{eqnnum1}{1}
\begin{equation}
\Sigma(\nu,u)\equiv(\nu^2-\frac{1}{4})\int_{1}^{\infty}d\alpha\psi(\alpha)
\alpha^{\frac{1}{4}}[\alpha^{\frac{\nu}{2}}+\alpha^{\frac{-\nu}{2}}]
e^{-\alpha u},
\end{equation}
then
\addtocounter{eqnnum1}{1}
\begin{equation}
M^{(\pm)}(\nu,k)=1+\int_{o}^{\infty}\Sigma(\nu,u)e^{\mp iku}du
\end{equation}

In fact we could have used Eqs. (6.29) and (6.30) as the starting 
definitions of $M^{(\pm)}(\nu,k)$.\\
v)  We define the S-matrix
\addtocounter{eqnnum1}{1}
\begin{equation}
S(\nu,k)=\frac{M^{(+)}(\nu,k)}{M^{(-)}(\nu,k)},
\end{equation}
and it follows that
\addtocounter{eqnnum1}{1}
\begin{equation}
\int_{-\infty}^{+\infty}|S(\nu,k)-1|^2dk<\infty.
\end{equation}
vi)  The reality condition does not hold for all $\nu$.  
If $\nu$ is purely imaginary, i.e. $\nu=i t$, then for real k, 
$(M^{(+)}(\nu,k))^*=M^{(-)}(\nu,k)$, and hence $|S(\nu,k)|=1$. 
 However, if $\nu$ is non-, i.e. $Re\nu\neq0$, then the above 
 relation does not hold.   However, we still have for real $k$
\addtocounter{eqnnum1}{1}
\begin{equation}
|S(\nu,k)|=1+O(\frac{1}{|\nu|^2}).
\end{equation}

For an arbitrary $\nu$, $\nu\epsilon\it{S}(T_o)$, 
we can still carry out the inverse scattering program by properly handling 
the one zero in the upper half $k$ plane.  We will do that in Section VIII, 
where the resulting potential is complex.
\newpage
\vspace{.25in}
\setcounter{chanum}{7}
\setcounter{eqnnum1}{1}
\noindent{\bf\Large{VII. The Limit Case, $|\nu|\rightarrow\infty$}}

Before we proceed to the main proof, we shall solve exactly the limiting 
case $|\nu|\rightarrow\infty$.  This result will be extremely useful 
in the rest of this paper.

We start with
\begin{equation}
M^{(\pm)}(\nu,k)M^{(\pm)}_o(k),~~|\nu|\rightarrow\infty.
\end{equation}
where from Eq. (6.13) we have
\addtocounter{eqnnum1}{1}
\begin{equation}
M^{(\pm)}_o(k)=\frac{-k[k\mp i(2+a_1)]}{(1\pm ik)^2}.
\end{equation}

We thus have Jost functions which are rational in $k$.  This is the case 
first studied by Bargmann$^{23}$ in the paper which gave the famous phase
 equivalent potentials.

The S-matrix is also rational, 
\addtocounter{eqnnum1}{1}
\begin{equation}
S_o(k)\equiv\frac{M^{(+)}_o}{M^{(-)}_o}=[\frac{k-i(2+a_1)}{k+i(2+a_1)}]
(\frac{1-ik}{1+ik})^2,
\end{equation}
where $(2+a_1)>1$, $a_1=-1+8\psi(1)$.  One can use Bargmann's method to 
uniquely determine a potential $V_o(x)$ which has the S-matrix given here.  
But we prefer to determine $V_o$ by using Marchenko's method.

The Marchenko kernel $F_o$ is
\addtocounter{eqnnum1}{1}
\begin{equation}
F_o(x)=\frac{1}{2\pi}\int_{-\infty}^{\infty}(S_o(k)-1)e^{ikx}dk.
\end{equation}
This Fourier transform converges in the mean, $(S_o-1)\rightarrow O(1/k)$ 
as $k\rightarrow\pm\infty$.

By contour integration 
\addtocounter{eqnnum1}{1}
\begin{equation}
F_o(x)=\lambda_oe^{-x}+\lambda_1xe^{-x},
\end{equation}
where
\addtocounter{eqnnum1}{1}
\begin{equation}
\lambda_o=\frac{8a_1+4a_1^2-4}{(3+a_1)^2},
\end{equation}
and
\addtocounter{eqnnum1}{1}
\begin{equation}
\lambda_1=\frac{-4(1+a_1)}{(3+a_1)}.
\end{equation}
Both $\lambda_o$ and $\lambda_1$ are negative.

The Marchenko equation is,
\addtocounter{eqnnum1}{1}
\begin{equation}
A_o(x,y)=F_o(x+y)+\int_{x}^{\infty}A_o(x,u)F_o(u+y)du;
\end{equation}
and $A_o(x,y)\equiv0$ for $y<x$.  With $F_o$ as defined by Eq. (7.5), 
one can easily obtain the exact solution of the integral equation (7.8).

From Bargmann's paper$^{23}$ it is clear that we have the ansatz,
\addtocounter{eqnnum1}{1}
\begin{equation}
A_o(x,y)\equiv[B(x)+(y-x)C(x)]e^{-(y-x)}.
\end{equation}
Substituting this trial solution in Eq. (7.8) and carrying out the 
integration over $u$ we get
\addtocounter{eqnnum1}{1}
\begin{eqnarray}
A_o(x,y)&=&e^{-(y-x)}\{e^{-2x}[(\frac{B(x)}{2}+\frac{C(x)}{4})
(\lambda_o+\lambda_1x)\\ \nonumber
&+&\frac{(B(x)+C(x))}{4}\lambda_1+\lambda_o+x\lambda_1] 
+ye^{-2x}[\frac{B(x)\lambda_1}{2}+\frac{C(x)\lambda_1}{4}+\lambda_1]\}.
\end{eqnarray}
But from Eq. (7.9) we also have
\addtocounter{eqnnum1}{1}
\begin{equation}
A_o(x,y)=[(B-xC)+yC]e^{-(y-x)}.
\end{equation}
Comparing the coefficients of $y$ in Eqs. (7.10) and (7.11) we get
\addtocounter{eqnnum1}{1}
\begin{equation}
e^{2x}C(x)=\frac{B(x)\lambda_1}{2}+\frac{C(x)\lambda_1}{4}+\lambda_1,
\end{equation}
and the terms to zero order in $y$ give
\addtocounter{eqnnum1}{1}
\begin{equation}
(\frac{B}{2}+\frac{C}{4})(\lambda_o+\lambda_1x)+\frac{B\lambda_{1}}
{4}+\frac{C\lambda_1}{4}+\lambda_o+x\lambda_1=e^{2x}[B-xC].
\end{equation}
These last two equations determine $B(x)$ and $C(x)$ giving,
\addtocounter{eqnnum1}{1}
\begin{equation}
C(x)=\frac{\lambda_1-(\lambda^2_1/4)e^{-2x}}{[e^{2x}-(\frac{\lambda^2_1}
{16})e^{-2x}-\frac{1}{2}(\lambda_o+\lambda_1)-\lambda_1x]},
\end{equation}
and
\addtocounter{eqnnum1}{1}
\begin{equation}
B(x)=\frac{\lambda_o+2\lambda_1x+(\frac{\lambda^2_1)}{4})e^{-2x}}
{[e^{2x}-(\frac{\lambda^2_1}{16})e^{-2x}-\frac{1}{2}(\lambda_o+\lambda_1)
-\lambda_1x]},
\end{equation}
Since $\lambda_1<0$, $\lambda_o<0$, and $(\lambda_1^2/16)<<1$, 
the denominators in Eqs. (7.14) and (7.15) do not vanish for any $x\geq 0$.

One can simplify Eqs. (7.14) and (7.15) by defining
\addtocounter{eqnnum1}{1}
\begin{equation}
\rho\equiv-\frac{\lambda_1}{4}=(\frac{1+a_1}{3+a_1})=0.1474.
\end{equation}
Then its is easy to show that
\addtocounter{eqnnum1}{1}
\begin{equation}
-\frac{1}{2}(\lambda_o+\lambda_1)=1-\rho^2.
\end{equation}

We obtain
\addtocounter{eqnnum1}{1}
\begin{equation}
C(x)=\frac{-(4\rho+4\rho^2e^{-2x})}{[e^{2x}-\rho^2e^{-2x}
+(1-\rho^2)+4\rho x]},
\end{equation}
and
\addtocounter{eqnnum1}{1}
\begin{equation}
B(x)=\frac{(2\rho^2+4\rho-2)-8\rho x+4\rho^2e^{-2x}}{[e^{2x}
-\rho^2e^{-2x}+(1-\rho^2)+4\rho x]}.
\end{equation}
From Eq. (7.9) we see that the potential, $V_o(x)$, is
\addtocounter{eqnnum1}{1}
\begin{equation}
V_o(x)=-2\frac{dA_o}{dx}(x,x)=-2\frac{dB(x)}{dx}.
\end{equation}
This leads to
\addtocounter{eqnnum1}{1}
\begin{equation}
V_o(x)=\frac{(4-2\rho^2-16\rho)e^{2x}-4(1-\rho^2)\rho^2e^{-2x}
-16\rho x(e^{2x}+\rho^2e^{-2x})-64\rho^2x-16\rho^2}{[e^{2x}-\rho^2e^{-2x}
+(1-\rho^2)+4\rho x]^2}.
\end{equation}
Note that $V_o(x)=O(e^{-2x})$ as $x\rightarrow+\infty$.

The two Jost solutions $f^{(\pm)}_o$ are given by
\addtocounter{eqnnum1}{1}
\begin{equation}
f^{(\pm)}_o(k,x)=e^{\mp ikx}+
\int_{x}^{\infty}dy A_o(x,y)e^{\mp iky}.
\end{equation}
Substituting Eq. (7.9) for $A_o$, we get
\addtocounter{eqnnum1}{1}
\begin{equation}
f_o^{\pm}=e^{\mp ikx}\{1+\frac{B(x)}{(1\pm ik)}+\frac{C(x)}{(1\pm ik)^2}\}
\end{equation}

One can now check directly that
\addtocounter{eqnnum1}{1}
\begin{equation}
\frac{-d^2f_o^{\pm}}{dx^2}+V_o(x)f^{\pm}_o=k^2f_o^{\pm},
\end{equation}
also $f_o^{(\pm)}\rightarrow M^{\pm}_o(k)$ as $x\rightarrow 0$.

The results of this section can also be obtained by using the technique
 developed by Bargmann$^{23}$ which preceded the results of refs. 6 and 7. 
  The Jost functions defined in Eq. (7.2) do indeed determine uniquely the 
  potential $V_o(x)$ and its solutions $f_o^{(\pm)}(k,x)$.  
  The fact that $M_o^{(\pm)}(k)=O(k)$ as $k\rightarrow 0$, 
  leads to the degenerate case using Bargmann's method but 
  the final results agree with those by Marchenko's method$^{24}$.

Finally, it should be remarked that the full scattering amplitude 
for $V_o(x)$ has a pole at $k=0$.  However, this is not part of the
 point spectrum.  There is no $L^2(0,\infty)$ solution of the Schrodinger 
 equation with $V_o(x)$ for $k=0$.
\newpage
\vspace{.25in}
\setcounter{chanum}{8}
\setcounter{eqnnum1}{1}
\noindent{\bf\Large{VIII. The Marchenko Equation}}

We define an $S$-matrix as in Eq. (6.34)
\begin{equation}
S(\nu,k)=\frac{M^{(+)}(\nu,k)}{M^{(-)}(\nu,k)},
\end{equation}
where $\nu\in{\cal S}(T_o)$.  $S(\nu,k)$ is, for any $\nu$, analytic 
in $k$ in the strip $-1<Imk<+1$.  Next we \underline{define} a Marchenko
 kernel for $S(\nu,k)$,
\addtocounter{eqnnum1}{1}
\begin{equation}
F(\nu;x)=\frac{1}{2\pi}\int_{L}dk[S(\nu,k)-1]e^{ikx}, ~~x>0,
\end{equation}
where $L$ is a line $Im k=\delta>0$, $1>\delta>0$, and without loss 
of generality we fix $\delta,\delta=1/4$.  This Fourier transform is 
convergent in the mean, $(S-1)=O(1/k)$ as $Re k\rightarrow\pm\infty$.

Actually we can also perform an integration by parts on (8.2) for 
any $x>\varepsilon>0$.  Since $dS/dk$ is bounded 
and $(dS/dk)=O(1/k^2)$ as $k\rightarrow\pm\infty$, 
we have absolute convergence for any $x>0$.

It is important to note here that as shown in Eq. (6.32), 
$[S(\nu,k)-1]\in L_2(-\infty,+\infty)$, along the line $Im k=1/4$.

Of course Eq. (8.2) is not the standard definition of the Marchenko kernel.
  In the standard case one integrates along the real $k$-axis. 
  If we move the contour in (8.2) to the real axis, then there could be an 
  extra contribution from the pole produced by the zero of 
  $M^{-)}(\nu,k)$ when $Re\xi(\nu+1/2)<0$.  But for a Martin 
  type S-matrix all the scattering data, including that coming 
  from the point spectrum, is contained in the discontinuity across
   the branch cut on the imaginary k-axis, see references 20,21, and 22.

We prove the following lemma:

{\underline{Lemma 8.1:}}

$F(\nu,x)$ is a.) continuous and differentiable in $x$, $x\in[0,\infty)$; 
b.) $F(\nu,x)=O(e^{-x})$as $x\rightarrow+\infty$; c.)  Both $F(\nu,0)$ 
and $F'(\nu,0)$ are finite, and
\addtocounter{eqnnum1}{1}
\begin{eqnarray}
\int_{o}^{\infty}|F(\nu,x)|dx<\infty,\\ \nonumber
\int_{o}^{\infty}|F'(\nu,x)|dx<\infty
\end{eqnarray}

d.)  $F(\nu,x)$ is analytic in $\nu$, for $\nu\epsilon{\cal {S}}
(T_0)$, and fixed $x\geq0$.
 
{\underline{Proof:}}

In the Appendix, we prove that $F(\nu,x)$ can be written as,
\addtocounter{eqnnum1}{1}
\begin{equation}
F(\nu,x)=\frac{1}{\pi}\int_{1}^{\infty}D(\nu,\alpha)e^{-\alpha x}d\alpha,
\end{equation}
where
\addtocounter{eqnnum1}{1}
\begin{equation}
D(\nu,\alpha)=\frac{\pi(\nu^2-1/4)\psi(\alpha)\alpha^{\frac{1}{4}}
[\alpha^\frac{\nu}{2}+\alpha^{\frac{-\nu}{2}}]}
{M^{(-)}(\nu,i\alpha)}.
\end{equation}
This result is obtained by deforming the contour in Eq. (8.2) and
 using the original representation (3.10) for $M^{(\pm)}(\nu,k)$. 
  We note that $S(\nu,k)$ is analytic for $Imk>0$, except on the 
  cut $k=i\tau$; $1\leq\tau<\infty$.  Next we note that
\addtocounter{eqnnum1}{1}
\begin{equation}
|M^{(-)}(\nu;i\alpha)|\geq|\frac{\alpha(\alpha+(2+a_1))}
{(1+\alpha)^2}|-|g||M^{(-)}_1(i\alpha)+R^{(-)}_1(\nu;i\alpha )|.
\end{equation}

This follows from Eqs. (6.4) and (6.13).  Since $(2+a_1)>1$,we get
\addtocounter{eqnnum1}{1}
\begin{equation}
|M^{(-)}(\nu,i\alpha)|\geq\frac{1}{2}+\frac{1}{2}[\frac{2+a_1}
{(1+\alpha)}]-\frac{C_2}{T^2_o};~~~~~\alpha\geq1;
\end{equation}
where the last term comes from lemma (6.1).  Hence we have
\addtocounter{eqnnum1}{1}
\begin{equation}
|M^{(-)}(\nu,i\alpha)|\geq\frac{1}{2}.
\end{equation}
Finally we obtain from (8.5), with $|Re \nu|<1/2$,
\addtocounter{eqnnum1}{1}
\begin{equation}
|D(\nu,\alpha)|\leq2\pi(\nu^2-\frac{1}{4})e^{-\pi\alpha}\alpha^{\frac{1}{2}}
, ~~ \alpha\geq 1.
\end{equation}
This bound guarantees the absolute and uniform convergence of the Laplace 
transform in Eq. (8.4) for all $x\in[0,\infty)$, and hence all the assertions
 a.), b.), and c.) of our lemma are true.  Finally, d.) is also true,
  given the lower bound in Eq. (8.8) and the uniform bound 
  on $(\alpha^{\nu/2}+\alpha^{-\nu/2})$ for $\nu\epsilon{\cal{S}}(T_o)$.

The Marchenko Equation can now be defined as
\addtocounter{eqnnum1}{1}
\begin{equation}
A(\nu;x,y)=F(\nu;x+y)+\int_{x}^{\infty}du A(\nu;x,u)F(\nu;u+y)
\end{equation}
with
\addtocounter{eqnnum1}{1}
\begin{equation}
A(\nu;x,y)\equiv 0,~~~~y<x.
\end{equation}

The integral equation (8.10) is of Fredholm type and the Hilbert-Schmidt
 norm of $F$ is finite,
\addtocounter{eqnnum1}{1}
\begin{equation}
\thickbar  F \thickbar^2=\int_{x}^{\infty}du\int_{x}^{\infty}dv\mid 
F(\nu;u+v)\mid^2<\infty
\end{equation}
This follows from lemma 8.1.

We first prove the following lemma:

{\underline{Lemma 8.2:}}

For all $\nu\epsilon {\cal{S}}(T_o)$ and $x\geq0$, we have 
\addtocounter{eqnnum1}{1}
\begin{equation}
|F(\nu;x)-F_o(x)|<\frac{C}{|Im\nu|^2}e^{-x/4},
\end{equation}
where the constant $C$ is bounded
\addtocounter{eqnnum1}{1}
\begin{equation}
C~<~2\times 10^3
\end{equation}

{\underline{Proof:}}

From Eqs. (6.4) and (6.5) we have 
\addtocounter{eqnnum1}{1}
\begin{equation}
M^{(\pm)}(\nu,k)-M_0^{(\pm)}(k)=gM_1^{(\pm)}(k)+gR_1^{(\pm)}(\nu;k),
\end{equation}
where $g=(\nu^2-1/4)^{-1}$.  From the definitions of $F(\nu,x)$ 
and $F_o(x)$ we get
\addtocounter{eqnnum1}{1}
\begin{equation}
F(\nu,x)-F_o(x)=\frac{1}{2\pi}\int_L dk[\frac{M^{(+)}(\nu,k)}
{M^{(-)}(\nu,k)}-\frac{M^{(+)}_o(k)}{M_o^{(-)}(k)}]e^{ikx}.
\end{equation}
This gives
\addtocounter{eqnnum1}{1}
\begin{equation}
F-F_o=(\Delta F)_1+(\Delta F)_2,
\end{equation}
with
\addtocounter{eqnnum1}{1}
\begin{equation}
(\Delta F)_1\equiv\frac{g}{2\pi}\int_L e^{ikx}\frac{(M^{(-)}_oM^
{(+)}_1-M^{(+)}_oM^{(-)}_1)}{M^{(-)}_oM^{(-)}}.
\end{equation}
and
\addtocounter{eqnnum1}{1}
\begin{equation}
(\Delta F)_2\equiv\frac{g}{2\pi}\int_L e^{ikx}\frac{(M^{(-)}_oR^{(+)}_1
-M^{(+)}_oR^{(-)}_1)}{M^{(-)}_oM^{(-)}}.
\end{equation}
Again here $L$ is the line $k=(\lambda+\frac{i}{4})$, 
and $-\infty<\lambda<+\infty$.

In order to get bounds on $(\Delta F)_{1,2}$, we need to separate
 out the terms which are only conditionally convergent, 
 i.e. $O(1/k)$ as $k\rightarrow\infty$, from those which are 
 absolutely convergent and hence easier to handle.

From Eq. (5.36) and Eq. (6.5), we write
\addtocounter{eqnnum1}{1}
\begin{equation}
M^{(\pm)}_1(k)=\frac{b_1}{(1\pm ik)}+\hat{M}^{(\pm)}_1(k);
\end{equation}
and
\addtocounter{eqnnum1}{1}
\begin{equation}
R^{(\pm)}_1(\nu,k)=\int_1^{\infty}d\alpha\frac{\Phi_1^{(2)}(\alpha)}
{(\alpha\pm ik)}(\alpha^{\nu/2}+\alpha^{-\nu/2})+\hat{R}^{(\pm)}_1(\nu,k);
\end{equation}
where
\addtocounter{eqnnum1}{1}
\begin{equation}
\hat{M}^{(\pm)}_1(k)=\sum_{\ell=2}^{4}\frac{b_{\ell}}{(1\pm ik)^{\ell}};
\end{equation}

\addtocounter{eqnnum1}{1}
\begin{equation}
\hat{R}_1^{(\pm)}(\nu,k)=\int_{1}^{\infty}d\alpha\sum_{\ell=2}^{5}
\frac{\Phi_{\ell}^{(2)}(\alpha)}{(\alpha\pm ik)^{\ell}}[\alpha^
{\nu/2}+\alpha^{-\nu/2}].
\end{equation}
Both $\hat{M}_1$ and $\hat{R}_1$ are $O(1/k^2)$ as $|k|\rightarrow\infty$.

For $(\Delta F)_1$, we can write
\addtocounter{eqnnum1}{1}
\begin{equation}
(\Delta F)_{1}\equiv(\Delta F)_{11}+(\Delta F)_{12},
\end{equation}
with
\addtocounter{eqnnum1}{1}
\begin{equation}
(\Delta F)_{11}\equiv\frac{gb_1}{2\pi}\int_L dk e^{ikx}[\frac{M_o^{(-)}(k)}
{(1+ik)}-\frac{M_o^{(+)}(k)}{(1-ik)}](\frac{1}{M_o^{(-)}M^{(-)}}),
\end{equation}
and
\addtocounter{eqnnum1}{1}
\begin{equation}
(\Delta  F)_{12}=\frac{g}{2\pi}\int_{L}dk e^{ikx}\left[\frac{M^{(-)}_0(k)
\hat{M}^{(+)}_1(k)-M^{(+)}_0(k)\hat{M}^{(-)}_1(k)}{M^{(-)}_0(k)M^{(-)}(\nu,k)}
\right]
\end{equation}

The integral in (8.25) is conditionally convergent, 
$|M^{\pm}_o(k)|\rightarrow 1$ as $|k|\rightarrow\infty$, 
and $|M^{(-)}(\nu,k)|$ is bounded from below for all $k$ 
with $Imk\geq1/4$.  Also $|M^{(-)}(\nu,k)|\rightarrow1$ 
as $|k|\rightarrow\infty$.

To obtain a bound on $(\Delta F)_{11}$ we first note the following:
\addtocounter{eqnnum1}{1}
\begin{equation}
\int_{L}dk e^{ikx}\left\{(\frac{M^{(-)}_0(k)}{(1-ik)})\frac{1}{M^{(-)}_0(k)M^
{(-)}(\nu,k)}\right\}=0.
\end{equation}
This follows from Jordan's lemma.  The integrand in (8.27) is analytic 
for $Imk\geq 0$, and the bracketed term is $O(1/k)$ for large $|k|$.

Adding twice the $\ell.h.s$ of Eq. (8.27) to Eq. (8.25), one obtains
\addtocounter{eqnnum1}{1}
\begin{equation}
(\Delta F)_{11}=\frac{gb_1}{2\pi}\int_{L}dke^{ikx}\left[\frac{2M^{(-)}_0(k)}
{(1+k^2)}-\frac{(M^{(+)}_0(k)-M^{(-)}_0(k))}{(1-ik)}\right](\frac{1}
{M_0^{(-)}(k)M^{(-)}(\nu,k)})
\end{equation}

To obtain an upper bound on $|(\Delta F)_{11}|$, we first need lower 
bounds on $M^{(-)}_0(k)$ for $k\in L$.

From Eq. (5.13) we have,
\addtocounter{eqnnum1}{1}
\begin{equation}
|M^{(-)}_0(k)|\geq 1-\frac{|a_1|}{|1-ik|}-\frac{|a_2|}{|1-ik|^2},
\end{equation}
with $|a_1|=0.654$, and $|a_2|=0.346$, and $k=\lambda +i/4$,
 $-\infty<\lambda<+\infty$, we finally have
\addtocounter{eqnnum1}{1}
\begin{equation}
|M^{(-)}_0(k)|\geq 0.255>1/4~;~ k\in L.
\end{equation}

Lemma 6.1 will now give us a lower bound for $M^{(-)}(\nu,k), 
\nu\in{\cal{S}}$ and $k=\lambda+i/4$.  Since $|M^{(-)}-M^{(-)}_0|\leq 
C_2/T_o^2$ we get
\addtocounter{eqnnum1}{1}
\begin{equation}
|M^{(-)}(\nu,k)|\geq 1/4,
\end{equation}
for $k=\lambda+i/4$.

Finally, we need to bound $M^{(+)}_0-M^{(-)}_0$, where from Eq. (5.13)
\addtocounter{eqnnum1}{1}
\begin{equation}
M^{(+)}_0-M^{(-)}_0=\frac{-2ika_1}{(1+k^2)}-\frac{4ika_2}{(1+k^2)^2},
\end{equation}
But for $k=\lambda+i/4$, $|k/(1+ik)|<1$, we obtain
\addtocounter{eqnnum1}{1}
\begin{equation}
|M^{(+)}_0(k)-M^{(-)}_0(k)|\leq\frac{\delta}{|1-ik|},
\end{equation}
with
\addtocounter{eqnnum1}{1}
\begin{equation}
\delta=2|a_1|+\frac{64}{15}|a_2|\cong 2.78.
\end{equation}
The above bounds lead us immediately to
\addtocounter{eqnnum1}{1}
\begin{equation}
|\Delta F|_{11}\leq\frac{2|g||b_1|e^{\frac{-x}{4}}}{\pi}\int_{-\infty}
^{+\infty}d\lambda\left\{\frac{2}{|1+(\lambda+i/4)^2|}+\frac{4\delta}
{|1-ik|^2}  \right\},
\end{equation}
\addtocounter{eqnnum1}{1}
\begin{equation}
\leq \frac{c_{11}}{|Im\nu|^2}e^{-x/4},
\end{equation}
where
\addtocounter{eqnnum1}{1}
\begin{equation}
c_{11}={\frac{4|b_1|}{\pi}(2+4\delta}).\int_{0}^{\infty}\frac{d\lambda}
{(\frac{16}{15}+\lambda^2)}\leq290.
\end{equation}
The bound for $(\Delta F)_{12}$ is easier to calculate, from (8.26)
\addtocounter{eqnnum1}{1}
\begin{equation}
|(\Delta F)_{12}|\leq\frac{|g|e^{-x/4}}{2\pi}\int_{-\infty}^
{+\infty}d\lambda\left\{4|\hat{M}_1^{(+)}(k)|+4|\hat{M}^
{(-)}_1(k)|.|\frac{M^{(+)}_0}{M^{(-)}_0}|\right\},
\end{equation}
For $k=\lambda+i/4$, a simple calculation gives
\addtocounter{eqnnum1}{1}
\begin{equation}
|\frac{M^{(+)}_0(k)}{M^{(-)}_0(k)}|\leq 5/3.
\end{equation}

Using Eq. (8.22) we get
\addtocounter{eqnnum1}{1}
\begin{equation}
|(\Delta F)_{12}|\leq\frac{c_{12}e^{-x/4}}{|Im\nu|^2},
\end{equation}
with
\addtocounter{eqnnum1}{1}
\begin{equation}
c_{12}=\frac{4}{\pi}(\sum_{\ell=2}^{4}|b_{\ell}|)
   \int_{0}^{\infty}d\lambda
   [\frac{1}{[(\frac{3}{4})^2+\lambda^2]}
    +(\frac{5}{3})\frac{1}{[(\frac{5}{4})^2+\lambda^2]}]
   \cong 112.
\end{equation}
The values of $b_{\ell}$ are given in Eq. (5.40).

To estimate $(\Delta F)_2$ we again split $R^{(\pm)}
(\nu,k)$ into two terms as in Eq. (8.21).  We write
\addtocounter{eqnnum1}{1}
\begin{equation}
(\Delta F)_2\equiv(\Delta F)_{21}+(\Delta F)_{22}.
\end{equation}
where
\addtocounter{eqnnum1}{1}
\begin{equation}
(\Delta F)_{21}
    = \frac{g}{2\pi} \int_{L}dk e^{ikx}
      \frac{[M^{(-)}_0(k)r_1^{(+)}-M^{(+)}_0(k)r^{(-)}_1]}
      {M^{(-)}_0(k)M^{(-)}(\nu,k)};
\end{equation}
and
\addtocounter{eqnnum1}{1}
\begin{equation}
(\Delta F)_{22}=\frac{g}{2\pi}\int_{L}dke^{ikx}[\frac 
{M^{(-)}_0(k)\hat{R}_1^{(+)}-M^{(+)}_0\hat{R}^{(-)}_1]}
{M^{(-)}_0(k)M^{(-)}(\nu,k)};
\end{equation}
with $\hat{R}_1^{(\pm)}$ given in Eq. (8.23) and
\addtocounter{eqnnum1}{1}
\begin{equation}
r_1^{(\pm)}(\nu,k)=\int_{1}^{\infty}d\alpha\frac{\Phi^{(2)}_1(\alpha)}
{(\alpha\pm ik)}[\alpha^{\nu/2}+\alpha^{-\nu/2}].
\end{equation}

Equation (8.43) can be rewritten as
\addtocounter{eqnnum1}{1}
\begin{equation}
(\Delta F)_{21}
   =\frac{g}{2\pi} \int_{L}dk e^{ikx}
   \{\frac{M^{(-)}_0(r^{(+)}_1+r^{(-)}_1)}{M^{(-)}_0           
     M^{(-)}}
   -\frac{(M^{(+)}_0-M^{(-)}_0)r^{(-)}_1}{M^{(-)}_0                              M^{(-)}}\}.
\end{equation}
Here we have again used Jordan's lemma, which implies
\addtocounter{eqnnum1}{1}
\begin{equation}
\int_{L}dk                                                    
e^{ikx}[\frac{M^{(-)}_0r^{(-)}_1}{M^{(-)}_0M^{(-)}}]\equiv 0.
\end{equation}

From Eqs. (8.45) and (8.46) we get
\addtocounter{eqnnum1}{1}
\begin{equation}
|\Delta F)_{21}|\leq\frac{|g|e^{-x/4}}{2\pi}[\int_{-\infty}^
{+\infty}d\lambda\int_{1}^{\infty} d\alpha(2\alpha^{1/4})
|\Phi^{(2)}_1(\alpha)|\{\frac{8\alpha}{|\alpha^2+k^2|}
+\frac{16|M^{(+)}_0-M^{(-)}_0|}{|\alpha+1/4-i\lambda|}\}
\end{equation}
where $k=\lambda+i/4$.  Using the bound (8.33) we obtain
\addtocounter{eqnnum1}{1}
\begin{equation}
|(\Delta F)_{21}|\leq\frac{c_{21}e^{-x/4}}{|Im\nu|^2},
\end{equation}
with
\addtocounter{eqnnum1}{1}
\begin{eqnarray}
c_{21}&=&\frac{1}{2\pi}\int_{1}^{\infty}d\alpha(2\alpha^{1/4})
|\Phi^{(2)}_1(\alpha)|\\ \nonumber
&.&\int_{-\infty}^{+\infty}d\lambda[\frac{8\alpha}{(\alpha^2
-\frac{1}{16})+\lambda^2}+\frac{16\delta}{|\alpha+1/4-i\lambda||\frac{5}{4}
-i\lambda|}]
\end{eqnarray}
This leads to
\addtocounter{eqnnum1}{1}
\begin{equation}
c_{21}\leq\int_{1}^{\infty}d\alpha\alpha^{1/4}|\Phi^{(2)}_1(\alpha)
|[\frac{128}{15}+\frac{64}{5}\delta],\leq(45).\int_{1}^{\infty}d\alpha 
\alpha^{1/4}|\Phi^{(2)}_1(\alpha)|.
\end{equation}
Using the definition of $\Phi^{(2)}_1(\alpha)$ in Eq. (5.26),
 table I, and the bounds on $\psi^{(j)}(\alpha)$ given in Eq. 
 (6.10), one can easily get a rough numerical bound on the above integral, 
 $\int_{1}^{\infty}d\alpha\alpha^{1/4}|\Phi^{(2)}(\alpha)|\leq 10$, and hence
\addtocounter{eqnnum1}{1}
\begin{equation}
c_{21}\leq 450.
\end{equation}
$(\Delta F)_{22}$.  From Eqs. (8.44), (8.30), (8.31), and (8.23), we get
\addtocounter{eqnnum1}{1}
\begin{equation}
|(\Delta F)_{22}|\leq
\frac{4|g|}{4}e^{-x/4}\sum_{\ell=2}^{5}\int_{1}^{\infty}
d\alpha \alpha^{1/4}|\Phi^{(2)}_{\ell}(\alpha)|\int_{-\infty}^
{+\infty}d\lambda\{\frac{1}{|\alpha+ik|^{\ell}}+|\frac{M^{(+)}_0}
{M^{(-)}_0}|\frac{1}{|\alpha-ik|^{\ell}}\}
\end{equation}
with $k=\lambda+i/4$.  Using the fact that $|\alpha\pm ik|
\geq[(\alpha\mp 1/4)^2+\lambda^2]^{1/2}$, and $|M^{(+)}_0/
M^{(-)}_0|<(5/3)$ for $k\in L$, we have
\addtocounter{eqnnum1}{1}
\begin{equation}
|(\Delta F)_{21}|\leq\frac{c_{22}}{|Im\nu|^2}.e^{-x/4}
\end{equation}
with
\addtocounter{eqnnum1}{1}
\begin{equation}
c_{22}\leq    \frac{32}{3\pi}\sum_{\ell=2}^{5}\int_{1}^
{\infty}d\alpha \alpha^{1/4}|\Phi^{(2)}_{\ell}(\alpha)|(\frac{1}
{\alpha-1/4})^{\ell-1}.\beta_{\ell},
\end{equation}
where
\addtocounter{eqnnum1}{1}
\begin{equation}
\beta_{\ell}\equiv\int_{-\infty}^{+\infty}\frac{du}{(1+u^2)^{\ell/2}}                           .
\end{equation}
For $\alpha\geq1$, $|(\alpha/(\alpha-1/4)|<4/3$ and hence
\addtocounter{eqnnum1}{1}
\begin{equation}
c_{22}\leq\frac{32}{3\pi}\sum_{\ell=2}^{5}A_{\ell}.\beta_{\ell}.
(4/3)^{\ell}
\end{equation}
with
\addtocounter{eqnnum1}{1}
\begin{equation}
A_{\ell}=\int_{1}^{\infty}d\alpha \alpha^{5/4-\ell}|\Phi^{(2)}_{\ell}
(\alpha)|
\end{equation}
A simple numerical estimate will give
\addtocounter{eqnnum1}{1}
\begin{equation}
c_{22}\leq 10^3.
\end{equation}
This completes the proof of our Lemma with $C=c_{11}+c_{12}+c_{21}
+c_{22}<2\times 10^3$.

Lemma (8.2) guarantees that for $\nu\in{\cal{S}}$, $\thickbar 
F\thickbar =\thickbar F_o\thickbar +O(\frac{1}{T^2_0})$.  
Indeed we have for the Hilbert-Schmidt norms
\addtocounter{eqnnum1}{1}
\begin{equation}
\thickbar F \thickbar\leq\thickbar F_0 \thickbar+\thickbar F-F_0 
\thickbar
\end{equation}
However, from the lemma
\addtocounter{eqnnum1}{1}
\begin{eqnarray}
\thickbar F-F_0\thickbar ^2&=&\int_{x}^{\infty}du\int_{x}^
{\infty}dv|F(\nu,u+v)-F_0(u+v)|^2\\ \nonumber
&\leq&\frac{C^2}{|\nu|^4}\int_{x}^{\infty}du\int_{x}^{\infty}dve^
{\frac{-(u+v)}{2}}\\ \nonumber
&\leq&\frac{C^2}{|\nu|^4}\int_{0}^{\infty}w e^{\frac{-w}{2}}dw.
\end{eqnarray}
Thus
\addtocounter{eqnnum1}{1}
\begin{equation}
\thickbar F-F_0 \thickbar \leq\frac{2C}{T^2_0}\leq\frac{2}{T_0}.
\end{equation}

We have the exact expression for $F_0(x)$ given in Eq. (7.5) and we 
can calculate $\thickbar F_0\thickbar$ exactly.  The Hilbert-Schmidt 
norm for $\thickbar F_0\thickbar$ is
\addtocounter{eqnnum1}{1}
\begin{equation}
\thickbar F_0\thickbar^2_x=\int_{x}^{\infty}du\int_{x}^{\infty}dv[F_0(u+v)]^2.
\end{equation}
Note that here $x$ appears as a parameter (see Eq. (8.10)).  
As $x$ increases the norm of $F_0$ tends to zero.  From Eq. (7.5) we get
\addtocounter{eqnnum1}{1}
\begin{equation}
\thickbar F_0\thickbar ^2_{x=0}=\int_{0}^{\infty}dw[\lambda^2_0w+2\
lambda_0\lambda_1w^2+\lambda_1w^3]e^{-2w},
\end{equation}
with $\lambda_0$ and $\lambda_1$ given by Eqs. (7.6) and (7.7).  
After some algebra we obtain,
\addtocounter{eqnnum1}{1}
\begin{equation}
\thickbar  F_0\thickbar ^2_{x=o} =1+\frac{\lambda^2_1}{16}~~,
\end{equation}
with $\lambda_1=-0.590$.  Thus the Hilbert-Schmidt norm for 
$x=0$ is slightly bigger than one, however, it is easy to show 
that for $x>x_o\cong0.1$,
\addtocounter{eqnnum1}{1}
\begin{equation}
\thickbar F_0\thickbar <1.
\end{equation}

Hence the iterative series for both $F$ and $F_o$ converges for $x>0.1$. 
 But to prove the existence of $A(\nu;x,y)$ for all $x\geq0$, we will use 
 Eq. (8.61) and proceed in another way.

In section VII we gave an explicit solution of the integral equation,
\addtocounter{eqnnum1}{1}
\begin{equation}
A_o=F_o+A_oF_o.
\end{equation}
We can write
\addtocounter{eqnnum1}{1}
\begin{equation}
I_o\equiv (1-F_o)^{-1},
\end{equation}
and
\addtocounter{eqnnum1}{1}
\begin{equation}
A_o=(1-F_o)^{-1}F_o=I_oF_o.
\end{equation}
This leads to
\addtocounter{eqnnum1}{1}
\begin{equation}
A_o+1=I_o.
\end{equation}
The kernel $A_o(x,y)$ is given explicitly in Eq. (7.9) and (7.18-19), 
and thus $I_o(x,y)$ is known.

The full Marchenko equation (8.10) can now be written as
\addtocounter{eqnnum1}{1}
\begin{equation}
A=F+AF.
\end{equation}
We define $I$ as
\addtocounter{eqnnum1}{1}
\begin{equation}
I=(1-F)^{-1},
\end{equation}
and
\addtocounter{eqnnum1}{1}
\begin{equation}
A=(1-F)^{-1}F.
\end{equation}

We prove that both I and A exist and in  (8.70) have a finite norm.  We have
\addtocounter{eqnnum1}{1}
\begin{equation}
A+1=I=(1-F)^{-1}.
\end{equation}
To show that I(and A) exist, we note first,
\addtocounter{eqnnum1}{1}
\begin{equation}
I=(1-F)^{-1}=(1-F_o-\Delta)^{-1},
\end{equation}
where
\addtocounter{eqnnum1}{1}
\begin{equation}
\Delta\equiv F-F_o.
\end{equation}
Using the fact that $(1-F_o)I_o=1$ we get
\addtocounter{eqnnum1}{1}
\begin{eqnarray}
I&=& [(1-F_o)I_o\{(1-F_o)-\Delta\}]^{-1},\\ \nonumber
&=& (1-I_o\Delta)^{-1}I_o.
\end{eqnarray}
Using Eq. (8.74) we get
\addtocounter{eqnnum1}{1}
\begin{eqnarray}
A&=&(1-I_o\Delta)^{-1}I_o-1 \\ \nonumber
&=&(1-I_o\Delta)^{-1}(A_o+I_o \Delta).
\end{eqnarray}
where $A_o+1=I_o$.

Next we define the operator $K$ as
\addtocounter{eqnnum1}{1}
\begin{equation}
K\equiv I_o\Delta=(A_o+1)\Delta,
\end{equation}
and obtain
\addtocounter{eqnnum1}{1}
\begin{equation}
A=A_o+(1-K)^{-1}.(1+A_o)K.
\end{equation}

The Hilbert-Schmidt norm of $\Delta$ is small.  Indeed using 
Eq. (8.61) we have
\addtocounter{eqnnum1}{1}
\begin{equation}
\thickbar \Delta\thickbar^2\equiv\thickbar F-F_o \thickbar^2\leq\frac{C^2}
{|\nu|^4}\leq\frac{2}{T^2_o}<<1.
\end{equation}
However, $K\equiv A_o \Delta+\Delta$, and we get
\addtocounter{eqnnum1}{1}
\begin{equation}
\thickbar K\thickbar\leq\thickbar A_o\thickbar .\thickbar 
\Delta\thickbar +\thickbar \Delta\thickbar.
\end{equation}
$A_o$ is known and $\thickbar A_o\thickbar^2<5$, thus
\addtocounter{eqnnum1}{1}
\begin{equation}
\thickbar K\thickbar \leq\frac{3C}{|\nu|^2}<<1,
\end{equation}
for all $\nu\in{\cal{S}}(T_o)$.  Thus the inverse$(1-K)
^{-1}$ is given by an absolutely convergent series,
\addtocounter{eqnnum1}{1}
\begin{equation}
(1-K)^{-1}=\sum^{\infty}_{n=0}K^n,
\end{equation}
and has a bounded norm, $\thickbar (1-K)^{-1}\thickbar\leq 2$.

The fina result for $A$ is,
\addtocounter{eqnnum1}{1}
\begin{equation}
A=A_o+H+A_oH,
\end{equation}
where
\addtocounter{eqnnum1}{1}
\begin{equation}
H\equiv\sum_{n=1}^{\infty}K^n;
\end{equation}
and $\thickbar H\thickbar\leq 2\thickbar K\thickbar<<1$.  
Indeed we have $\thickbar A-A_o\thickbar \leq\tilde{C}=/|\nu|^2$.

The kernel$K(\nu;x+y)$ can be written as
\addtocounter{eqnnum1}{1}
\begin{equation}
K(\nu;x,y)=F(\nu;x,y)-F_o(x+y)+\int_{x}^{\infty}duA_o(x,u)
[F(\nu;u+y)-F_o(u+y)]
\end{equation}

The properties of the kernel $K(\nu;x,y)$ are similar to 
those of $F(\nu;x,y)$.  We have the following:

{\underline{Lemma 8.3:}}

$K(\nu;x,y)$ is for $y\geq x\geq 0$, a.)  analytic 
for $\nu\in{\cal{S}}(T_o)$; b.)  differentiable in 
both $x$ and $y$; c.) analytic for $Re x\geq 0$ and 
$Re y\geq 0$ when $\nu\in{\cal{S}}(T_o)$; and d.) 
$|K(\nu;x,y)|\leq\frac{C}{|\nu|^2}e^{-(-\frac{x+y}{4})}$.

{\underline{Proof:}}

These results follow from Eq. (8.88), lemma (8.1), and 
the exact result (7.11) for $A_o(x,u)$.  We note that the 
denominators appearing in the expressions (7.18) and (7.19) 
for $B(x)$ and $C(x)$ do not vanish for $Re x\geq 0$.  
Both B and C  are thus analytic in the half plane $Re x\geq 0$.

The full expression for $A(\nu;x,y)$ is
\addtocounter{eqnnum1}{1}
\begin{equation}
A(\nu;x,y)=A_o(x,y)+H(\nu;x,y)+\int_{x}^{\infty}A_o(x,u)H(\nu;u,y).
\end{equation}
where
\addtocounter{eqnnum1}{1}
\begin{equation}
H(\nu;x,y)=\sum_{n=1}^{\infty}K^{(n)}(\nu;x,y),
\end{equation}
and
\addtocounter{eqnnum1}{1}
\begin{equation}
K^{(n)}(\nu;x,y)=\int_{x}^{\infty}du_1...\int_{x}^{\infty}du_{n-1}
K(\nu;x,u_1)K(\nu;u_1,u_2)....K(\nu;u_{n-1},y)
\end{equation}
The series in (8.89) is absolutely and uniformly convergent 
for $y\geq x\geq0$, and all $\nu\in{\cal{S}}(T_o)$.

The properties of $A(\nu;x,y)$ can be summarized in the following lemma:

{\underline{Lemma 8.4:}}

a.)  For $y\geq x\geq 0$, $A(\nu;x,y)$ is analytic in $\nu$ 
for $\nu\in{\cal{S}}(T_o)$; b.) $A(\nu;x,y)$ is differentiable in both 
$x$ and $y$, $y\geq x$.  Also $A(\nu,0,0)$, and 
$[\frac{d}{dx}(A(\nu;x,x)]_{x=0}$ are finite; c.)  
For fixed $\nu$, $\nu\in{\cal{S}}(T_o)$, $A(\nu;x,y)$ 
is analytic in $x$ and $y$ for $Re x\geq 0$, 
$Re y\geq Re x\geq 0$; and d.)  For all $\nu\in{\cal{S}}(T_o)$, 
we have the bound
\addtocounter{eqnnum1}{1}
\begin{equation}
|A(\nu;x,y)-A_o(x,y)|\leq\frac{\tilde{C}}{|\nu|^2}.e^
{-(\frac{x+y}{4})}.
\end{equation}

{\underline{Proof:}}

These results follow immediately from Eq. (8.89) and 
lemmas (8.1) and (8.3).  The bound (8.91) follows from the 
bound (8.13) of lemma (8.2).  The constant $\tilde{C}$ is 
certainly such that, $\tilde{C}<10^4$, which is sufficient for 
our purposes at this stage, but can be improved with more 
careful estimates.

The next step is to \underline{define} two functions 
$U(\nu;x)$ and $f^{(\pm)}(\nu;k,x)$ as follows:
\addtocounter{eqnnum1}{1}
\begin{equation}
U(\nu;x)=-2\frac{d}{dx}A(\nu:x,x) ~,~x\geq 0;
\end{equation}
and
\addtocounter{eqnnum1}{1}
\begin{equation}
f^{\pm}(\nu;k,x)=e^{\mp ikx}+\int_{x}^{\infty}dy A(\nu;x,y)e^{\mp iky}.
\end{equation}

Without recourse to the standard methods of inverse scattering, 
one can prove directly the next lemma.

{\underline{Lemma 8.5}}:

For any $\nu\in{\cal S}(T_0)$, $f^{\pm}$ satisfy a Schrodinger 
equation with $U(\nu,x)$ as the potential
\addtocounter{eqnnum1}{1}
\begin{equation}
-\frac{d^2f^{\pm}}{dx^2}+U(\nu;x)f^{\pm}=k^2f^{\pm}.
\end{equation}

{\underline{Proof:}}

From a.) and b.) in Lemma (8.4) it follows that $U(\nu;x)$ is 
analytic in $\nu$ for $\nu\in{\cal{S}}(T_o)$ and $x\geq 0$.  
Similarly, from Eq. (8.94) it follows that $f^{(-)}(\nu;k,x)$ 
with $x\geq 0)$ and $Im k\geq 0$ is also analytic in $\nu$ in 
the truncated strip.  Similarly, $f^{(+)}$ with $Imk\leq 0$ 
is analytic.  The same is true for $\frac{d}{dx}(f^{(\pm)})$, 
and $\frac{d^2}{dx^2}(f^{(\pm)})$ since absolute and uniform 
conveergence allows us to differentiate under the integral 
sign in (8.93).

In the next section we will prove the validity of Eq. (8.94) 
on the line $\nu= it$, $t\geq T_o$.  Hence by analytic 
continuation, the Schrodinger equation (8.94) holds 
for all $\nu\in{\cal{S}}(T_o)$.  

In the Appendix will give a more direct proof of Eq. (8.94) 
and also show that one does indeed recover the original 
Jost function from the potential $U(\nu;x)$. 

\newpage
\vspace{.25in}
\setcounter{chanum}{9}
\setcounter{eqnnum1}{1}
\noindent{\bf\Large{IX.  The Case, $\nu=it$.}}

For purely imaginary $\nu$, our S-matrix, $S(\nu,k)$, is 
unitary and satisfies all the properties needed for the 
standard inverse scattering methods of Gelfand, Levitan 
and Marchenko to be applicable.  We sketch some relevant 
results in this section.

First, we define $S$ as:
\begin{equation}
S(it,k)\equiv\frac{M^{(+)}(it,k)}{M^{(-)}(it,k)},\quad t>\pi^2.
\end{equation}
For real $k$, it follows from Eq. (3.10) that
\addtocounter{eqnnum1}{1}
\begin{equation}
[M^{(+)}(it,k)]^*=M^{(+)}(it,-k)=M^{(-)}(it,k);
\end{equation}

$S(it,k)$ satisfies all the conditions given in 
Faddeev's$^{14}$ review paper, which are sufficient to 
guarantee that the Marchenko equation will lead to a 
unique real potential $U(it,x)$.  We can easily check 
that for real $k$,
\addtocounter{eqnnum1}{1}
\begin{equation}
|S(it,k)|=S(it,0)=S(it,\infty)=1;
\end{equation}
and
\addtocounter{eqnnum1}{1}
\begin{equation}
[S(it,k)]^*=S(it,-k).
\end{equation}

The number of discrete eigenvalues of $S$ for fixed $t>\pi^2$ is at 
most one (see Section VI).  In the physicist's language, we have either 
one bound state or one antibound state.  This is evident from 
Eq. (6.28) and the fact that $\xi(it+1/2)$ is real. 
 We will discuss this point in more detail at the end of this section.

The Marchenko kernel is now given by
\addtocounter{eqnnum1}{1}
\begin{equation}
F(it,x)=\frac{1}{2\pi}\int_{-\infty}^{\infty}dk[S(it,k)-1]e^{ikx}; x>0;
\end{equation}
for the case,
\addtocounter{eqnnum1}{1}
\begin{equation}
\xi(it+\frac{1}{2})\geq0.
\end{equation}
and
\addtocounter{eqnnum1}{1}
\begin{equation}
F(it,x)=\frac{1}{\pi}\int_{-\infty}^{+\infty}dk[S(it,k)-1]
e^{ikx}+c_oe^{-\tau_ox}.
\end{equation}
for the case where $\xi(it+1/2)<0$, i.e. with a bound
 state at $E=-\tau^2_o$.  Both Fourier transforms in 
 Eqs. (9.5) and (9.6) are convergent in the mean 
 since $[S-1]=O(1/k)$ for large $k$.  Also it is 
 clear that $F(it,x)$ is real.

As noted previously $\xi(it+1/2)=O(t^pe^{-\frac{\pi t}{4}})$
and hence small for, $t>\pi^2$, this makes $\tau_o<<1/4$ 
and the bound state is very shallow for $t>\pi^2$. 
One can now move the contour of integration up in
both Eqs. (9.5) and (9.6) to obtain
\addtocounter{eqnnum1}{1}
\begin{equation}
F(it,x)=\frac{1}{2\pi}\int_{L}dk[S(it,k)-1]e^{ikx};\quad x>0;
\end{equation}
for both cases.  Here $L$ is the line $Imk=1/4$.  
In the case of Eq. (9.6), the contribution from the
pole at $k=i\tau_o$ exactly cancels the second 
term on the r.h.s., see ref. 22.

The solution of the Marchenko equation, $A(it;x,y)$ exists, 
is real, and differentiable for $y\geq x>0$.  The resulting 
potential, $U(it,x)$, is real, continuous for all $x\geq 0$, 
and $O(e^{-2x})$ for large $x$.

We close this section by calculating the position of the bound 
state or antibound state for fixed $\nu=it$, $t>T_o$.

Rewriting Eq. (6.26) for $\nu=it$, we get 
\addtocounter{eqnnum1}{1}
\begin{eqnarray}
M^{(-)}(it,k)&=&2\xi(it+\frac{1}{2})+M_o^{(-)}(k)+gM^{(-)}_1(k)\\ \nonumber
&+&2g\int_{1}^{\infty}d\alpha\left(\sum_{\ell=1}^{5}\left\{\frac{1}
{(\alpha-ik)^{\ell}}-\frac{1}{\alpha^{\ell}}\right\}\Phi^{(2)}_{\ell}
(\alpha)[\cos(\frac{t}{2}\log\alpha)]\right),
\end{eqnarray}
where
\addtocounter{eqnnum1}{1}
\begin{equation}
g=\frac{-1}{t^2+1/4}
\end{equation}
Now, $\xi(it+1/2)$ is real and exponentially small for large $t$.  
The three other terms on the r.h.s. of Eq. (9.8) are all $O(k)$ for 
small $k$.

From Eq. (6.13) we have,
\addtocounter{eqnnum1}{1}
\begin{equation}
M^{(-)}_o(k)=-ik(2+a_1)+O(k^2),
\end{equation}
with
\addtocounter{eqnnum1}{1}
\begin{equation}
a_1=-1+8\psi(1)=-0.6543.
\end{equation}
Thus for $t>T_0$, the one zero of $M^{(-)}(it,k)$ will occur only 
when $k=i\tau$ and
\addtocounter{eqnnum1}{1}
\begin{equation}
\tau=-\frac{2\xi(it+1/2)}{(2+a_1)+0(\frac{1}{t^2})}+O(\tau^2),
\end{equation}
where the $O(1/t^2)$ term is real, and the same fore the $O(\tau^2)$ term.  
Note that $M^{(-)}_o(i\tau)$ and $M^{(-)}_1(i\tau)$ are both real as is the 
integral in (9.8) for $k=i\tau$.

The resulting potential, or one parameter family of potentials,
\addtocounter{eqnnum1}{1}
\begin{equation}
U(it,x)\equiv V(g,x),
\end{equation}
has a remarkable property as $t$ increases, $t>T_o$.  
It will have exactly one bound state when $\xi(it+1/2)<0$, 
with energy $E_o=-\tau^2_o$,
\addtocounter{eqnnum1}{1}
\begin{equation}
E_o=-\frac{4[\xi(it+1/2)]^2}{[(2+a_1)+O(\frac{1}{t^2})]^2}
+O([\xi(it+1/2)]^3).
\end{equation}
Then as we pass a Riemann zero and $\xi(it+1/2)>0$, there 
will be no bound state until $t$ reaches the next Riemann zero.

As $t\rightarrow+\infty$, the potential, $U(it,x)$, presents us 
with a seemingly puzzling situation.  The bound state, i.e. a point 
spectrum of one, appears and then as $t$ increases disappears, with 
this process repeating as $t$ increases until as $t\rightarrow\infty$ 
we reach $V_o(x)$ which has no point spectrum.

Schwinger's theorem relating the number of bound states to the number 
of nodes of the zero energy regular solution, $\phi(it;0;x)$, defined 
in Eq. (2.8), is instructive for the present case.

From ref. 13, we have an integral equation for $\phi$,
\addtocounter{eqnnum1}{1}
\begin{equation}
\phi(it;k;x)=\frac{sinkx}{k} + \int_{o}^{x}\frac{sink(x-x')}
{k}V(x')\phi(it;k;x')dx',
\end{equation}
where clearly $\phi(it;k;0)\equiv0$.  The zero energy $\phi$ is 
given by
\addtocounter{eqnnum1}{1}
\begin{equation}
\phi(it;0;x)=[1+\int_{o}^{n}V((x')\phi(it;0;x')dx']x-      
\int_{o}^xV(x')\phi(it;0,x')dx'
\end{equation}

For $x$ not large, $x\leq T_o$, we can approximate $\phi$ by 
the $t\rightarrow\infty$ solution
\addtocounter{eqnnum1}{1}
\begin{equation}
\phi(it;k;x)=\phi_o(k,x)+O(\frac{1}{t^2}),
\end{equation}
where $\phi_o$ is defined as in (2.8) but with $f^{\pm}$ replaced 
by $f^{\pm}_o$ and $M^{\pm}$ replaced by $M^{\pm}_o$.  It is easy 
to check that $\phi_o(0,x)$ is positive for $x$ not large, 
and $\phi_o(0;0)=0$.  But from Eq. (9.17) we see that the 
large $x$ behavior is
\addtocounter{eqnnum1}{1}
\begin{equation}
\phi(it;0;x)\rightarrow[C(t)+o(1)]+x[1+\int_{o}^{\infty}V(x')
\phi(it;0;x')dx'+o(1)]
\end{equation}
where
\addtocounter{eqnnum1}{1}
\begin{equation}
C(t)=-\int_{o}^{\infty}x'V(x')\phi(it;0;x')dx'.
\end{equation}
But under the integral sign we can replace $V$ by $V_o$ 
and $\phi$ by $\phi_o$, and obtain
\addtocounter{eqnnum1}{1}
\begin{equation}
C(t)=-\int_{o}^{\infty}x'V_o(x')\phi_o(it;0,x')dx'+O(\frac{1}{t^2}).
\end{equation}
where $V_o$ and $\phi_o$ are known exactly from section VII.  
It is a simple matter to check that
\addtocounter{eqnnum1}{1}
\begin{equation}
C(t)>0, \:\;\;t>T_o.
\end{equation}
Next in ref. 13, we have the result relating the Jost 
function to $\phi$, and it gives
\addtocounter{eqnnum1}{1}
\begin{equation}
M^{(-)}(it;k)=1+\int_{o}^{\infty}dx'V(it;x')\phi(it;k;x').
\end{equation}
Taking the $k\rightarrow 0$ limit we have from Eq. (9.19)
\addtocounter{eqnnum1}{1}
\begin{equation}
\phi(it;0;x)\rightarrow C(t)+2\xi(it+\frac{1}{2})x, ~~  x\rightarrow\infty.
\end{equation}
We can only have a node in $\phi$ if $\xi(it+\frac{1}{2})<0$, otherwise 
there is no node and no bound state.  For $t>T_o$ and large, the node 
occurs at large values of $x$,
\addtocounter{eqnnum1}{1}
\begin{equation}
x_o\cong\frac{C(t)}{2\xi(it+\frac{1}{2})}\cong O(e^{\frac{\pi t}{4}}).
\end{equation}

This discussion shows that while our asymptotic estimates for $f$, $\phi$, 
and V, are good for low values of $x$, $x<T_o$, one cannot use them for 
large values of $x$ except in estimating integrals as in Eq. (9.21).

The well established results for a unitary $S$ and real potentials 
now guarantee that Schrodinger's equation holds for $f^\pm it;k;x)$ 
and $U(it;x)$, i.e.
\addtocounter{eqnnum1}{1}
\begin{equation}
-\frac{d^2}{dx^2}f^\pm(it;k;x)+U(it;x)f^\pm=k^2f^\pm,~~ t>T_o.
\end{equation}
We are also guaranteed that
\addtocounter{eqnnum1}{1}
\begin{equation}
f^\pm(it;k;0)=M^{(\pm)}(it;k),
\end{equation}
where $M^{(\pm)}$ is the original Jost-function we started with
\addtocounter{eqnnum1}{1}
\begin{equation}
M^{(\pm)}(it;k)=1-(t^2+1/4)\int_{1}^{\infty}\frac{\psi(\alpha)\
alpha^{\frac{1}{4}}[\alpha^{\frac{it}{2}}+\alpha^{\frac{-it}{2}}]}
{[\alpha\pm ik]}d\alpha.
\end{equation}
Thus by analytic continuation the Jost solutions $f^\pm(\nu;k;x)$ 
given in the previous sections will also give the original Jost 
functiona, i.e. (9.28) with $it$ replaced by $\nu$, and $\nu\in{\cal{S}}
(T_o)$.
\addtocounter{eqnnum1}{1}

\newpage
\vspace{.25in}
\setcounter{chanum}{10}
\setcounter{eqnnum1}{1}
\noindent{\bf\Large{X.  Asymptotic Expansion in Powers of g}}

In this section we carry out the asymptotic expansion of the 
kernels $F$ and $A$ and the potential $V$ in powers of g, 
where $g\equiv(\nu^2-1/4)^{-1}$, and $\nu\epsilon S(T_0)$ with 
$T_o>10^3$.

We start with the definition of the Marchenko kernel $F(\nu;x)$ 
given in Eq. (8.2),
\begin{equation}
F(\nu;x)=\frac{1}{2\pi}\int_{L}dk[S(\nu,k)-1]e^{ikx}, x>0,
\end{equation}
where $L$ is the line $Imk=1/4$.

Using the asymptotic expansion of $M^{(\pm)}(\nu,k)$ given in 
Eq. (5.47), we have
\addtocounter{eqnnum1}{1}
\begin{equation}
S(\nu,k)-1=\frac{\displaystyle{\sum_{n=0}^{N}g^nM^{(+)}_n(k)
+g^NR^{(+)}_N(\nu,k)}}{\displaystyle{\sum_{n=0}^{N}}g^nM^{(-)}_n(k)
+g^NR^{(-)}_N(\nu,k)}-1,
\end{equation}
where $R^{(\pm)}(|nu,k)$ are both $0(g)$ for small g.

Next we recall the lower bound obtained in section VIII 
for $M^{(-)}_o(k)$ when $k=\lambda+i(1/4)$, and $-\infty<\lambda<+\infty$. 
This is given in Eq. (8.30)
\addtocounter{eqnnum1}{1}
\begin{equation}
\mid M^{(-)}_o(k)\mid\geq0.255,k\epsilon L.
\end{equation}
From lemma 6.1, we get also a lower bound on $|M^{(-)}(\nu,k)|$ 
for $\nu\epsilon S(T_o)$, and $k\epsilon L$ given in Eq. (8.31)
\addtocounter{eqnnum1}{1}
\begin{equation}
\mid M^{(-)}(\nu,k)\mid\geq1/4~~,~~T_o>10^3.
\end{equation}
The last two bounds guarantee that the denominator in Eq. (10.2) does 
not vanish for any $k\epsilon L$ and $\nu\epsilon S(T_o)$.  
We can then proceed to expand $[S(\nu,k)-1]$ in powers of $g$ for 
any $k\epsilon L$ and get
\addtocounter{eqnnum1}{1}
\begin{equation}
[S(\nu,k)-1]=\sum_{n=0}^{N}g^nH_n(k)+g^NH^{(N)}_R(g,k)
\end{equation}
where from Eq. (10.2) we get
\addtocounter{eqnnum1}{1}
\begin{equation}
H_o(k)\equiv S_o(k)-1=\frac{M^{(+)}_o(k)}{M^{(-)}_o(k)}-1;
\end{equation}
\addtocounter{eqnnum1}{1}
\begin{equation}
H_1(k)\equiv\frac{1}{M^{(-)}_o(k)}\{M^{(+)}_1(k)
-\frac{M^{(+)}_o(k)M^{(-)}_1(k)}{M^{(-)}_o(k)}\},
\end{equation}
and
\addtocounter{eqnnum1}{1}
\begin{eqnarray}
H_2(k)&\equiv&\frac{1}{M^{(-)}_o(k)}\{M^{(+)}_2(k)
-\frac{M^{(+)}_1(k)M^{(-)}_1(k)}{M^{(-)}_o(k)}
-\frac{M^{(-)}_2(k)M^{(+)}_o(k)}{M^{(-)}_o(k)}\\ \nonumber
&+&\frac{M^{(+)}_o(k)[M^{(-)}_1(k)]^2}{[M^{(-)}_o(k)]^2}\}
\end{eqnarray}
with similar expressions for $H_n(k)$, $n>2$, 
which we will not need in this paper.  The remainder 
term $H^{(N)}_R$ is $O(g)$ as $g\rightarrow 0$.  
All $H_n(k)$ are rational functions of $k$.

Eq. (10.5) immediately gives us the asymptotic expansion for 
the kernel $F(\nu;x)$,
\addtocounter{eqnnum1}{1}
\begin{equation}
F(\nu;x)=F_o(x)+gF_1(x)+g^2F_2(x)+ ...+g^NF^{(N)}_R(g;x)
\end{equation}
where
\addtocounter{eqnnum1}{1}
\begin{equation}
F_n(x)=\frac{1}{2\pi}\int_{L}dkH_n(k)e^{ikx}, x\geq0.
\end{equation}

This last Fourier transform is conditionally convergent 
since $|H_n(k)|=O(1/k)$ for large $k$ (note that $L$ is the 
line $Im k=1/4$).

Next we stress that all the $H_n(k)$ are analytic for $Im k>0$.  
The denominators in Eqs. (10.6)-(10.8), do not vanish in $Imk\geq0$, 
except at $k=0$.

This also holds for $H^{(N)}_R(g,k)$.  Thus we can shift back the 
contour in Eq. (10.10) to the real $k$-axis and obtain
\addtocounter{eqnnum1}{1}
\begin{equation}
F_n(x)=\frac{1}{2\pi}\int_{-\infty}^{+\infty}dk H_n(k)e^{ikx}, x\geq0.
\end{equation}
for real $k$, it follows from Eq. (5.49) 
that $[M^{(+)}_n(k)]^*=M^{(-)}(k)$, and that $M^{(+)}_n(-k)=M^{(-)}_n(k)$.  
This leads us to,
\addtocounter{eqnnum1}{1}
\begin{equation}
H_n^*(k)=H_n(-k), \quad {\rm{for\;k\;real}}.
\end{equation}

Thus it immediately follows from Eq. (10.11) that 
all $F_n(x)$, $n=0, 1,2, ... ,N,$ are real functions.  
However, $F^{(N)}_R(g,x)$, is certainly not real 
for $\nu\epsilon {\cal{S}}(T_o)$.

We have explicitly calculated $F_o(x)$ in section VII, 
and obtained
\addtocounter{eqnnum1}{1}
\begin{equation}
F_o(x)=\lambda_oe^{-x}+\lambda_1xe^{-x},
\end{equation}
with $\lambda_o$ and $\lambda_1$ real and given in Eqs. (7.6) and (7.7).

One can also easily calculate explicitly $F_1(x)$ by contour integration.
\addtocounter{eqnnum1}{1}
\begin{equation}
F_1(x)=\frac{1}{2\pi}\int_{-\infty}^{+\infty}dk
[\frac{M^{(+)}_1(k)}{M^{(-)}_o(k)}-\frac{M^{(+)}_o(k)M^{(-)}_1(k)}
{[M^{(-)}_o(k)]^2}]e^{ikx}.
\end{equation}
The result is
\addtocounter{eqnnum1}{1}
\begin{equation}
F_1(x)=(\sum_{n=0}^3\sigma_nx^n)e^{-x},
\end{equation}
where the constants $\sigma_n$ are explicitly given as 
functions of $a_1$, and $b_j, j=1, ..., 4$.  Here we will 
only give the numerical value of the $\sigma_n's$
\addtocounter{eqnnum1}{1}
\begin{eqnarray}
\sigma_o&=&26.5228,\\ \nonumber
\sigma_1&=&1.7901,\\ \nonumber
\sigma_2&=&-9.3291,\\ \nonumber
\sigma_3&=&-2.3582.
\end{eqnarray}
The result for $F_2(x)$ will be similar,
\addtocounter{eqnnum1}{1}
\begin{equation}
F_2(x)\equiv e^{-x}(\sum_{n=0}^{5}\beta_nx^n).
\end{equation}
We will not give its explicit value as it is not needed.

The Marchenko Equation (8.10), for $\nu\epsilon S(T_o)$ can 
be written in operator form
\addtocounter{eqnnum1}{1}
\begin{equation}
A=F+AF
\end{equation}

Writing
\addtocounter{eqnnum1}{1}
\begin{equation}
A=A_o+gA_1+g^2A_2+g^2A^{(2)}_R,
\end{equation}
and using the expansion for $F$ given in Eq. (10.9) we get 
by comparing terms,
\addtocounter{eqnnum1}{1}
\begin{equation}
A_o=F_o+A_oF_o,
\end{equation}
an equation we solved explicitly in section VII.  In addition we have
\addtocounter{eqnnum1}{1}
\begin{equation}
A_1=(F_1+A_oF_1)+A_1F_o,
\end{equation}
as the integral equation for $A_1$, and
\addtocounter{eqnnum1}{1}
\begin{equation}
A_2=(F_2+A_1F_1+A_oF_2)+A_2F_o,
\end{equation}
for $A_2$.  It is obvious that the integral equations 
for $A_n, n=0,1,2,...$ all have the same kernel $F_o$.  
Thus they are all explicitly solvable.  Given our solution, 
$A_o$, for Eq. (10.20) we get,
\addtocounter{eqnnum1}{1}
\begin{equation}
A_o=(1-F_o)^{-1}F_o,
\end{equation}
and hence
\addtocounter{eqnnum1}{1}
\begin{equation}
(1-F_o)^{-1}=A_o+1.
\end{equation}
This leads to solutions for $A_1, A_2$, etc. with
\addtocounter{eqnnum1}{1}
\begin{equation}
A_1=F_1+2A_oF_1+A_o(A_oF_1),
\end{equation}
and, given $A_1$, we can now get $A_2$ explicitly as:
\addtocounter{eqnnum1}{1}
\begin{equation}
A_2=F_2+A_1F_1+2A_oF_2+A_o(A_1F_1)+A_o(A_oF_2).
\end{equation}
It is now obvious that all the $A_n$'s are real and continuously 
differentiable, for\\
$y\geq x\geq0$, since from Eq. (7.11),
\addtocounter{eqnnum1}{1}
\begin{equation}
A_o(x,y)=[B(x)+(y-x)C(x)]e^{-(y-x)},
\end{equation}
with $B(x)$ and $C(x)$ given by Eqs. (7.18) and (7.19) 
and $B,C$ are $O(e^{-2x})$ as $x\rightarrow\infty$.  The 
kernels $F_n(x)$ are of the form
\addtocounter{eqnnum1}{1}
\begin{equation}
F_n(x)=e^{-x}[\sum_{j=0}^{2n+1}\sigma_j^{(n)}x^j].
\end{equation}
The potential $U(\nu;x)$ is given by
\addtocounter{eqnnum1}{1}
\begin{equation}
U(\nu;x)=-2\frac{d}{dx}A(\nu;x,x), x\geq0.
\end{equation}

Using the variable $g\equiv(\nu^2-1/4)^{-1}$, we write
\addtocounter{eqnnum1}{1}
\begin{equation}
U(\nu;x)\equiv V(g;x).
\end{equation}
The expansion of $A$ in powers of $g$, given in Eq. (10.19), now gives us,
\addtocounter{eqnnum1}{1}
\begin{equation}
V(g;x)=V_o(x)+gV_1(x)+g^2V_2(x)+ ...+g^NV^{(N)}_R(g,x),
\end{equation}
where
\addtocounter{eqnnum1}{1}
\begin{equation}
V_n(x)=-2\frac{d}{dx}A_n(\nu;x,x),
\end{equation}

\addtocounter{eqnnum1}{1}
\begin{equation}
V^{(N)}_R(g,x)=-2\frac{d}{dx}A^{(N)}_R(\nu;x,x).
\end{equation}
and $V^{(N)}_R$ is $O(g)$.

All the $V_n$'s are real, continuous for $x\in[0,\infty)$, 
and $O(e^{-2x})$ for large $x$.  $V^{(N)}_R(g,x)$ is complex 
for $\nu\epsilon S(T_o)$ but also continuous and $O(e^{-2x})$ 
for large $x$.

It is now clear why one can refer to $"g"$ as a coupling 
constant specially for large values of $Im\nu$, i.e., 
small $g$, $g=O(|Im\nu|^{-2})$.

We have already calculated the first term in the expansion, 
$V_o(x)$, and it is given explicitly in Eq. (7.21).  Later we 
will need to have $V_1(x)$ and we proceed to calculate it here.

From Eq. (10.25) we get,
\addtocounter{eqnnum1}{1}
\begin{eqnarray}
A_1(x,x)&=&F_1(2x)+2\int_{x}^{\infty}dz A_o(x,z)F_1(z+x)\\ \nonumber
&+&\int_{x}^{\infty}dz_1\int_{x}^{\infty}dz_2A_o(x;z_1)
A_o(x;z_2)F_1(z_1+z_2),
\end{eqnarray}
where $A_o$ and $F_1$ are given in Eqs. (10.27) and (10.25).  
It is now evident that $A_1(x,x)$ is continuously differentiable and
\addtocounter{eqnnum1}{1}
\begin{equation}
V_1(x)=-2\frac{\partial A_1}{\partial x}(x,x),
\end{equation}
can be explicitly calculated in terms of $B(x)$, $C(x)$, 
and the real constants $\sigma_j$.  The resulting $V_1(x)$ 
is continuous, finite at $x=0$, and $V_1=0(x^3e^{-2x})$ for large $x$.  
We do not write down the full result, but exhibit a graph of $V_1(x)$ in 
figure 1.

In closing we comment on the asymptotic expansion (10.31) 
for $V(g;x)$.  It of course is not convergent no matter how 
small $|g|$ is.  This follows from the fact that the expansions 
for $M^{\pm}(\nu;k)$ are also divergent.  There is an essential 
singularity at $g=0$.  The constants 
$\chi_\ell^{(n)}$, $\ell=1, ... ,2n+2$, given 
in eqs. (5.32) and (5.33) grow fast.  However, Eq. (10.31)
can still give an extremely good estimate for $V(g;x)$ as 
long as $N$ is $O(1)$.  Indeed it is possible to get a uniform 
bound on $V^{(N)}_R(g,x)$ which is
\addtocounter{eqnnum1}{1}
\begin{equation}
|V_R^{(N)}(g,x)|\leq\frac{C(N)}{|Im\nu|^2}e^{-x}, x>0,                      
\end{equation}
where $C(N)$ grows fast with $N$.  For the purposes of 
this paper we need at most $N=2$ or 3.  For $T_o=10^4$, 
$(C(2)/T_o^2)<<1$.  This makes $(g^2V_R^{(2)})$ smaller  
than $(\frac{o(1)}{|Im\nu|^4})e^{-x}$, and thus $V_o+gV_1+g^2V_2$, 
give an excellent estimate of $V(g,x)$ for all $\nu\epsilon{\cal S}
(T_o)$, and $x$ not large.  However, this estimate is not good for 
large $x$ where both V and the error are small.

Finally, we will need an important result on the phase 
of $V_R^{(N)}$ near the critical line $\nu=it$.

As we have shown in section IX, $F(it;x)$, $A(it;x;y)$, 
and $U(it;x)$ are all real, for $x\epsilon [0,\infty)$ 
and $y\geq x$.  In addition, in the asymptotic expansion 
given in Eq. (10.31) all the coefficients $V_n(x)$ are real.  
But the remainder term, $V_R^{(N)}(g,x)$, is in general complex 
for $\nu=\omega+it$, and $\omega\neq0$.  However, for $\omega=0$, 
we again have reality,
\addtocounter{eqnnum1}{1}
\begin{equation}
U^{(N)}_R(it;x)\equiv V_R^{(N)}(-(t^2+\frac{1}{4})^{-1};x)=(V_R^{(N)})^*.
\end{equation}

In Section VIII we proved that both $A(\nu;x,y)$ and $U(\nu;x)$ are 
analytic in $\nu$ for $\nu\in{\cal{ S}}(T_o)$.  Hence so is 
$U^{(N)}_R(\nu;x)$.  This leads us to the following lemma.

{\underline{Lemma 10.1}}:

For $\nu \epsilon S(T_o)$, and $\nu=\delta+it$, we have
\addtocounter{eqnnum1}{1}
\begin{equation}
|Im(V_R^{(N)}(g;x))|=|ImU^{(N)}_R(\delta+it;x)|<C(x).|\delta|+O(\delta^2),
\end{equation}
where
\addtocounter{eqnnum1}{1}
\begin{equation}
C(x)\leq\frac{c_1}{t^2}e^{-x}.
\end{equation}

{\underline{Proof:}}
\addtocounter{eqnnum1}{1}
\begin{equation}
[ImU^{(N)}_R(\nu;x)]=Im\frac{dU^{(N)}_R(\nu;x)}{d\nu}|_{\nu=it}(\nu'-\nu) 
+O((\nu'-\nu)^2)
\end{equation}
The derivative is finite, and setting $\nu'=\delta+it$, $\delta<<\frac{1}
{2}$, we get equation (2.38).

The bound on $C(x)$ follows from our previous estimates.  We will not give 
the proof here.

\newpage
\vspace{.25in}
\setcounter{chanum}{11}
\setcounter{eqnnum1}{1}
\noindent{\bf\Large{XI.  The Zeros of $M^{(-)}(\nu;k)$ for Fixed $k$}}

In this section we shall study the properties of the infinite set of zeros, 
$\nu_n(k)$, of $M^{(-)}(\nu,k)$ for fixed $k$.  We prove three lemmas for 
$\left\{\nu_n(k)  \right\}$.

For convenience and without loss of generality, we set $k=i\tau$, and 
$\tau\geq0$.  We write
\begin{equation}
M(\nu;\tau)\equiv M^{(-)}(\nu,i\tau),
\end{equation}
and
\addtocounter{eqnnum1}{1}
\begin{equation}
M_n(\tau)\equiv M^{(-)}_n(i\tau).
\end{equation}

It is clear from the equation defining $M(\nu,\tau)$, i.e. Eq. (3.10) 
with $k=i\tau$, that $M(\nu,\tau)$, fixed $\tau>0$, is an entire function 
of $\nu$ with order the same as $\xi(\nu+1/2)$, i.e. order 1.  
Hence $M(\nu,\tau)$ has an infinite set of zeros, $\nu_n(\tau)$, 
with $|\nu_n(\tau)|\rightarrow \infty$ as $n\rightarrow\infty$.

Next it follows from section VI that, $\nu_n(\tau)$, will all be 
outside the truncated critical strip, $S(T_o)$, for $\tau>T^N_o
e^-{\frac{\pi T_o}{4}}$.  As we decrease $\tau$, $\nu_n(\tau)$, 
will start appearing in ${\cal{S}}(T_o)$ for 
$\tau<O(T^pe^-{\frac{\pi T}{4}})$.

Our first lemma is the following:

{\underline{Lemma 11.1:}}
  
As $\tau\rightarrow 0$,
\addtocounter{eqnnum1}{1}
\begin{equation}
lim_{\tau\rightarrow0}\nu_n(\tau)=\nu_n,
\end{equation}
where $(\nu_n+\frac{1}{2})$ is a zero of the zeta function i.e.
\addtocounter{eqnnum1}{1}
\begin{equation}
\xi(\nu_n+1/2)=0.
\end{equation}
{\underline{Proof:}}

From the asymptotic expansion of $M(\nu;\tau)$ given in Eq. (5.47), we have
\addtocounter{eqnnum1}{1}
\begin{equation}
M(\nu;\tau)=2\xi(\nu+\frac{1}{2})+M_o(\tau)+gM_1(\tau)+g^2M_2(\tau)
+g^2\tilde{R}_2(\nu,\tau),
\end{equation}
where
\addtocounter{eqnnum1}{1}
\begin{equation}
\tilde{R}_2(\nu,\tau)=\int_{1}^{\infty}d\alpha
(\sum_{\ell=1}^{7}\Phi_{\ell}^{(3)}(\alpha)\left\{\frac{1}
{(\alpha+\tau)^{\ell}}-\frac{1}{\alpha^{\ell}}\right\})
[\alpha^{\frac{\nu}{2}}+\alpha^{\frac{-\nu}{2}}].
\end{equation}
In getting Eqs. (11.4) and (11.5) we have used Eq. (5.46) 
for the $\xi$-function.

By definition we have,
\addtocounter{eqnnum1}{1}
\begin{equation}
M(\nu_n(\tau),\tau)\equiv 0.
\end{equation}
Hence we get
\addtocounter{eqnnum1}{1}
\begin{eqnarray}
-2\xi(\nu_n(\tau)+\frac{1}{2})&=&M_o(\tau)+g_n(\tau)M_1(\tau)
+g^2_n(\tau)M_2(\tau)\\ \nonumber
&+&g^2_n(\tau)\tilde{R}_2(\nu_n(\tau);\tau),
\end{eqnarray}
with
\addtocounter{eqnnum1}{1}
\begin{equation}
g_n(\tau)=\frac{1}{[\nu^2_n(\tau)-\frac{1}{4}]}.
\end{equation}
Now all the terms on the r.h.s. of Eq. (11.8) are $O(\tau)$ as 
$\tau\rightarrow 0$.  Hence we get
\addtocounter{eqnnum1}{1}
\begin{equation}
lim_{\tau\rightarrow0}\xi(\nu_n(\tau)+\frac{1}{2})=\xi(\nu_n(0)
+\frac{1}{2})=0,
\end{equation}
and therefore
\addtocounter{eqnnum1}{1}
\begin{equation}
\nu_n(0)=\nu_n.
\end{equation}
The next lemma gives us as estimate of $(\nu_n(\tau)-\nu_n(0))$ 
as $\tau\rightarrow 0$

{\underline{Lemma 11.2}}:

If $\nu_n$ is a first order zero of $\xi(\nu+1/2)$, then as 
$\tau\rightarrow 0$
\begin{equation}
(\nu_n(\tau)-\nu_n)=O(\tau),
\end{equation}
and if $\nu_n$ is of order $p$ then
\addtocounter{eqnnum1}{1}
\begin{equation}
(\nu_n(\tau)-\nu_n)=O(\tau^{1/p}).
\end{equation}

{\underline{Proof:}}

Since $\xi(\nu+\frac{1}{2})$ is entire we can write for any $\nu_n$
\addtocounter{eqnnum1}{1}
\begin{equation}
\xi(\nu_n(\tau)+\frac{1}{2})=\xi(\nu_n+\frac{1}{2})
+\frac{d\xi}{d\nu}|_{\nu=\nu_n}(\nu_n(\tau)-\nu_n)
+O[(\nu_n(\tau)-\nu_n)^2].
\end{equation}
But the first term on the right is zero, and $(\xi')_{\nu=\nu_n}\neq0$ 
for a first order zero, and we get from Eq. (11.7)
\addtocounter{eqnnum1}{1}
\begin{equation}
-2(\xi')_{\nu=\nu_n}(\nu_n(\tau)-\nu_n)=\tau[(2+a_1)
+O(g_n(\tau))]+O(\tau^2).
\end{equation}

This gives our proof for a first order zero.

For a zero of multiplicity $p$, we have by definition 
$(\xi^{(j)})_{\nu=\nu_n}=0$, for $j=1,2,...,p-1$, and 
$(\xi^{(p)})_{\nu=\nu_n}\neq0$.  Hence we get,
\addtocounter{eqnnum1}{1}
\begin{equation}
-2[\xi^{(p)}]_{\nu=\nu_n}(\nu_n(\tau)-\nu)^p=\tau[(2+a_1)
+O(g_n(\tau))]+O(t^2).
\end{equation}

This leads to
\addtocounter{eqnnum1}{1}
\begin{equation}
(\nu_n(\tau)-\nu_n)=O(\tau^{1/p}), \tau\rightarrow0.
\end{equation}

So far we have shown that every $\nu_n(\tau)$ 
approaches a Riemann zero as $\tau\rightarrow0$, but 
have not established the converse, i.e. that any $\nu_n$ is 
the limit of a $\nu_n(\tau)$ as $\tau\rightarrow0$.

To do this we first define a rectangular region $R(T_o,T)$ as follows:
\addtocounter{eqnnum1}{1}
\begin{equation}
R(T_o,T)=\{\nu|-3/2\leq Re\nu\leq 3/2; T_o\leq Im\nu\leq T\};
\end{equation}
with $T>>T_o$.

We now prove our third lemma.

{\underline{Lemma 11.3}}:

Let $N_{\xi}(T_o,T)$ be the number of zeros of 
$\xi(\nu+\frac{1}{2})$ for $\nu\epsilon R(T_o,T)$, 
and $N_M(T_o,T;\tau)$ be the number of zeros, $\nu_n(\tau)$, 
of $M(\nu,\tau)$, with $\nu_n(\tau)\epsilon R(T_o,T)$, 
then for sufficiently small $\tau$
\addtocounter{eqnnum1}{1}
\begin{equation}
|N_M(T_o,T;\tau)-N_\xi(T_o,T)|<1.
\end{equation}
There exist a small interval in $\tau$, $0\leq\tau\leq\tau_o(T)$, such that
\addtocounter{eqnnum1}{1}
\begin{equation}
N_M(T_o,T;\tau)=N_\xi(T_o,T).
\end{equation}
{\underline{Proof}}:

We start with the standard expression:	
\addtocounter{eqnnum1}{1}
\begin{equation}
N_M-N_\xi=\frac{1}{2\pi i}\oint_{\Gamma_R}d\nu\{\frac {M'(\nu;\tau)}
{M(\nu;\tau)}-\frac{\xi'(\nu+1/2)}{\xi(\nu+1/2)}\},
\end{equation}
where $\Gamma_R$ is the boundary of the rectangle $R$.  
We also choose $T$, and $T_o$ such that they both lie between 
the abscissa of successive zeros $\nu_n$; i.e. 
$Im\nu_{n_1}<T<Im\nu_{N_1+1}$, and $Im\nu_{n_o}<T_o<Im\nu_{n_o+1}$.  
Thus $\Gamma_B$ never has a zero of $\xi$ on it.  The prime in (11.20) 
denotes $(d/d\nu)$.
We follow the method used to prove theorem 9.3 in ref. 18.

Using the symmetry of $\xi$ in $\nu$ we have
\addtocounter{eqnnum1}{1}
\begin{equation}
N_\xi=\frac{1}{\pi i}\int_{\frac{3}{2}+iT_o}^{\frac{3}{2}
+iT}d\nu\frac{\xi'}{\xi}+\frac{1}{\pi i}\int_{\frac{3}{2}+iT}^{iT} 
d\nu \frac{\xi'}{\xi}+\frac{1}{\pi i}\int_{iT_o}^{\frac{3}{2}+iT_o}
d\nu\frac{\xi'}{\xi},=\frac{1}{\pi}\{\bigtriangleup arg\xi(\nu+1/2)\};
\end{equation}
where $\Delta$ denotes the variation from 
$iT_o$ to $\frac{3}{2}+iT_o$ then from $(\frac{3}{2}
+iT_o)$ to $\frac{3}{2}+iT$, and thence to $iT$.

But from Eq. (11.5) we get
\addtocounter{eqnnum1}{1}
\begin{equation}
\{M(\nu;\tau)/\xi(\nu+\frac{1}{2})\}=1
+\frac{(2+a_1)\tau+cg\tau}{2\xi(\nu+\frac{1}{2})}+O(\tau^2);
\end{equation}

On the horizontal parts of $\Gamma_R$, we can use theorem 9.7 
of ref. 18 to obtain a lower bound on $|\xi(\nu+\frac{1}{2})|$.  
Indeed, there is a constant,$A$, such that each interval $(T,T+1)$ 
contains a value of t for which,
\addtocounter{eqnnum1}{1}
\begin{equation}
|\zeta(\nu+\frac{1}{2})|>t^{-A}, -\frac{3}{2}\leq \omega\leq\frac{3}{2},
\end{equation}
where $\nu=\omega+it$.  On the vertical parts of 
$\Gamma_B$,$|\zeta(\nu+\frac{1}{2})|$ is obviously bounded 
from below.  Using the standard asymptotic expression for 
$\Gamma(\frac{1}{4}+\frac{\nu}{2})$ as $t\rightarrow\infty$, we finally get
\addtocounter{eqnnum1}{1}
\begin{equation}
|[argM(\nu;\tau)-arg\xi(\nu+\frac{1}{2})]|\leq\frac{\tau}{|\xi(\nu
+\frac{1}{2})|},
\end{equation}
We can choose $\tau=T^{-A-N}(e^{\frac{-\pi T}{4}})$ with $N\geq3$, 
and obtain
\addtocounter{eqnnum1}{1}
\begin{equation}
\Delta(argM(\nu,\tau)-arg\xi(\nu+\frac{1}{2}))\leq \frac{1}{T^{N-1}}.
\end{equation}
Hence as $T\rightarrow\infty$, we get
\addtocounter{eqnnum1}{1}
\begin{equation}
N_M-N_\xi=0
\end{equation}

This completes our proof.

We stress one important fact that is a consequence of the results of 
this section.  Namely, we are now not limited to the study of the zero 
energy zeros, $\nu_n(0)$.  One can consider the case for small enough 
$\tau$, but $\tau>0$, and obviously if there is an interval $0<\tau<\tau_n$ 
for which $[Re\nu_n(\tau)]=0$, this will be sufficient for the validity of 
the Riemann hypothesis.  In the next section we will see the importance of 
this remark.

\newpage
\vspace{.25in}
\setcounter{chanum}{12}
\setcounter{eqnnum1}{1}
\noindent{\bf\Large{XII.  The Potential $V(g,x)$ and the 
Riemann Hypothesis}}

Following the notation of the previous section we define the 
Jost solution $f(g;\tau,x)$ as
\begin{equation}
f(g;\tau;x)=f^{(-)}(g;i\tau,x),
\end{equation}
with $k=i\tau$, and $\nu\epsilon{\cal S}(T_o)$.  We also 
recall the result given in Faddeev's review, ref. 14,
\addtocounter{eqnnum1}{1}
\begin{equation}
|f-e^{-\tau x}|\leq Ke^{-\tau x}\int_{x}^{\infty}u|V(g,u)|du,
\end{equation}
hence $f=O(e^{-\tau x})$ as $x\rightarrow\infty$, $\tau\geq0$.

We write the Schrodinger equation for $f$ and $f^*$,
\addtocounter{eqnnum1}{1}
\begin{equation}
\frac{-d^2f}{dx^2}+V(g,x)f=-\tau^2f,
\end{equation}
and
\addtocounter{eqnnum1}{1}
\begin{equation}
\frac{-d^2f^*}{dx^2}+V^*(g,x)f^*=-\tau^2f^*.
\end{equation}
Multiplying the first equation by $f^*$ and the second by $f$, 
integrating from $x=0$ to $x=\infty$, and subtracting the two 
equations we get
\addtocounter{eqnnum1}{1}
\begin{equation}
2i\int_{o}^{\infty} dx|f(g;\tau;x)|^2[ImV(g,x)]=
M(\nu,\tau)\Lambda^*(\nu,\tau)-M^*(\nu,\tau)\Lambda(\nu,\tau),
\end{equation}
where
\addtocounter{eqnnum1}{1}
\begin{equation}
\Lambda(\nu,\tau)=[\frac{df(g;\tau;x)}{dx}]_{x=0},
\end{equation}
and $g=(\nu^2-1/4)^{-1}$.  The derivative exists for 
$x\rightarrow 0$ in our present case, since $V(g,0)$ is finite.  
One can also check this from the expression for $f$ in terms of 
$A(\nu;x,y)$,
\addtocounter{eqnnum1}{1}
\begin{equation}
f(g;\tau,x)=e^{-\tau x}+\int_{x}^{\infty}dyA(\nu;x,y)e^{-\tau y}.
\end{equation}
From section VIII, we know that $A(\nu;0,0)$ is finite, 
and also $(\partial A/ \partial x)$, for $x\geq0$, $y\geq x$ 
exists and is integrable.

Next we set $\nu=\nu_n(\tau)$, and $g=g_n(\tau)$ in Eq. (12.3), 
and we get
\addtocounter{eqnnum1}{1}
\begin{equation}
\int_{o}^{\infty}dx|f(g_n(\tau);\tau;x)|^2[ImV(g_n(\tau);x)]=0
\end{equation}
This last integral is absolutely and uniformly convergent 
since $V=O(e^{-2x})$ as $x\rightarrow\infty$.
Thus we can take the limit $\tau\rightarrow0$ and obtain,
\addtocounter{eqnnum1}{1}
\begin{equation}
\int_{o}^{\infty}dx|f(g_n;0;x)|^2[ImV(g_n;x)]=0.
\end{equation}

Now $V(g,x)$ has an asymptotic expansion in $g$.
\addtocounter{eqnnum1}{1}
\begin{equation}
V(g;x)=V_o(x)+gV_1(x)+g^2V_2(x)+g^2V^{(2)}_R(g,x).
\end{equation}
Also from the expansion for $A(\nu;x,y)$ we get for the Jost solution
\addtocounter{eqnnum1}{1}
\begin{equation}
f(g;\tau;x)=f_o(\tau;x)+gf_1(\tau;x)+g^2f_2(\tau;x)+g^2f_R(g;\tau;x).
\end{equation}
Both $V^{(2)}_R$ and $f^{(2)}_R$ are $O(g)$ and have 
bounds in $x$.  It is sufficient, for the validity of the 
Riemann hypothesis for a constant $c>0$, $c=O(1)$, to exist 
such that,
\addtocounter{eqnnum1}{1}
\begin{equation}
\int_{o}^{\infty}dx|f_o(x)|^2V_1(x)=c>0;
\end{equation}
where $f_o(x)=f_o(0,x)$.

To prove this last statement, we first note that setting 
$\nu_n=\nu_n(0)$ we have
\addtocounter{eqnnum1}{1}
\begin{equation}
\nu_n=\omega_n+it_n,\:\:\: t_n>T_o;
\end{equation}
and
\addtocounter{eqnnum1}{1}
\begin{equation}
Img_n=\frac{2\omega_nt_n}{[t^2_n-\omega^2_n+1/4]^2};\;\;\;\omega^2_n\leq1/4.
\end{equation}
The vanishing of $Img_n$ with $t_n>T_o>0$, 
implies $\omega_n=0$ and hence $s_n=1/2+it_n$.

Next we write from Eq. (12.10),
\addtocounter{eqnnum1}{1}
\begin{eqnarray}
ImV(g_n;x)&=&(Img_n)V_1(x)+(Img^2_n)V_2(x)\\ \nonumber
&+&(Img^2_n)(ReV_R^{(2)}(g_n;x))+(Reg^2_n)(ImV^{(2)}_R(g_n;x)).
\end{eqnarray}
Also in the integrand in (12.0) we can write
\addtocounter{eqnnum1}{1}
\begin{equation}
|f(g_n;0;x)|^2=|f(0;0;x)|^2+O(\frac{1}{t^2_n}).
\end{equation}
In Eq. (12.15) we note that $(Img_n)=O(\frac{\omega_n}{t^3_n})$ 
and $Img^2_n=O(\frac{\omega_n}{t^5_n})$, while$(Reg^2_N)=
O(\frac{1}{t^4_n})$.  On substituting Eqs. (12.15) and 
(12.16) in (12.9), and using the assumption (12.12) we see 
at first that consistency requires $\omega_n$ to be small, 
i.e. $\omega_n=O(\frac{1}{t^3_n})$.  Here we use the fact 
that $|V^{(2)}_R|=O(g)$ for small $g$.  However, we have 
more information on $V^{(2)}_R$ and specifically its phase 
for small $\omega_n$.  This was given in lemma 10.1, where it 
was shown that $ImV_R^{(2)}=O(\omega)$ for small $\omega$ and 
that $V^{(2)}_R$ is real for $\omega=0$.  Given the bound on 
$ImV_R^{(2)}$ from this lemma we see that the leading contribution 
from (12.15) to Eq. (12.9) must come from $[(Img_n)V_1]$ and cannot 
be cancelled by the other three terms.  We have
\addtocounter{eqnnum1}{1}
\begin{equation}
(Img_n)c+(Img^2_n)X_1+(Img^2_n)X_2+(Reg^2_n)X_3=0.
\end{equation}
with
\addtocounter{eqnnum1}{1}
\begin{eqnarray}
|X_1|&\leq& 2\int_{o}^{\infty}|f_o(x)|^2|V_2(x)|dx;\\ \nonumber
|X_2|&\leq& 2\int_{o}^{\infty}|f_o(x)|^2|ReV^{(2)}_R(g_n;x)|;\\ \nonumber
|X_3|&\leq& 2\int_{o}^{\infty}|f_o(x)|^2|ImV^{(2)}_R(g_n,x)|dx\\ \nonumber
&\leq&\frac{2\omega_nc_1}{t^2_n}\int_{o}^{\infty}|f_o(x)|^2e^{-x}dx
\end{eqnarray}

Hence we have constants $B_j$ such that $|X_j|<B_j;j=1,2$, 
and $|X_3|<(\frac{\omega_n}{t^2_n})B_3$.  
Note that in the last inequality we used lemma 10.1.  
We take $T_o$ large enough such that 
$\frac{|B_j|}{T^2_o}<<c,  \;\;\;j=1,2,3;$ 
and with $c=O(1)$.  Next we rewrite Eq. (12.17) as
\addtocounter{eqnnum1}{1}
\begin{equation}
(Img_n)[c+2(Re g_n)X_1+2(Reg_n)X_2+(Reg_n^2).t_n\hat{X}_3]=0;
\end{equation}
where now $|\hat{X}_3|<2B_3$.  The term in the square 
bracket cannot vanish, and hence we obtain
\addtocounter{eqnnum1}{1}
\begin{equation}
[Img_n]c=[Img_n]\int_{o}^{\infty}|f_o(x)|^2V_1(x)dx=0,
\end{equation}
and thus, if $c\neq0$, then for all $t_n>T_o$, we get
\addtocounter{eqnnum1}{1}
\begin{equation}
Img_n=0; or \;\; s_n=\frac{1}{2}+it_n.
\end{equation}

We have already calculated $f_o(x)$ exactly in 
section VII, and $V_1(x)$ in section X, and we 
can compute the integral in (12.12) directly, the result is
\addtocounter{eqnnum1}{1}
\begin{equation}
\int_{o}^{\infty}|f_o(x)|^2V_1(x)=0.
\end{equation}
This can be checked numerically, and indeed can be 
rigorously proved.  So we have no information on $(Img_n)$ 
from (12.17).  However, the proof of (12.22) suggests to us 
how we can proceed further.

To prove (12.22) we use the Schrodinger equation and the 
expansions (12.10) and (12.11).  We obtain
\addtocounter{eqnnum1}{1}
\begin{equation}
\frac{-d^2f_o}{dx^2}(\tau,x)+V_o(x)f_o(\tau,x)=-\tau^2f_o(\tau,x),
\end{equation}
and
\addtocounter{eqnnum1}{1}
\begin{equation}
\frac{-d^2f_1}{dx^2}(\tau,x)+V_o(x)f_1(\tau,x)+
V_1(x)f_o(\tau,x)=-\tau^2f_1(\tau,x).
\end{equation}
Multiplying the first equation above by $f_1$ 
and the second by $f_o$, integrating from zero 
to infinity, and subtracting, we have
\addtocounter{eqnnum1}{1}
\begin{equation}
\int_{o}^{\infty}[f_o(\tau,x)]^2V_1(x)dx=
-[f_1(\tau,0)f'_o(\tau,0)-f_o(\tau,0)f'_1(\tau,0)],
\end{equation}
where the prime denotes $(d/dx)$.

By definition, we have
\addtocounter{eqnnum1}{1}
\begin{equation}
f_o(\tau,0)=M_o(\tau),
\end{equation}
and
\addtocounter{eqnnum1}{1}
\begin{equation}
f_1(\tau,0)=M_1(\tau).
\end{equation}
Both $M_o(\tau)$ and $M_1(\tau)$ are $O(\tau)$ 
for small $\tau$ and vanish as $\tau\rightarrow0$.  
We then get after taking the limit,
\addtocounter{eqnnum1}{1}
\begin{equation}
\int_{o}^{\infty}|f_o(0,x)|^2V_1(x)dx\equiv0.
\end{equation}
The crucial factor here is the fact that 
$M(\nu,0)=2\xi(\nu+1/2)$, and for large 
$|Im\nu|$, $\nu\varepsilon\it{S}(T_o)$, 
$M(\nu,0)=2\xi(\nu+1/2)=O(e^{\frac{-\pi|Im\nu|}{4}})$.  
This forces all the coefficients, $M_n(\tau)$, 
in the asymptotic expansion of $M(\nu,\tau)$ 
in powers of $g$ to vanish as $\tau\rightarrow0$.  
The culprit is the factor $\Gamma(\frac{\nu}{2}+1/4)$ 
in the Eq. (3.4) which relates $\xi(s)$ to $\zeta(s)$.  
We will also see below how this fact hinders us 
in treating the case $\tau\neq0$, but $\tau$ small.

From the results of sec. XI, it is evident 
that it is sufficient to prove that 
$Re\nu_n(\tau)=0$ in an interval 
$0<\tau<\tau_o(n)$, where $\tau_o(n)=
O(exp(\frac{-\pi|Im\nu_n|}{4}))$.  
From Eq. (12.19) one can prove that
\addtocounter{eqnnum1}{1}
\begin{equation}
\int_{o}^{\infty}[f_o(\tau,x)]^2V_1(x)dx=K\tau+O(\tau^2),
\end{equation}
where $K$ is a constant, $K=O(1)$.  
The integral does \underline{not} vanish if $\tau>0$.

This suggests trying a double expansion 
in powers of $g$ and $\tau$.  However, 
again this will not lead to any restriction 
on $Img_n$.  The main problem is the relevant 
domain in $\tau$ is small, i.e. $\tau=
O(e^\frac{-\pi t}{4})$, and terms of order 
$g^2$ are much larger than terms of order $\tau$.

To proceed further along the lines suggested by 
this paper one has to do two things:

I.  First, find an even function $h(\nu)$, 
analytic for $\nu\epsilon S(T_o)$, and having 
no zeros in $\it{S}(T_o)$, such that if we define, $\chi(\nu)$ as
\addtocounter{eqnnum1}{1}
\begin{equation}
\chi(\nu)\equiv\xi(\nu+1/2)h(\nu),
\end{equation}
we have
\addtocounter{eqnnum1}{1}
\begin{equation}
\chi(\nu)=O([t^2]^{-p}), \:\:1<p<3/2.
\end{equation}
where $\nu=\omega+it$, $t>T_o$.  The point 
here is that $\chi(\nu)$ is small but not smaller than $O(g^2)$.

This first step is achievable.  For example, we can define $\chi$ as
\addtocounter{eqnnum1}{1}
\begin{equation}
\chi(\nu)\equiv\frac{\xi(\nu+\frac{1}{2})[cos\frac{\pi}{4}\nu]}
{(\nu^2-\frac{1}{4})^{2+\delta}},\;\;\;\frac{1}{4}>\delta>0.
\end{equation}
This will give
\addtocounter{eqnnum1}{1}
\begin{equation}
\chi(\nu)=O([t^2]^{-(1+\delta)}).
\end{equation}
The second requirement is much harder to achieve:

II.  One has to construct Jost functions, 
$M_{\chi}^{(\pm)}(\nu,k)$, preferably of the Martin type, such that
\addtocounter{eqnnum1}{1}
\begin{equation}
lim_{k\rightarrow0}M_{\chi}^{(\nu)}=\chi(\nu+1/2).
\end{equation}
and for small $g$, $\nu\varepsilon{\cal S}(T_o)$,

In addition $M^{\pm}_\chi$ has to be of the 
Martin type and it must have an asymptotic 
expansion in powers of $g=(\nu^2-\frac{1}{2})^{-1}$.

\noindent{\bf\Large{Acknowledgements}}

The author wishes to thank James Liu and H.C. Ren 
for untiring help in checking much of the algebraic 
manipulations in this paper including the use of Mathematica 
to produce tables 1 and 2 and carry out other numerical work.  
This work was supported in part by the U.S. Department of Energy 
under grant number DOE91ER40651 TaskB.

\newpage
\renewcommand{\theequation}{A.\arabic{eqnnum1}}
\setcounter{eqnnum1}{1}
\vspace{.25in}

\noindent{\bf\Large{Appendix}}

In this Appendix we first give a proof of the 
Laplace transform representation for the Marchenko kernel, 
$F(\nu;x)$, $x>0$.

Starting with the definition (8.2)
\begin{equation}
F(\nu;x)=\frac{1}{2\pi}\int_{L}(S(\nu;k)-1)e^{ikx}dk,
\end{equation}
where $L$ is the line $Imk=\delta$, with $\frac{1}{4}\leq\delta<1$, 
we note that for $\nu\in{\cal{S}}(T_o)$ we have $S(\nu;k)$ analytic 
in $k$ for $Imk>\delta>O(e^{\frac{-\pi T_o}{4}})$, except for the cut 
along the positive imaginary k-axis, $1\leq Imk<\infty$.  
Second, in this region, we have a bound for large $|k|$
\addtocounter{eqnnum1}{1}
\begin{equation}
|S(\nu,k)-1|<\frac{C}{|k|}, |k|\rightarrow\infty.
\end{equation}
This bound holds along any radial direction that excludes the cut.

We can deform the contour $L$ from along the line 
$Imk=1/4$, to a contour surrounding the cut, i.e.
\addtocounter{eqnnum1}{1}
\begin{equation}
F(\nu;x)=\frac{1}{2\pi}\int_{C}[{\cal{S}}(\nu;k)-1]e^{ikx}dk, x>0.
\end{equation}
where C starts at $(-\varepsilon,+i\infty)$ and descends 
to $(-\varepsilon,+i)$, turns around the point $k=i$, 
and then extends from $(+\varepsilon,+i)$ to $(+\varepsilon,
+i\infty)$.  The contribution from the large semicircle, $|k|=K$, 
vanishes as $K\rightarrow\infty$.  For 
$|K|^{-1/2}<arg k<\pi-|K|^{-1/2}$, the contribution is 
$O(exp(-|K|^{1/2}x))$, and vanishes for $x>0$.  
Here we use the bound (A.2).  For the regions, 
$0\leq argk\leq|K|^{-1/2}$, and 
$\pi-|K|^{-1/2}\leq arg k\leq\pi$, 
the contribution from the semicircle to (A.1) 
will be $O(|K|^{-1/2})$ and also vanishes as 
$|K|\rightarrow\infty$.

From (A.3) we finally obtain
\addtocounter{eqnnum1}{1}
\begin{equation}
F(\nu;x)=\frac{1}{\pi}\int_{1}^{\infty}D(\nu;\alpha)e^{-\alpha x}da, x>0;
\end{equation}
where $D$ is given in Eq. (8.5), and from Eq. 
(4.2) is just the discontinuity of $S(\nu;k)$ along the cut.  
Noting that $D=O(e^{-\pi\alpha})$ as $\alpha\rightarrow\infty$, 
we see that (A.4) will also hold for $x=0$.

The next task for this Appendix is to give a direct 
proof of the fact that $U(\nu;x)$ and $f^{(\pm)}(\nu;k;x)$ 
as defined in Eqs. (8.93) and (8.94) do indeed satisfy the 
Schrodinger equation, i.e. to prove lemma (8.5) directly.  
We will also give an explicit expression for $U(\nu;x)$.

Following ref. 22 we define an operator, $Q(\nu;x)$, 
depending on two parameters, $\nu$ and $x$, with $Re x\geq 0$, 
and $\nu\in{\cal{S}}(T_o)$.  $Q$ acts on functions 
$u(\beta)$, $1\leq\beta<\infty$, with $u\in L_2(1,\infty)$.  
We define $Q$ as:
\addtocounter{eqnnum1}{1}
\begin{equation}
[Q(\nu;x)u](\alpha)\equiv\int_{1}^{\infty}Q(\nu;x;\alpha,\beta)
u(\beta)d\beta,
\end{equation}
where
\addtocounter{eqnnum1}{1}
\begin{equation}
Q(\nu;x;\alpha,\beta)=\frac{1}{\pi}\frac{D(\nu;\beta)e^{-2\beta x}}
{[\alpha+\beta]}, Re x\geq 0,
\end{equation}
with $D(\nu;\beta)$ given by Eqs. (4.3) and (4.4).  
$Q$ will have a finite Hilbert-Schmidt norm
\addtocounter{eqnnum1}{1}
\begin{equation}
\thickbar Q(\nu;x)\thickbar \leq\Lambda e^{-2x},
\end{equation}
where $\Lambda$ depends on $\nu$.

We introduce a new integral equation,
\addtocounter{eqnnum1}{1}
\begin{equation}
W(\nu;x;\alpha)=1+\int_{1}^{\infty}Q(\nu;x;\alpha,\beta)W(\nu;x;\beta)d\beta
\end{equation}
with $W\in L_2(1,\infty)$.

This integral equation is equivalent to the Marchenko equation (8.10).  
To see that we write
\addtocounter{eqnnum1}{1}
\begin{equation}
Z(\nu;x;\alpha)\equiv\frac{1}{\pi}D(\nu;\alpha)e^{-\alpha x}W(\nu;x;\alpha)
\end{equation}
$\tilde{A}(\nu;x,y)$ is
\addtocounter{eqnnum1}{1}
\begin{equation}
\tilde{A}(\nu;x,y)\equiv\int_{1}^{\infty}Z(\nu;x;\alpha)e^{-\alpha y}d\alpha.
\end{equation}
The Laplace transform exists since $D=O(e^{-\pi\alpha})$ and $W\in L_2$.  
Substituting (A.9) and (A.10) in Eq. (A.8) and using (A.4) 
we obtain for $\tilde{A}$
\addtocounter{eqnnum1}{1}
\begin{equation}
\tilde{A}(\nu;x,y)=F(\nu;x+y)+\int_{x}^{\infty}\tilde{A}
(\nu;x,u)F(\nu;u+y)a
\end{equation}
But this is just the Marchenko equation which we have 
shown in section VIII does have a unique solution $A$.  
Hence $\tilde{A}=A$ for all $\nu\in{\cal{S}}(T_o)$.  
Thus we conclude that Eq. (A.8), which is of the Fredholm type, 
has a unique solution, since the homogeneous equation $W=QW$ cannot 
have a solution for that will lead to the existence of a solution 
for $A=AF$ which we have shown in section VIII is not possible.

Given the function $W(\nu;x;\alpha)$ we can easily get 
expressions for $U(\nu;x)$ and $f^{\pm}(\nu;k;x)$.

The potential is given by
\addtocounter{eqnnum1}{1}
\begin{equation}
U(\nu;x)=-2\frac{d}{dx}[\frac{1}{\pi}\int_{1}^{\infty}D(\nu,\alpha)
e^{-2\alpha x}W(\nu;x;\alpha)da], Re x\geq 0.
\end{equation}
Similarly, from (A.9) and (A.10) we get
\addtocounter{eqnnum1}{1}
\begin{equation}
f^{(\pm)}(\nu;k,x)~=e^{\mp ikx}+e^{\mp ikx}(\frac{1}
{\pi}\int_{1}^{\infty}\frac{D(\nu;\alpha)e^{-2\alpha x}
W(\nu;x;\alpha)}{(\alpha\pm ik)} d\alpha.
\end{equation}

To check that we recover the same Jost functions 
we started with, we write
\addtocounter{eqnnum1}{1}
\begin{equation}
f^{(\pm)}(\nu;k;0\equiv)M^{(\pm)}(\nu;k)\equiv 1
+(\frac{1}{\pi})\int_{1}^{\infty}\frac{D(\nu;\alpha)
W(\nu;0;\alpha)}{(\alpha\pm ik)} d\alpha.
\end{equation}
However the integral equation for $W$ for $x=0$ is 
trivially soluble.  From Eq. (A.8) we have
\addtocounter{eqnnum1}{1}
\begin{equation}
W(\nu;0;\alpha)=1+\frac{1}{\pi}\int\frac{D(\nu;\beta)W(\nu;0;\beta)}
{(\alpha+\beta)} d\beta.
\end{equation}
Setting
\addtocounter{eqnnum1}{1}
\begin{equation}
W(\nu;0;\alpha)\equiv M^{(-)}(\nu;i\alpha),
\end{equation}
and using the Eqs. (4.3) and (4.4) for $D(\nu;\alpha)$ we have
\addtocounter{eqnnum1}{1}
\begin{equation}
M^{(-)}(\nu;i\alpha)=1+(\nu^2-1/4)\int_{1}^{\infty}
d\beta\frac{\psi(\beta)\beta^{1/4}[\beta^{\frac{\nu}{2}}
+\beta^{\frac{-\nu}{2}}]}{(\beta+\alpha)}.
\end{equation}
This is our original expression for $M^{(-)}$.

The operator $Q$ is a Fredholm type operator, and for 
$Re x\geq 0$, and $\nu\in{\cal{S}}(T_o)$, we proved that 
there are no non-trivial solutions of the homogeneous equation 
$u=Qu$.  Hence the determinant, $Det(1-Q)$, cannot vanish for 
any $Re x\geq 0$.  This determinant can be calculated explicitly,
as was done in ref. 22, and we obtain
\addtocounter{eqnnum1}{1}
\begin{equation}
Det(1-Q(\nu;x))=
1+\sum_{n=1}^{\infty}\frac{1}{(n!)}
\int_{1}^{\infty}d\alpha_1...
\int_{1}^{\infty}d\alpha_n(\prod_{j=1}^{n}
\frac{D(\nu,\alpha_j)e^{-2\alpha_jx}}{2\pi\alpha_j})
\prod_{i<j}^{n}\frac{(\alpha_i-\alpha_j)^2}{(\alpha_i+\alpha_j)^2}
\end{equation}
This series is absolutely convergent for all $Re x\geq0$,~ 
$\nu\in{\cal{S}}(T_o)$, and $|Im\nu|<\infty$.  
Also as shown in ref. 22 the potential $U(\nu;x)$ 
is now given by Dyson's$^{25}$ formula, i.e.
\addtocounter{eqnnum1}{1}
\begin{equation}
U(\nu;x)=-2\frac{d^2}{dx^2}\{log[Det(1-Q(\nu;x))]\}.
\end{equation}

Finally, we give here a direct check on the validity 
of the Schrodinger equation for $f^{\pm}$ and $U(\nu;x)$.

From Eq. (A.7) giving a bound on the norm of $Q$ 
we see that  for some $x_o$, $\thickbar Q\thickbar <1$ for 
all $(Re x)>x_o$, $x_o\geq \frac{log\Lambda}{2}$.  
Thus the iterative series for $W(\nu;x;\alpha)$ is 
absolutely convergent for all $Re x> x_o$,
\addtocounter{eqnnum1}{1}
\begin{equation}
W=1+\sum_{n=1}^{\infty}Q^n
\end{equation}
Using Eqs. (A.6), (A.12), and (A.13), we obtain
\addtocounter{eqnnum1}{1}
\begin{equation}
U(\nu;x)=4\sum_{n=0}^{\infty}(\frac{1}{\pi})^{n+1}\int_{1}^{\infty}
d\alpha_o ...\int_{1}^{\infty}d\alpha_n\frac{(\prod_{j=0}^{n}
D(\nu;\alpha_j)e^{-2\alpha_jx})}{(\prod_{j=0}^{n-1}(\alpha_j
+\alpha_{j+1}))}(\sum_{j=0}^{n}\alpha_j);~~ Re x> x_o.
\end{equation}
A similar series holds for $f^{\pm}$
\addtocounter{eqnnum1}{1}
\begin{equation}
f^{(\pm)}(\nu;k;x)=e^{\mp ikx}+e^{\mp ikx}\sum_{n=0}^{\infty}
(\frac{1}{\pi})^{n+1}\int_{1}^{\infty}d\alpha_o ...\int_{1}^{\infty}
d\alpha_n\frac{(\prod_{j=0}^{n-1}D(\nu;\alpha_j)e^{-2\alpha_j x})}
{[\prod_{j=0}^{n-1}(\alpha_j+\alpha_{j+1})](\alpha_o\pm ik)}    
\end{equation}

Using these series we can check directly for $Re x>x_o$, that 
$U$ and $f^{\pm}$ give a potential and its unique Jost solutions.

We define $h^{(\pm)}$,
\addtocounter{eqnnum1}{1}
\begin{equation}
h^{(\pm)}=e^{\pm ikx} f^{(\pm)}
\end{equation}
The Schrodinger equation for $h^{(\pm)}$ is now
\addtocounter{eqnnum1}{1}
\begin{equation}
\frac{d^2}{dx^2}h^{\pm}(\nu;k;x)\mp 2ik\frac{dh^\pm}{dx}(\nu;k;x)=
U(\nu;x)h^\pm(\nu;k;x)
\end{equation}
Substituting, the expressions (A.21) and (A.22) in the above, 
we see, after some algebra, that for $x >x_o$, (A.24) is satisfied 
if the following algebraic identity holds 
\addtocounter{eqnnum1}{1}
\begin{equation}
[\sum_{n=0}^{r-1}\{(\alpha_n+\alpha_{n+1})(\sum_{j=0}^{n}\alpha_j)\}]=
(\sum_{j=0}^{n}\alpha_j)^2-\alpha_r(\sum_{j=0}^{r}\alpha_j).
\end{equation}
But this is equivalent to
\addtocounter{eqnnum1}{1}
\begin{equation}
\sum_{n=0}^{r}\alpha_n(\sum_{j=0}^{n}\alpha_j)
+\sum_{n=0}^{r-1}(\alpha_{n+1})(\sum_{j=0}^{n}\alpha_j)=
[\sum_{j=0}^{r}\alpha_j]^2..
\end{equation}
This last equation is an identity and can be proved by induction.

The Schrodinger equation thus is valid for all $x>x_o$.  
But again using analytic continuation, now in $x$ we easily 
see that it must hold for all $x\geq 0$.  All the terms in 
(A.24) are analytic in the half plane $Re x\geq 0$.

\newpage
\noindent{\bf\LARGE{References}}

\noindent  1. F.J. Dyson, J. Math. Phys.{\underline{3}}, 
140 (1962).\\ 
2.  H.L. Montgomery, Proc. Symp. Pure Math., 
{\underline{24}}, Amer. Math. Soc., Providence, R.I., 1973, pp. 181-193.\\
3.  M.V. Berry, ``Riemann's zeta function: a model of quantum chaos."  
Lecture Notes in Physics {\underline{262}}, Springer-Verlag, 1986.\\ 
4.  K. Chadan, private communication.  See also K. Chadan and M. Musette, 
C.R. Acad. Sci. Paris, vol. {\underline{316}}, Series II, pp. 1-6, 1993.  
In this paper an example is given with some 
important properties of the zeta function demonstrated.\\
5.  K. Meetz, J. of Math. Phys., {\underline{3}}, 690 (1962).\\
6.  I.M. Gelfand and B.M. Levitan, Izvest. Akad. Nauk. S.S.S.R. Ser. mateon, 
{\underline{15}}, 309 (l951).\\
7.  V.A. Marchenko, Dokl. Akad. Nauk S.S.S.R., 
{\underline{104}}, 695 (l955).  [Math. Rev. {\underline{17}}, 740 (l956)].\\
8.  A. Martin, Nuovo Cim. {\underline{19}}, 1257 (l961).\\
9.  R. Jost, Helvetica Physica Acta, {\underline{20}}, 256 (l947).\\
10. N. Levinson, Kgl. Danske Videnskab. Selskab, Math.-fys. Medd. 
{\underline{25}}, No. 9 (l949).\\
11. V. Bargmann, Proc. Nat. Acad. Sci. U.S.A. {\underline{38}}, 961 (l952).\\
12. R. Jost and A. Pais, Phys. Rev. {\underline{82}}, 840 (1951).\\
13. K. Chadan and R.C. Sabatier, 
"Inverse Problems in Quantum Scattering Theory", 
Second Edition, Springer Verlag, New York, 1989).\\
14. L.D. Faddeev, J. of Math. Phys. {\underline{4}}, 72 (l963).\\
15. R. Blankenbecler, M.L. Goldberger, N.N. Khuri, 
and S.B. Treiman, Annals of Phys. {\underline{10}}, 62 (l960).\\
16. T. Regge, Nuovo Cim. {\underline{14}}, 5, 951 (l959).\\
17. See ref. 8 above and: A. Martin, Nuovo Cimento, 
{\underline{14}}, 403 (l959).\\
18. E.C. Titchmarsh,"The Theory of the Riemann Zeta-function", 
Second Edition, Revised by D.R. Heath-Brown (Oxford University Press, 
Ozford, 1986).\\
19. See refs. 8 and 13.\\
20. D.J. Gross and B.J. Kayser, Phys. Rev. {\underline{152}, 1441 (l966).\\
21. H. Cornille, J. of Math. Phys. {\underline{8}}, 2268 (l967).\\
22. N.N. Khuri, "Inverse Scattering Revisited: Explicit Solution of the 
Marchenko-Martin Method," in "Rigorous Methods in Particle Physics," 
edited by S. Ciulli, F. Scheck, and W. Thirring, 
(Springer Verlag, Berlin, 1990), pp. 77-97.\\
23. V. Bargmann, Rev. of Mod. Phys. {\underline{21}}, 488 (l949).\\
24. We thank H.C. Ren for clarifying and checking this point.\\
25. F.J. Dyson, in "Studies in Mathematical Physics"; 
E. Lieb, B. Simon and A.S.   Wightman, editors; 
(Princeton Univ. Press, Princeton, N.J., 1976) pp. 151-167.

%%%%%%%%%%%%%%%%%%%%%%%%%%%%%%%%%%
\begin{center}
{\bf\Large Tables}
\end{center}

\begin{table}[h]

\begin{tabular}{r|r}
$C^{(0)}$&$j=0$\\
\hline
$l=1$&1
\end{tabular}

\vspace{12pt}

\begin{tabular}{r|rrr}
$C^{(1)}$&$j=0$&1&2\\
\hline
$l=1$&6&14&4\\
2&$-14$&$-8$\\
3&8
\end{tabular}

\vspace{12pt}

\begin{tabular}{r|rrrrr}
$C^{(2)}$&$j=0$&      1&       2&       3&       4\\
\hline
$l=1$&      36&     364&     500&     176&      16\\
  2&      $-364$&   $-1000$&    $-528$&     $-64$\\
  3&      1000&    1056&     192\\
  4&     $-1056$&    $-384$\\
  5&       384
\end{tabular}

\vspace{12pt}

\begin{tabular}{r|rrrrrrr}
$C^{(3)}$&$j=0$&      1&       2&       3&       4&       5&       6  \\
\hline
$l=1$&     216&    7784&   29152&   29128&   10448&    1440&      64\\
  2&     $-7784$&  $-58304$&  $-87384$&  $-41792$&   $-7200$&    $-384$\\
  3&     58304&  174768&  125376&   28800&    1920\\
  4&   $-174768$& $-250752$&  $-86400$&   $-7680$\\
  5&    250752&  172800&   23040\\
  6&   $-172800$&  $-46080$\\
  7&     46080
\end{tabular}

\caption{Values of $C^{(n)}(l;j)$ for $n\le 3$.}

\end{table}

\newpage
\begin{table}

\begin{center}
\begin{tabular}{r|r|r|r}
$\chi^{(n)}_l$&$n=1$&$n=2$&$n=3$\\
\hline
$l=1$& 10.9973 &$-460.8231$&   28967.9828 \\
   2 &  4.7050 &$-309.0434$&   36560.3049 \\
   3 &$-7.4045$&  451.5522 & $-18626.4002$\\
   4 &$-8.2977$&  910.6755 &$-114291.8885$\\
   5 &         &   71.4580 & $-76110.8714$\\
   6 &         &$-663.8194$&  131495.9200 \\
   7 &         &           &  123526.6033 \\
   8 &         &           &$-111521.6508$\\
\end{tabular}
\end{center}

\caption{Values of $\chi^{(n)}_l$ for $n = 1,2,3$.}

\end{table}

\newpage
\begin{center}
{\bf\Large Figures}
\end{center}

\begin{figure}[h]
\centerline{\epsffile{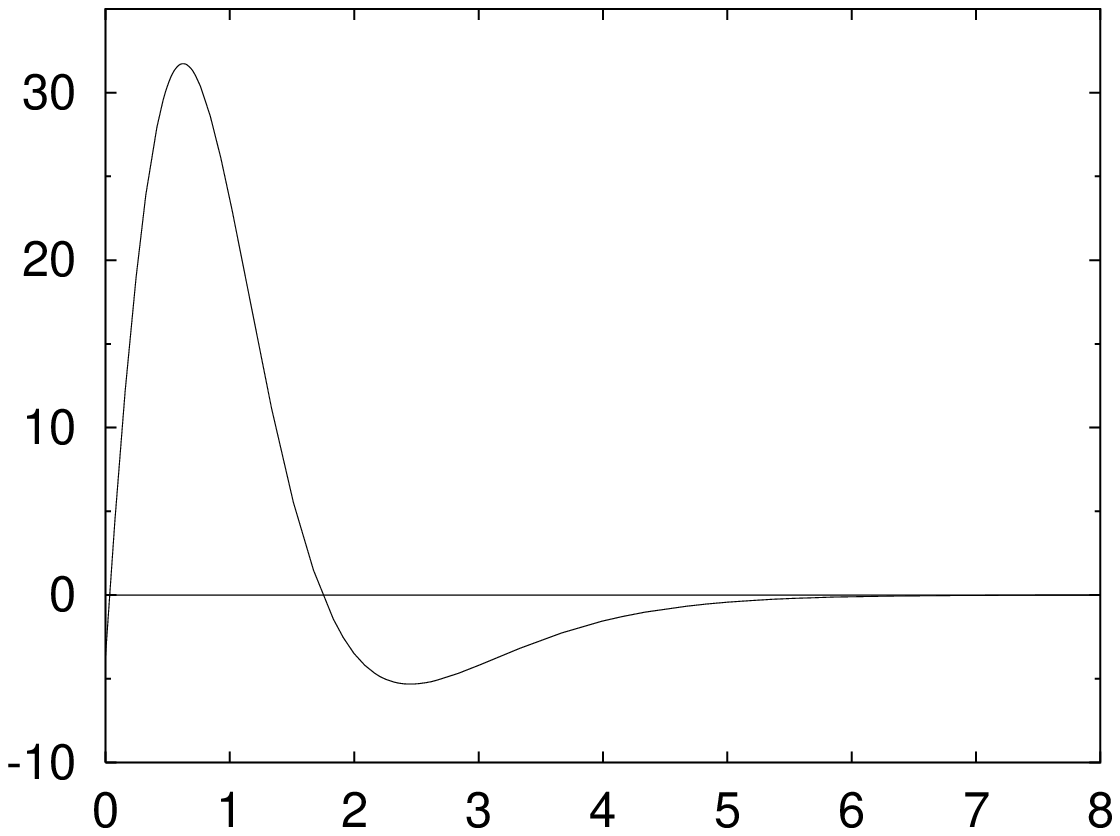}}                                                                                                                     
\caption{Plot of $V_1(x)$.}
\end{figure}

\end{document}